\documentclass{aa}
\usepackage{graphicx}
\usepackage{xcolor}
\usepackage{txfonts}
\usepackage[draft]{hyperref}
\usepackage{placeins}
\usepackage{float}
\usepackage{caption}

\newcommand{\bceph}{$\beta$\,Cep}
\newcommand{\nGaia}{222}
\newcommand{\nlc}{216}
\newcommand{\nmc}{143} 
\newcommand{\nid}{148}
\newcommand{\nmodel}{119}
\newcommand{\nsplit}{34}

\begin{document}

\title{Mode identification and ensemble asteroseismology of \nmodel{} $\beta\,$Cep stars detected by \emph{Gaia} light curves and monitored by TESS
  \thanks{The full Tables \ref{tab:obs} and \ref{tab:tab_model} are only available in electronic form at the CDS via anonymous ftp to cdsarc.u-strasbg.fr (130.79.128.5) or via http://cdsweb.u-strasbg.fr/cgi-bin/qcat?J/A+A/}
  }

   \subtitle{}

   \author{D. J. Fritzewski\inst{1}
       \and M. Vanrespaille\inst{1}
          \and C. Aerts\inst{1,2,3}
          \and Z. Guo\inst{1}
          \and D. Hey\inst{4}
           \and J. De Ridder\inst{1}
          }

   \institute{Institute of Astronomy, KU Leuven, Celestijnenlaan 200D, 3001, Leuven, Belgium\\
   	\email{dario.fritzewski@kuleuven.be}
      \and
      Department of Astrophysics, IMAPP, Radboud University Nijmegen,
      PO Box 9010,
      6500 GL Nijmegen, The Netherlands
      \and
      Max Planck Institut für Astronomie, Königstuhl 17, 69117 Heidelberg, Germany
      \and
      Institute for Astronomy, University of Hawaii, 2680 Woodlawn Drive, Honolulu, HI 96822-1839, USA
   }

   \date{}
   
  \abstract
   {The \emph{Gaia} mission detected many new candidate $\beta$\,Cephei (\bceph{}) pulsators, whose variability classification has since been confirmed from Transiting Exoplanet Survey Satellite (TESS) space photometry of the nominal mission.}
   {We aim to analyse all currently available TESS data for these \bceph{} pulsators, of which 145 were new discoveries, in order to exploit their asteroseismic potential. Although they are of critical importance to improve evolution models of massive stars, \bceph{} stars are under-represented in the current space photometry revolution.}
   {We extracted light curves for \nlc{} stars from the TESS full-frame images and performed frequency analysis by means of pre-whitening. Based on \emph{Gaia} Data Release 3, we deduced stellar properties and compared them to those of known \bceph{} stars from the literature. We developed a methodology to identify the dominant pulsation modes of the \bceph{} stars from the detection of rotationally-split multiplets and \emph{Gaia} and TESS amplitude ratios. We used grid modelling to gain insights into the population of \bceph{} stars.}
   {Combining TESS and \emph{Gaia}, we successfully identified the mode degrees for \nid{} stars in our sample. We find the majority to have a dominant dipole non-radial mode. Many non-radial modes show splittings in their TESS frequency spectra, which we use to calculate their envelope rotation, spin parameter, and the level of differential envelope-to-surface rotation. For the latter, we find an upper limit of about 3. We also provide relative frequency asymmetries within the multiplets, ranging from -0.15 to 0.15 with most being positive. Based on grid modelling, we provide mass, convective core mass, and age distributions for \nmodel{} stars.}
   {Our sample enables asteroseismology of \bceph{} pulsators as a population. Our study prepares for future detailed modelling based on individual frequencies of identified modes leading towards a better understanding of these massive pulsators.}

   \keywords{Asteroseismology -- Stars: oscillations (including pulsations) -- Stars: massive --
   Stars: interiors -- Stars: evolution -- Stars: rotation   }

   \titlerunning{Asteroseismic analysis of \emph{Gaia}-TESS $\beta$\,Cep stars}
\authorrunning{Fritzewski et al.}

   \maketitle
%

\section{Introduction}
\label{sec:intro}
$\beta$\,Cephei (\bceph{}) stars are core-hydrogen burning massive stars ($8\lesssim M/M_\sun \lesssim 25$) pulsating in lower-order pressure (p-) or gravity (g-) modes. Despite being known for more than a century \citep{Frost1902, Frost1906, Guthnick1913} their mode excitation mechanism was only explained 30 years ago by \cite{Moskalik1992}, \citet{Gautschy1993}, and \cite{1993MNRAS.262..204D}. The pulsations are driven by the heat mechanism (also known as $\kappa$ mechanism) active in the partial ionisation zone of iron-like elements in the outer envelope. However, while mode excitation in \bceph{} stars is roughly understood, it remains incomplete. Indeed, more modes are detected in data than predicted by theory as revealed in some of the prototypical class members \citep{Pamyatnykh2004,Jagoda2010,Jagoda2013}. Their observations show g-modes while they are predicted to be stable modes (i.e.\ not excited). \citet{Moravveji2016-OP} provided a summary of observed modes in \bceph{} stars and illustrated the impact of higher opacities in the driving zone. Additional physical mechanisms may also be active in the interiors of such stars, while being absent in the current excitation theory. Microscopic atomic diffusion, notably radiative levitation, was recently shown to be one such missing ingredient \citep{Rehm2024}. 

In addition to shortcomings in the understanding of the mode excitation, mechanisms of mode selection are even less understood. It still remains unclear whether the mode selection is similar over evolutionary time scales and why certain modes are excited to observable amplitudes while others are not. The first step to understand the selection of modes is the identification of their geometrical configuration which remains challenging in \bceph{} stars, given that their mode frequencies are not in an asymptotic regime. The geometry of the low-order modes of most \bceph{} stars is well described by a spherical harmonic, characterised by $(l,m)$ representing the mode degree and azimuthal order. Identification of $(l,m)$ is a prerequisite for asteroseismic modelling \citep{Salmon2022}. Our current work aims to improve this aspect of mode identification for this class of massive pulsators, by relying on space photometry. Before diving into this topic, it is useful to recall the history and basic properties of \bceph{} star pulsations.

The interest in \bceph{} stars came in waves of discoveries. The earliest work focused on the period determination from radial velocity time series of the very few known stars similar to \bceph{}\footnote{Due to their differences in beating structure in the light and radial-velocity curves, some \bceph{} stars were initially classified as $\beta$ Canis Majoris stars \citep{vanHoof1957}. Eventually, \bceph{} stars became the accepted name of this class of variable stars.} and the lack of understanding of the instability mechanism \citep[e.g. review by][]{Struve1952}. In the following years, more \bceph{} stars were discovered and the efforts to understand the pulsation mechanism were continued, although the physical origin was still elusive \citep{Lesh1978}. The discovery and explanation of the mode excitation mechanism  \citep[][for a summary]{Pamyatnykh1999} sparked new interest in \bceph{} stars. \cite{Aerts1992, Aerts1994b} conducted successful mode identification from time series spectroscopy for a few bright class members, while \cite{Heynderickx1994} identified the degree of the dominant pulsation modes for a substantial number of class members from multi-colour photometry. While both methods worked, only a limited number of modes per star could be identified -- often just one or two. Recently, \cite{Shitrit2024} showed that ground-based multi-colour photometry is still a viable way to achieve mode identification for asteroseismology of \bceph{} stars, complementary to single-band modern space-based photometry.

Based on the numerous discovered \bceph{} stars, \cite{Stankov2005} composed a catalogue containing 93 genuine and 77 candidate \bceph{} stars. Yet, the majority of them lack a solid identification of their pulsation modes. The \bceph{} stars discovered from large ground-based surveys \citep[e.g.][]{Pigulski2008b,LabadieBartz2020} or dedicated observations of a few selected young open clusters \citep{Saesen2013,Mozdzierski2019} suffer from the same limitation in terms of seismic probing power: too few or no pulsation modes have been identified.

Inferences of a \bceph{} star interior were achieved thanks to the 21 year-long photometric monitoring of HD\,129929 \citep{Aerts2003,Aerts2004,Dupret2004}. These three dedicated studies kick-started \bceph{} asteroseismology from the identification of six modes: a radial mode, a rotationally split p-mode triplet and two components of a g-mode quintuplet. It led to the first seismic probing of the internal rotation of a main-sequence star other than the Sun, with the bottom of the envelope rotating about four-times faster than the outer envelope. Inspired by this result, in coordinated efforts months-long multi-site campaigns involving tens of astronomers worldwide were organised. As such, four of the brightest \bceph{} stars were scrutinised seismically from combined multi-colour photometric and spectroscopic time-series data. These stars are still the best modelled \bceph{} stars today, thanks to the unambiguous identification of their modes: $\nu\,$Eri \citep{Aerts2004-nuEri,DeRidder2004,Pamyatnykh2004}, $\theta\,$Oph \citep{Handler2005,Briquet2005,Briquet2007}, 12\,Lac \citep{Handler2006,Dziembowski2008,Desmet2009}, and V2052\,Oph \citep{Handler2012,Briquet2012}. The seismic estimates of the internal rotation, age (in terms of core hydrogen mass fraction), and core boundary mixing are summarised in \citet{Bowman2020}.

With the new era of space photometry, asteroseismology has gained a treasure trove of high-quality, high-cadence data for many pulsating stars \citep[see][for recent reviews]{Aerts2021-RMP,Kurtz2022}. For \bceph{} stars in particular, the earliest space-based observations with the Microvariability and Oscillations of Stars (MOST) satellite revealed multi-periodic behaviour of $\delta$\,Ceti, which was thought to be a mono-periodic radial pulsator until then \citep{Aerts2006b}. The CoRoT mission advanced our knowledge of only three more \bceph{} stars  \citep{Belkacem2009,Degroote2009,Briquet2011,Aerts2019} and the \emph{Kepler} space satellite did not reveal  any additional \bceph{} pulsators with identified modes \citep{Balona2011, Lehmann2011}. This led to a paucity of research on these stars with space-based time series photometry, although asteroseismology as a whole was revolutionised by the \emph{Kepler} mission.

Already with the first two sectors of the Transiting Exoplanet Survey Satellite \citep[TESS,][]{Ricker2015} mission it became obvious that massive star asteroseismology entered a new era \citep{Bowman2019, Pedersen2019, Burssens2020, Balona2020, Sharma2022}. \cite{Shi2024} provide a catalogue of all \bceph{} stars observed by TESS with 2\,min-cadence data. \cite{Southworth2022} identified eight \bceph{} pulsators in eclipsing binaries, while \cite{Eze2024} recently expanded this list to 78 systems. However, to best study these stars long-baseline time series observations are needed to resolve the anticipated close-by frequencies and identify the pulsation modes.

\cite{Burssens2023} analysed the 352\,d-long light curve of the bright \bceph{} star HD\,192575 located in the TESS Northern continuous viewing zone and were able to model its structure parameters with high precision as the fifth $\beta\,$Cep star with tight seismic constraints. They also estimated its internal rotation profile from forward modelling but encountered large uncertainty due to the use of only three rotationally split multiplets. Vanlaer et al. (in prep.) meanwhile analysed the 5-year TESS light curve (with two year-long gaps) and resolved some of the ambiguity in the mode identification, allowing the authors to perform a rotation inversion. 

Next to CoRoT, \textit{Kepler}, and TESS, the \emph{Gaia} mission has proven to be a highly successful gateway to discover pulsating stars. Despite not being designed for asteroseismology, the recurring visits to millions of stars in a quasi-random pattern allowed for the discovery of >100,000 new main-sequence pulsators \citep{DeRidder2023}. \cite{Hey2024} revisited these stars and analysed their first two years of TESS photometry, confirming the main pulsation frequency and classification for more than 80\% of these  \emph{Gaia}-identified pulsators. Given these findings, we are encouraged to focus on the \nGaia{} \bceph{} stars in the sample of \cite{DeRidder2023} and to follow up on the work of \cite{Hey2024} by considering all available TESS data. 

Here, we use TESS light curves with a considerably longer time base than \citet{Hey2024}, thus enabling improved frequency resolution. Our aim is to search for resolved low-(radial) order pressure and gravity modes and rotational splittings in the \nGaia{} \emph{Gaia}-detected \bceph{} stars, opening pathways to mode identification and seismic modelling from combined TESS photometry and \emph{Gaia} data, in a homogeneous way for an as large as possible ensemble of  \bceph\  stars with publicly available light curves assembled with both satellites.

\section{Data reduction}

We selected all \nGaia{} variable stars classified by \cite{DeRidder2023} as \bceph{} stars as our initial sample. Since our work focuses on the synergy between the TESS and \emph{Gaia} DR3 light curves, we only included these stars. Yet, we emphasise that the methods developed in this work could potentially be applied to TESS observations of known \bceph{} stars with \emph{Gaia} time series photometry. To prepare for the data reduction, we obtained for each star on our target list its position from the \emph{Gaia} archive and cross-matched it with the TESS input catalogue \citep{Stassun2019, Paegert2021} to obtain its TIC ID. Out of our initial sample of \nGaia{} stars, \nlc{} have been observed by TESS and we can obtain time series photometry for them. 

\subsection{TESS data reduction and light curve detrending}
\label{sec:reduction}

We employed the TESS data reduction pipeline \texttt{tessutils} \citep{tessutils} which enables the end-to-end creation of detrended light curves from TESS full-frame images (FFI) and is optimised for asteroseismic analysis. We briefly recall its work-flow here. For every TIC number, \texttt{tessutils} downloads cut-outs from the FFIs using \texttt{tesscut} \citep{2019ascl.soft05007B}. In the present work we chose a cut-out size of 25\,px by 25\,px. Subsequently, \texttt{tessutils} searches for the optimal aperture for each sector while trying to keep the contamination from neighbouring stars below 1\,\%. Afterwards, it extracts the light curve of that TESS sector as the total flux within the chosen aperture. The number of sectors in the light curves range from one to seven with a median of four visits by TESS over the current mission lifetime spanning about five years.

Fifteen of our stars are located in very crowded fields near the Galactic plane or in open clusters. Since the targeted \bceph{} stars are typically among the brightest stars in the field they provide the main flux contribution. To reduce the influence of neighbouring stars, we set the aperture for these stars to a single pixel centred on the target's position (see Appendix\,\ref{app:photometry} for an example). With this approach, we were able to obtain light curves for stars that would otherwise have (partially) been missed.

After aperture photometry, \texttt{tessutils} detrends the extracted light curves per sector with an automated principal component analysis with up to seven components to remove long-term trends in the data. This is justified for the asteroseismic analysis of \bceph{} stars as we are interested in variability with periodicities of typically several hours for the individual modes. The normalised light curves were binned to the longest cadence of the TESS photometry, that is 30\,min. This guarantees a homogeneous treatment of all TESS FFI data of the sample stars detected by {\it Gaia}, while it leads to excellent agreement among the dominant frequencies detected in the TESS and {\it Gaia\/} DR3 light curves, as we will show below. We tested that restricting to the higher-cadence TESS FFI data does not change the results for our \bceph\ population study. Finally, the data of all sectors were stitched to a single long-term light curve per star. However, despite the initial PCA detrending, we still noticed significant instrumental trends in many light curves. Therefore, we removed those parts of the light curves where instrumental effects were dominant and detrended the remaining light curve. The full description of the process and an example are given in Appendix \ref{app:photometry}. 

\subsection{Frequency extraction from pre-whitening}

For the extraction of frequencies, we used a pre-whitening procedure. This method consists in identifying the dominant signal frequency in the Fourier transform of the light curve, finding the amplitude and phase of it and subtracting the matching sinusoid from the light curve to construct a residual light curve. Subsequently, the next strongest frequency is identified in the residuals, a sinusoid is built, and subtracted, etc. This procedure is followed until a user-specified stop criterion is reached \citep[see, e.g. Chapter\,5 in][for details]{Aerts2010}.

In order to extract all significant frequencies up to the Nyquist frequency of 24\,d$^{-1}$, we used a customised version of the pre-whitening module of the publicly available frequency analysis code \texttt{STAR SHADOW} \citep{IJspeert2024}. While the reader is referred to that paper for details on the methodology, we point out that \texttt{STAR SHADOW}'s pre-whitening procedure is computationally efficient  because it does not rely on time-consuming non-linear regression for all extracted signals simultaneously. Rather, it only resorts to non-linear regression for those signals that strongly affect each other. On top of that, \texttt{STAR SHADOW} is well suited to analyse blended frequency peaks in Fourier spectra, which are quite common in our set of light curves due to the limited coverage of the baseline. 

In each pre-whitening step, \texttt{STAR SHADOW} demands the Bayesian Information criterion (BIC) \citep{Schwarz1978} decreases by at least 2 to accept a signal as significant. In order to discourage over-extraction of frequencies from the high-quality space photometry, we increased \texttt{STAR SHADOW}'s BIC-threshold in each pre-whitening step to 10. Additionally, we employed a hybrid method for the selection of the next signal to be pre-whitened. Herein, the pre-whitening initially uses amplitude hinting \citep[e.g.][]{VanBeeck2021} and swaps to signal-to-noise ratio (SNR) hinting once a frequency is rejected, where the noise level is computed over a range of 1\,d$^{-1}$ in the Fourier spectrum. If another potential signal is rejected, the pre-whitening is terminated. For the final significance assessment of each signal, \texttt{STAR SHADOW} demands the signal's SNR to be greater than the threshold defined in Eq.\,6 of \citet{Baran2021}. We increased the SNR-threshold by 0.25 for light curves with long gaps as recommended by \citet{Baran2021}.

\subsection{Extracted frequencies}

We found between four and 36 (median 13) independent signals in each of our TESS light curves. These values already show that \bceph{} stars pulsate with few modes. It also reveals that our detrending removed most of the instrumental effects, that would contaminate the Fourier domain and lead to more extracted yet spurious frequencies. 

\begin{figure*}
    \includegraphics[width=\textwidth]{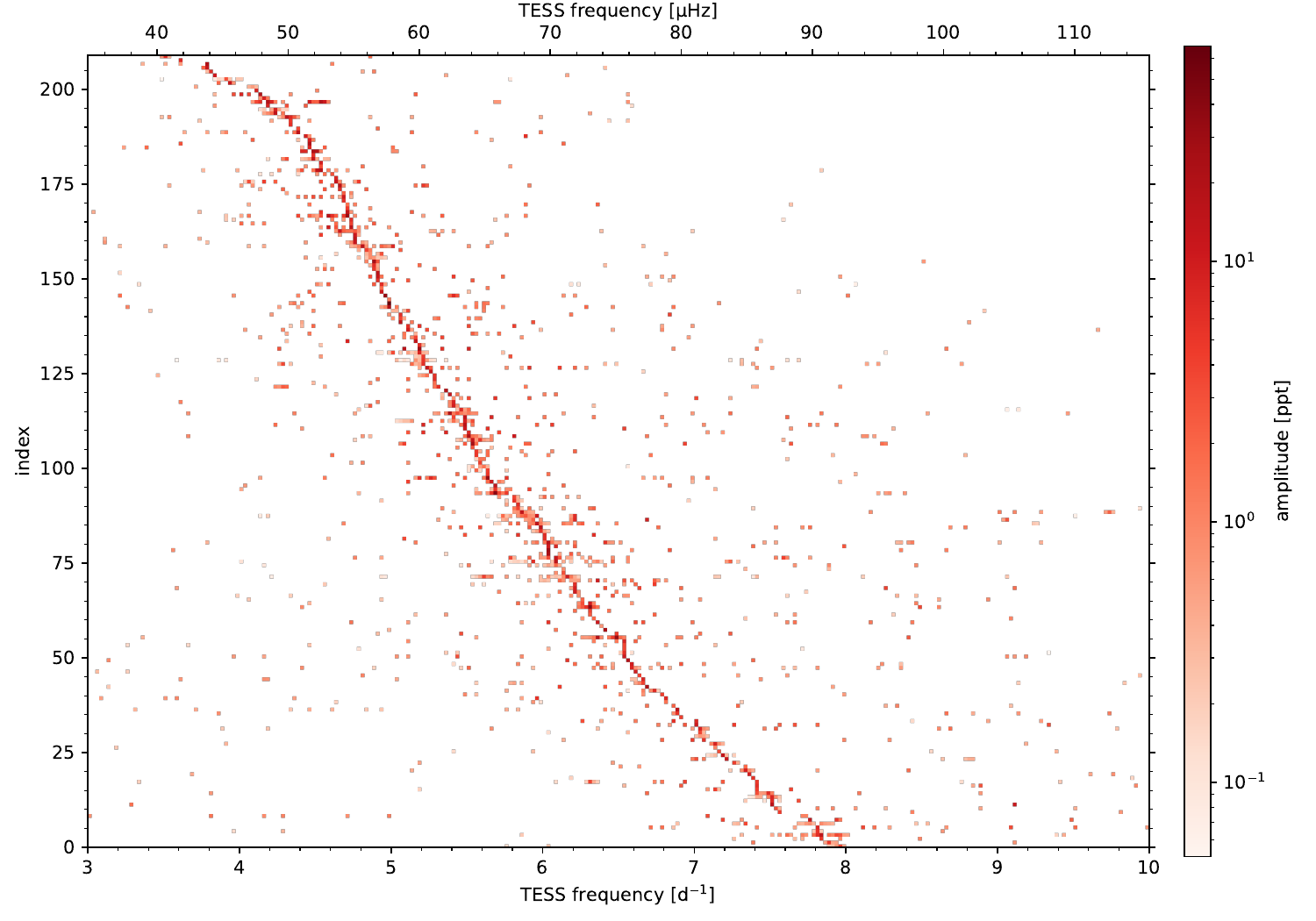}
    \caption{Stacked representation of the extracted independent, significant TESS frequencies for 212 of the \emph{Gaia}-detected \bceph{} stars, sorted by the TESS frequency with the largest amplitude. The colour-scale indicates the amplitude of each detected signal on a logarithmic scale. This figure includes only the stars whose main frequency occurs in the plotted \bceph{} range. Three stars have lower main frequencies and turn out to be SPB stars or \bceph{} and SPB hybrid pulsators with dominant high-order g-modes, or are dominated by other variability than pulsations.}
    \label{fig:stacked_pg}
\end{figure*}

To analyse a large number of pulsators, stacking and sorting the periodograms can reveal hidden structures \citep[e.g.][]{Gilliland2010PASP,Michel2017,GangLi2020,Hey2024,Read2024}. Figure~\ref{fig:stacked_pg} shows the extracted significant frequencies from the TESS photometry for our \emph{Gaia}-observed sample sorted by the frequency of highest amplitude. Since \bceph{} stars show typically few excited low-order modes that do not occur in an asymptotic regime of high or low frequencies,  no obvious structure emerges from this diagram. Nevertheless, this visual representation of the sample shows the range of detected significant frequencies. We observe that the distribution flattens towards low frequencies while declining more abruptly  towards high frequencies. This is fully in line with instability computations of low-order modes in young and evolved \bceph{}\ stars \citep{Pamyatnykh1999}. However, the sharp cut-off near 8\,d$^{-1}$  may also be partly due to the natural transition towards the $\delta$\,Sct star frequency regime adopted in the classification algorithms \citep{DeRidder2023}. Further, it becomes obvious from the figure that the observed significant frequencies for a given star depend on the position of its dominant frequency. 

\begin{figure}
    \includegraphics[width=\columnwidth]{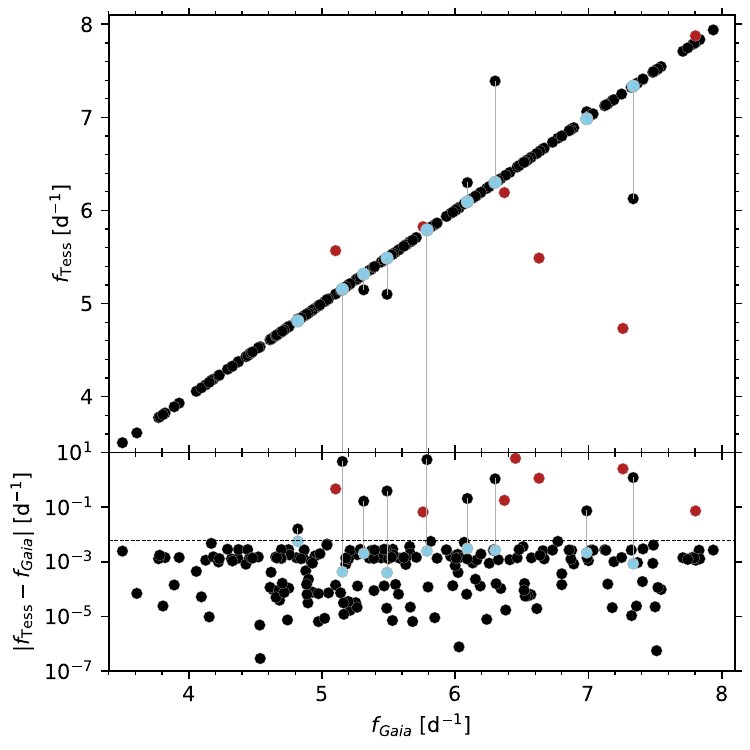}
    \caption{Comparison between the highest amplitude \emph{Gaia} and TESS frequencies for the 215 stars with significant TESS frequencies. Top: Dominant TESS frequency $f_\mathrm{TESS}$ against the \emph{Gaia} derived frequency $f_\emph{Gaia}$. Black symbols indicate stars for which we found a match in frequency, blue symbols show stars matching a frequency other than the dominant signal, and red symbols indicate the seven stars for which the \emph{Gaia} frequency does not match any frequency in the TESS data. Lines connect the offset frequencies to their matched counterpart. For three stars the main frequency is below the plotting range. Bottom: Difference between the \emph{Gaia} and TESS frequencies. The black dashed line indicates our chosen threshold for matching signals ($\Delta f < 0.006$\,d$^{-1}$).}
    \label{fig:f_compare}
\end{figure}

\subsection{Comparison to \emph{Gaia} frequencies}

\label{sec:freqcomp}
Figure~\ref{fig:f_compare} compares the main frequency from our TESS analysis with the frequencies derived from \emph{Gaia} DR3 by \cite{DeRidder2023}. As already found by \cite{Hey2024}, most of the pulsation frequencies agree within a tolerance of $\Delta f < 0.006$\,d$^{-1}$, highlighting the potential of the \emph{Gaia} time series photometry to discover non-radial pulsators at a level of 4\,mmag or above. For the twelve stars without a matching TESS primary frequency, we searched through other signals in order of decreasing amplitudes in the TESS data and compared them to the \emph{Gaia} frequency. Once a match within our permitted tolerance was found we stopped the search. For seven stars\footnote{The seventh star is not shown in Fig.~\ref{fig:f_compare} because its \emph{Gaia} frequency $f_\emph{Gaia}=21$\,d$^{-1}$ is outside the \bceph{} frequency range. Yet this star might still be a \bceph{} pulsator as the dominant TESS frequency $f_1=6.6$\,d$^{-1}$.}, we found no matching signal in the TESS data. The median difference between the TESS and \emph{Gaia} matched frequencies is $\Delta \tilde f = 0.0013$\,d$^{-1}$.

We limited the current analysis to the primary frequency detected with \emph{Gaia}. Although we possess measurements of the secondary frequency from the \emph{Gaia} DR3 time series photometry, we noticed that these frequencies match our extracted frequencies only for a minority of cases. Often, we find $f_\emph{Gaia}>10$\,d$^{-1}$, whereas we rarely extracted any significant frequency in this range from TESS. It was already discussed by \citet{DeRidder2023} that the \emph{Gaia} DR3 data suffer from instrumental effects at mmag level for the secondary frequencies, which may be connected to multiples of the satellite frequency of 4\,d$^{-1}$. Hence for the \bceph{} population study in our work, using the secondary frequency from \emph{Gaia} is not informative. We note that detailed work on certain stars might be able to derive additional constraints for the pulsation mode of the secondary frequency. Further, longer time series data delivered by future \emph{Gaia} data releases might provide more robust secondary frequencies in agreement with TESS.

\section{Stellar properties}
\subsection{\emph{Gaia} derived properties}

We aim to analyse the \bceph{} stars in our sample in a homogeneous way and hence chose to use stellar parameters from \emph{Gaia} DR3. Of particular interest are the effective temperature ($T_\mathrm{eff}$) and the luminosity ($\log L/L_\sun$). \cite{Shi2023} show that the \emph{Gaia} $T_\mathrm{eff}$ derived by the Extended Stellar Parametrizer for Hot Stars \citep[ESP-HS, ][]{Fouesneau2023} is reliable for pulsating hot main sequence stars. We follow their example and use these values as the basis for our analysis. Further, we extracted positions, parallaxes, and $G$ band magnitudes for every target. Where available, we also added stellar parameters from ESP-HS and the General Stellar Parametrizer from Photometry (GSPPHOT). Unfortunately, the \emph{Gaia} ESP-HS parameters are only available for 190 of the \nGaia{} stars. Hence, we are not able to place all the observed stars in the Hertzsprung-Russell diagrams (HRD). Further, this limits the sample of stars that we can model asteroseismically.

From the \emph{Gaia} data, we find the majority of our sample stars to be in the magnitude range $G=8.5-13.5$\,mag (c.f. Table~\ref{tab:obs}). Nine stars, including the well studied stars BW Vul and 12\,Lac, are brighter than $G=8.5$\,mag while no star is fainter than $G=13.6$\,mag. To derive the stellar luminosity, we require a bolometric correction. The \emph{Gaia} bolometric correction \citep{Creevey2023b} is only valid up to 10\,000\,K and \bceph{} stars are hotter with $15\,000\,K \le T_\mathrm{eff}\le 25\,000\,K$. \cite{Pedersen2020} derived bolometric correction functions for massive stars including the range of \bceph{} stars. We used these and calculated the luminosities $\log L/L_\sun$ based on the \emph{Gaia} DR3 absolute magnitudes taking into account the reddening based on the ESP-HS estimate and the distance as determined by the GSPPHOT pipeline for OB stars, the general GSPPHOT values, or the inverse of the parallax depending on the availability.

The uncertainties of the parameters in the astrophysical parameters table are often underestimated \citep{Fouesneau2023}. In particular the effective temperature has an uncertainty much larger than the typically given 500\,K ($\sim2$\,\%). For the purpose of this work, we assumed a conservative uncertainty of 10\,\% ($\Delta \log T_\mathrm{eff}= 0.043$\,dex) for all stars in the sample, which is a reasonable estimate given the comparison in Fig.~15 of \cite{Fouesneau2023}. The uncertainty of the luminosity is strongly dependent on the effective temperature through its use in the bolometric correction.

An estimate of the projected surface rotation frequency is useful when searching for rotationally split multiplets and examining differential rotation. The \emph{Gaia} ESP-HS parameters also include the projected surface rotational velocity $v\sin i$ calculated from the line broadening of the RVS spectra under the assumption that rotation determines the line broadening, which is a valid assumption \citep[cf.,][]{Aerts2023}. Together with the luminosity and effective temperature, which gives an estimate of the stellar radius, we can calculate the projected surface rotation rate of the stars. This parameter, $f_\mathrm{rot}\sin i$, is available for 167 stars. Although it is not very precise with a typical uncertainty $\sim 40\,\%$, it still proves useful in our subsequent analysis. 

\subsection{\emph{Gaia} data of known \bceph{} stars}

To compare the properties of the \emph{Gaia}-detected \bceph{} stars with these of known \bceph{} stars, we extracted their \emph{Gaia} DR3 data as well. As the literature sample, we use the genuine \bceph{} stars from the catalogues of \cite{Stankov2005}, \cite{Pigulski2008b}, \cite{LabadieBartz2020}, \cite{Shi2024}, and \cite{Eze2024}. For each of the catalogues, we derived the same \emph{Gaia} properties as described above. From the four catalogues in the literature, we find 582 stars of which 408 are classified as genuine \bceph{} stars. Among with the \nGaia{} \bceph{} candidates in our sample -- 145 are not listed in any of the literature catalogues -- the total number of \bceph{} stars in the combined sample is 553 stars.

\section{Population of \bceph{} stars}
\begin{figure*}
    \includegraphics[width=\textwidth]{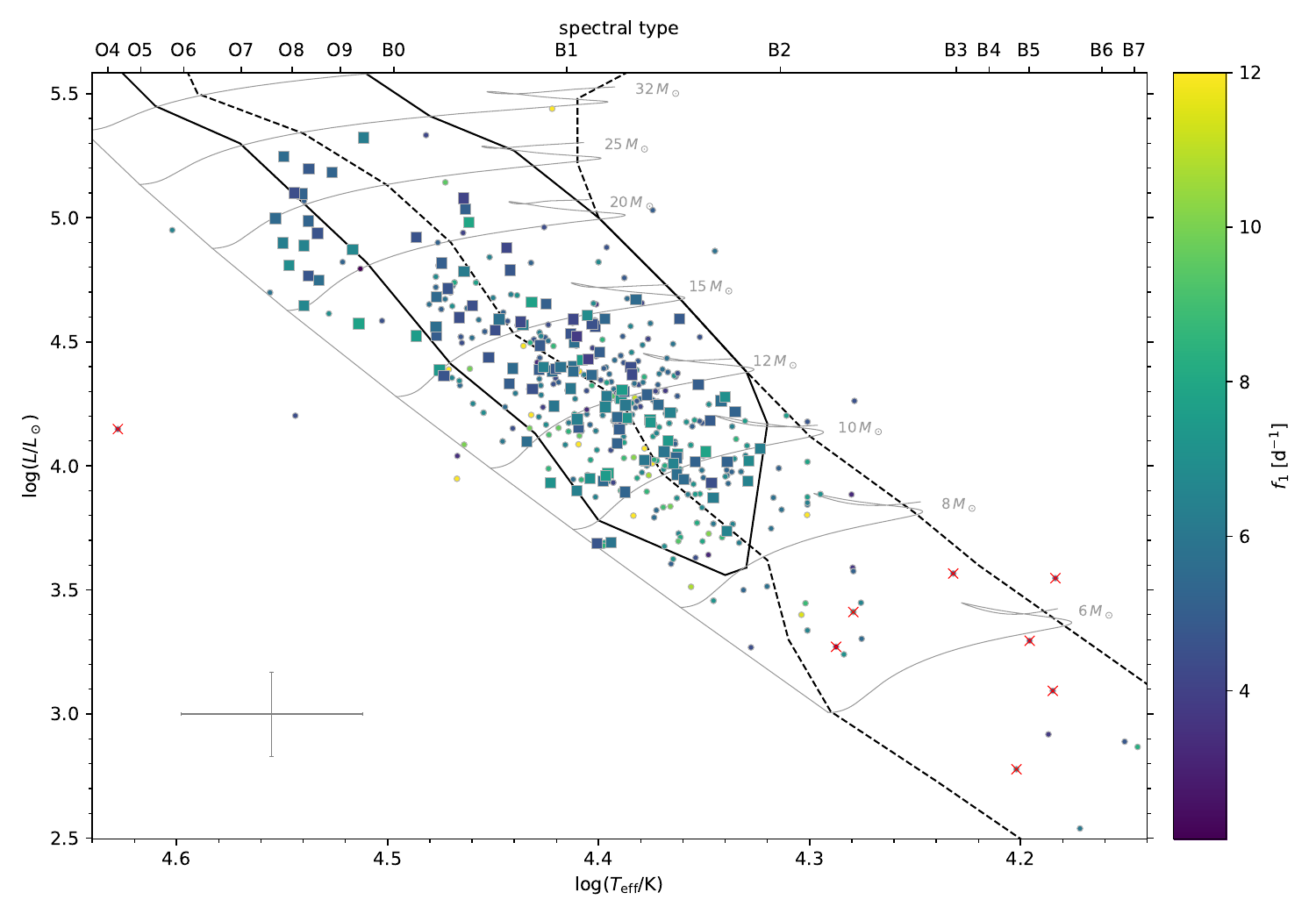}
    \caption{Hertzsprung-Russell diagram based on \emph{Gaia} data of all 321 \bceph{} stars from the literature (circles) and 121 new stars of our \emph{Gaia}-informed sample (squares, new detections). The typical uncertainty is indicated in the lower left. Black outlines show the instability regions for p-modes (solid, $\beta$\,Cep, $l=0-2$), and g-modes (dashed, SPB, $l=1-2$) based on the calculations from \cite{Burssens2020}. Main sequence evolutionary tracks of \cite{Burssens2020} are shown in grey. The colour-code gives the main pulsation frequency of the stars cut-off at 12\,d$^{-1}$. Stars marked with a red cross do not exhibit \bceph{} variability in TESS. The top axis marks the spectral type.
    }
    \label{fig:HRD_all}
\end{figure*}

Before analysing our \emph{Gaia}-detected sample in detail, we present the overall properties of the ensemble. Figure~\ref{fig:HRD_all} shows all 442 stars (out of 553) with adequate \emph{Gaia} ESP-HS parameters in an HRD. Most of the stars in our sample fall inside the instability strips for OB-type pulsators calculated by \cite{Burssens2020}. These are based on one particular choice of input physics (relying on exponential core overshoot with parameter $f_\mathrm{ov}=0.02$) at solar metallicity, so we expect some stars to occur outside that region, as also stressed by \cite{Aerts2024}, \cite{Rehm2024}, and \cite{Hey2024}.

Several stars from the literature occur below the \bceph{} instability region. We checked these stars with publicly available light curves from the TESS Quick Look Pipeline \citep[QLP,][]{Huang2020} using a classification similar to our main sample and outlined below in Sect.~\ref{sec:classification}. We find the majority of stars close to the instability region to be \bceph{} stars pointing either to imprecisions of the stellar parameters or a larger extend of the instability region as found for other pulsators in this area of the HRD too \citep{DeRidder2023,Mombarg2024c}. Stars further away from the instability region are either short-period eclipsing binary stars or SPB stars. The stars\footnote{TIC 17449899 (V2187\,Cyg), TIC 23548300 (HD\,55708), TIC 129364361 (NGC\,2483\,2), TIC 130826839 (HD\,108769), TIC 155571411 (HD\,69016), TIC 312931751 (HD\,116827), TIC 337165095 (HD\,42401), and TIC 427406031 (HD\,143605)} with obvious non-\bceph{} variability (and available light curves) are marked in Fig.~\ref{fig:HRD_all} with red crosses. All the new \emph{Gaia}-derived \bceph{} stars fall inside or close to the \bceph{} instability strip and can hence safely be assumed to be genuine \bceph{} stars from their HRD position and TESS variability characterisation (see below).

At $\log T_\mathrm{eff}\approx 4.5$, we find a gap between the bulk of the \bceph{} stars and a smaller group mostly containing stars from our new sample. With only the \emph{Gaia} ESP-HS data, we cannot be certain whether this gap is astrophysical or introduced by the intricate parameter estimation procedure. It is present in the entire ESP-HS data set. We note the position of the gap overlaps with the transition from spectral type O to B \citep{Pecaut2013}. In case the gap is physical the stars on its hotter side would be of great interest for follow-up observations and detailed asteroseismic modelling as these stars might be products of binary evolution \citep{Fabry2022,Fabry2023}.

\subsection{Detailed analysis of the \emph{Gaia} sample}
\label{sec:classification}

\begin{figure*}
    \includegraphics[width = \textwidth]{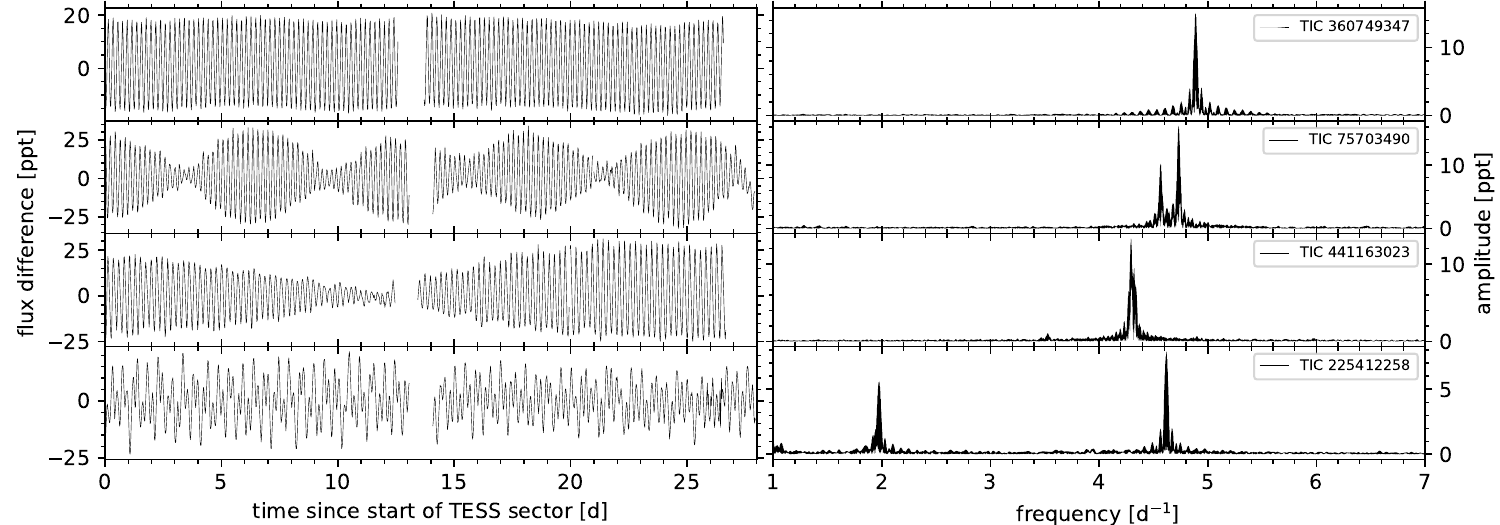}
    \caption{Examples of single sector TESS light curves (left) of \bceph{} variables of the different light curve sub-classes and their associated periodograms (right). The top light curve is a typical mono-periodic pulsator with a constant pulsation amplitude and a single peak in the frequency spectrum. The second and third panels show multi-periodic pulsators with different lengths of their beating patterns. The bottom panel shows a typical \bceph{} and SPB hybrid pulsator.}
    \label{fig:LCs}
\end{figure*}

After verifying that all of our sample stars with \emph{Gaia} ESP-HS parameters fall indeed into the parameter range of \bceph{} stars, we analyse their TESS light curves in more detail. In a visual inspection of the light curves and their Fourier transforms, three main sub-classes emerge (Fig.~\ref{fig:LCs}). The first class comprises of mono-periodic pulsators which show one dominant coherent oscillation in the entire light curve. The periodograms of these stars often reveal they pulsate in multiple frequencies, but their secondary frequencies have small amplitudes compared to the dominant mode. Clearly multi-periodic pulsators constitute the second class. Their light curves exhibit modulations due to mode beating with periods between one and tens of days, as already known from ground-based multisite campaigns prior to the space asteroseismology era. In some cases, the beating pattern is significantly longer than one TESS sector. Hence, multiple TESS sectors are needed to resolve the close frequency peaks. We distinguish between \bceph{} multi-periodic light curves (due to the presence of multiple low-order p- or g-modes) and  hybrid pulsators with SPB components as a third sub-class because their secondary signals have a much smaller frequency, resulting in a differently shaped light curve. (Fig.~\ref{fig:LCs}). We manually classified all \nlc{} light curves based on their shape and the content of their periodogram into these three main classes. 

In addition to the manual classification, we quantified the light curve shapes by calculating two measures of the amplitude variability. Firstly, we use the length of the main beating pattern given by
\begin{equation}
    P_\mathrm{beat}= \left|f_1-f_2\right|^{-1}.
\end{equation}
Secondly, we calculate the strength of amplitude variability
\begin{equation}
    A_\mathrm{var}=A_{90}/A_{95} \cdot 0.95/0.90,
\end{equation}
in which $A_{95}$ ($A_{90}$) is the difference between the 95th (90th) and 5th (10th) percentile of the light curve. The percentile differences ($A_{95}$) are known to robustly estimate stellar variability in space photometry \citep{Basri2011}. We expand on this idea and use the ratios of the percentile differences. By normalising this ratio, a pure sine variability (i.e. a mono-periodic pulsator) reaches unity, while multiperiodic stars can be found at lower values. We used the full light curve itself to compute $A_\mathrm{var}$ rather than the amplitudes of the extracted frequencies because it includes all variability beyond the two main frequencies in this measure.

\begin{figure}
    \includegraphics[width = \columnwidth]{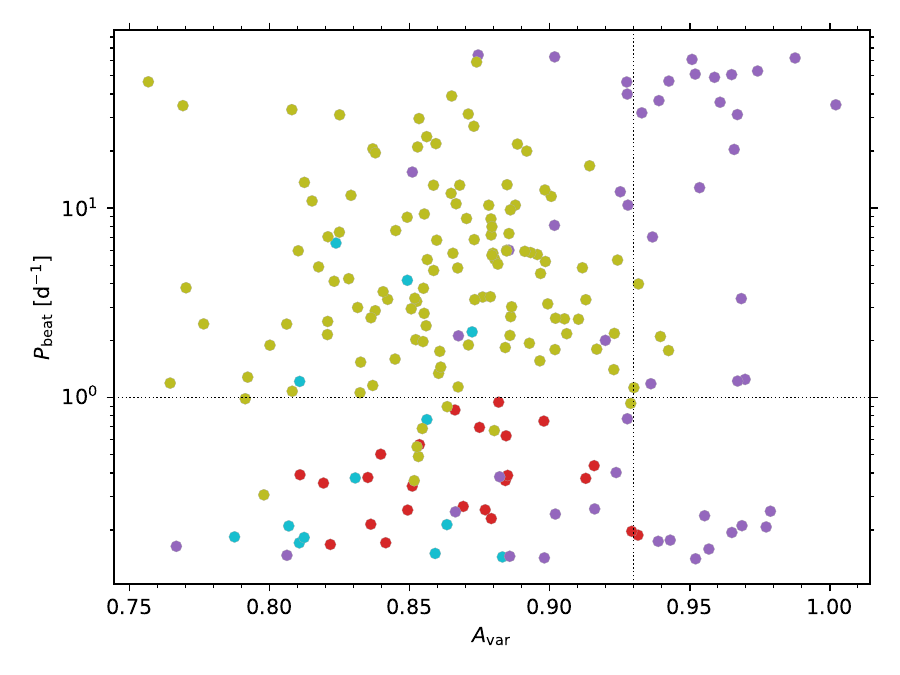}
    \caption{Beating period of the main two frequencies ($P_\mathrm{beat}$) against the normalised percentile amplitude ratio ($A_\mathrm{var}$). Light curves manually classified as mono-periodic are shown in purple, multi-periodic in olive, hybrids in red and other cases in cyan. The dashed lines indicate the approximate dividing lines of the groups of pulsating stars. Uncertainties on the beating period are within the symbol sizes ($A_\mathrm{var}$ has no formal uncertainty due to its definition).}
    \label{fig:amp_classes}
\end{figure}

Figure~\ref{fig:amp_classes} shows that we can clearly separate the above defined groups based on $A_\mathrm{var}$ and the beating period. The initial manual classifications are mostly confirmed. Mono-periodic stars fall towards the right. Multi-periodic pulsators populate the upper-left because they are not close to a single sinusoid, yet have a long beating pattern, while hybrids lie near the lower-left because their primary and secondary frequencies differ greatly. Some stars fall into regimes in Fig.~\ref{fig:amp_classes} that are not associated with their manual classification. Among the mono-periodic stars, several stars display additional rotational variability that modulated the light curve without creating a beating pattern, a feature that cannot be accounted for with the current parameters. Some stars have long beating patterns with low amplitudes which makes them appear mono-periodic. The grouping of mono-periodic stars at long beating periods indicate that these stars are indeed not truly mono-periodic but exhibit multi-periodicity with very small frequency differences, a well-known phenomenon of some classical \bceph{} stars (e.g. HD 129929, $\beta$\,CMa). Given the few differences between our manual classification and the regimes in Fig. \ref{fig:amp_classes}, we provide $A_\mathrm{var}$ and $P_\mathrm{beat}$ in Table~\ref{tab:obs} to enable quantified distinctions of the different groups.

Based on the manual classification, we find 49 (almost) mono-periodic, 126 multi-periodic, and 24 hybrid pulsators. The remaining 17 light curves show variability that does not fall into the three classes. Using $A_\mathrm{var}$ and $P_\mathrm{beat}$ the above introduced quantification for the number of multi-periodic stars stays constant at 124, while we find fewer true mono-periodic stars (33) and more hybrid pulsators (42). \cite{Stankov2005} classified the stars in their catalogue into similar categories and found a fraction 40\,\% of mono-periodic pulsators. However, they expected this fraction to be lower due to unobserved low amplitude pulsations in the ground-based light curves, which have much higher noise level than the TESS space photometry. Our fraction of 24\,\% (16.5\,\% when using the classification in Fig. \ref{fig:amp_classes}) of pulsators with one very dominant oscillation mode is in agreement with this expectation.

The stars of the three sub-classes are found across the whole instability region in the HRD. While most of the hybrid pulsators are close to or inside of the overlap region of the two instability strips, we find some notable exceptions. In particular the distinct group at high $T_\mathrm{eff}$ (see above), hosts two hybrid pulsators. A HRD with the stars colour-coded by the classifications is shown in the Appendix Fig.~\ref{fig:HRD_class}.

Among the 17 stars we exclude from the \bceph{} sub-classes, three stars (TIC\,22370803, TIC\,311328485, and TIC\,460977279 (HD\,137405)) are likely short period eclipsing binaries mimicking a pulsation signal in the right frequency domain. We remove these stars in the subsequent analysis. TIC\,234644092 might also be of this class but without compelling evidence, we kept the star in our analysis. In contrast, TIC\,285370131, TIC\,450383484, TIC\,329992262 (V1155\,Cas), TIC\,399436828, and TIC\,410100850 show \bceph{} pulsations but their light curves are dominated by long-term quasi-periodic behaviour. We postulate that they are either Be stars or strongly contaminated. TIC\,248275346 (HD\,159792) has additional high frequency components, which create a beating pattern unlike any other star in our sample. The three stars TIC\,326815593, TIC\,327282466, and TIC\,365623418 are noisy and modulated with additional non-periodic variability which stems likely from flux contamination of neighbouring stars. TIC\,217602397, TIC\,464732844, and TIC\,153657168 (HD\,114733) exhibit very noisy light curves with hidden periodic variability. Finally, TIC\,406696133 is likely contaminated by a nearby, very red ($G-J=4.3$\,mag) \emph{Gaia} long-period variable. In the TESS observations this contaminant is brighter than our target star.

The stars TIC\,90964721, TIC\,395218466, and TIC\,1861614014 (HD\,339003) are potentially (longer-period) eclipsing binaries with \bceph{} pulsations, making them interesting targets for detailed modelling. The latter of the three is also included in the sample of \cite{Eze2024}.

\subsection{Rotational properties}
\subsubsection{Surface rotation}

Rotation leaves a signature in a star's light curve by bringing chemical spots in and out of view. Some stars in our sample (e.g. TIC\,445256790, HD\,236535) show this kind of regular sinusoidal modulation in their light curves (Fig.~\ref{fig:all_lcs}) which is different from the typical multi-periodic pulsational behaviour. Rather, the amplitude of the pulsations is constant but the light curves reveal an added sinusoidal component with a period of a few days. When comparing the typical longer variability to the above derived estimate of the rotation frequency, we find them to match. Hence, the longer-term variability is caused by rotational modulation, a feature commonly observed in high-resolution spectroscopy and high-precision photometry among the brightest \bceph{} stars \citep[e.g. $\lambda$ and $\kappa$ Sco][]{Uytterhoeven2004,Uytterhoeven2005,Handler2013}.

In order to select all possible \bceph{} stars with rotational modulation, we searched for matching signals among the pre-whitened independent frequencies and the \emph{Gaia} ESP-HS spectroscopic rotation rate. For each star with $f_\mathrm{rot} \sin i$, we select the closest signal and accept it as the rotation frequency if it satisfies $0.9<f/f_\mathrm{rot}<2.5$, allowing for $1.1 \lesssim \sin i < 0.4$ to include the unknown projection effect of the rotational axis in the line-of-sight. Given the large uncertainty on the projected rotation rate the chosen range might exclude some true rotational signals but it ensures a clean selection. Overall, we find some variability within the probed range in 74 out of the 182 (41\%) stars with an estimate for $f_\mathrm{rot} \sin i$.

Most of the potential photometric rotational signals are close to the estimated $f_\mathrm{rot} \sin i$ indicating inclinations closer to edge-on. This skewed distribution stems from two selection effects favouring $\sin i\sim 1$ over $\sin i\sim 0$. Firstly, the inclination angles of least cancellation of low-degree tesseral and sectoral modes deliver $\sin i\in [0.8,1.0]$ \citep[see Appendix\,1 in ][for a definition and value of these viewing angles]{Aerts2010}. Moreover, for the fastest rotators the pulsation modes tend to be sectoral modes concentrated to a band around the equator, maximising the visibility of such pulsations from the equatorial region \citep[e.g.][]{Reese2006}. Secondly, stars with low $\sin i$ fall below the detection threshold in the \emph{Gaia} $v\sin i$ measurements and hence do not occur in the sub-sample of stars with this quantity available.

A substantial fraction (41\,\%) of our sub-sample features detectable rotational modulation, likely of  magnetic origin through chemical spots. This is a higher fraction than the typical $\sim$10\,\% of magnetic field detections in spectro-polarimetric observations of massive stars \citep[e.g.][]{Grunhut2017}. Hence, either the fraction of magnetic \bceph{} stars is larger than among non-pulsating massive stars \citep[as suggested e.g. by][]{Hubrig2009} or space photometry is a more sensitive detection diagnostic or the detected signals are not only connected to magnetism.

\subsubsection{Influence of internal rotation on pulsations}

Due to the shift between a reference frame rotating with the star and the observer's inertial reference frame, the observed pulsation frequencies of non-zonal modes get shifted with respect to their value in a frame of reference corotating with the star. We thus expect $f_1$ and $f_{\rm rot}$ to be correlated for sectoral ($l=|m| > 0$) or tesseral ($0 > |m| > l$) modes.  Depending on the mode's azimuthal order, a smaller or larger frequency shift of $m\langle\Omega\rangle$ occurs for the  mode's frequency between the two reference frames, where $\langle\Omega\rangle$ is the averaged internal rotation frequency throughout the stellar interior.  In general, the internal rotation of stars is not rigid and may hence differ substantially from the surface rotation $f_{\rm rot}$. Asteroseismology has shown the level of differentiality between the near-core region, the envelope, and the surface rotation is limited to about 10\% for most intermediate-mass main sequence single stars with a convective core \citep{Li2020,Aerts2021-RMP}. However, the few \bceph{} stars with such measurement show a large spread in the level of their differential rotation \citep{Burssens2023}. 

\begin{figure}
    \includegraphics[width=\columnwidth]{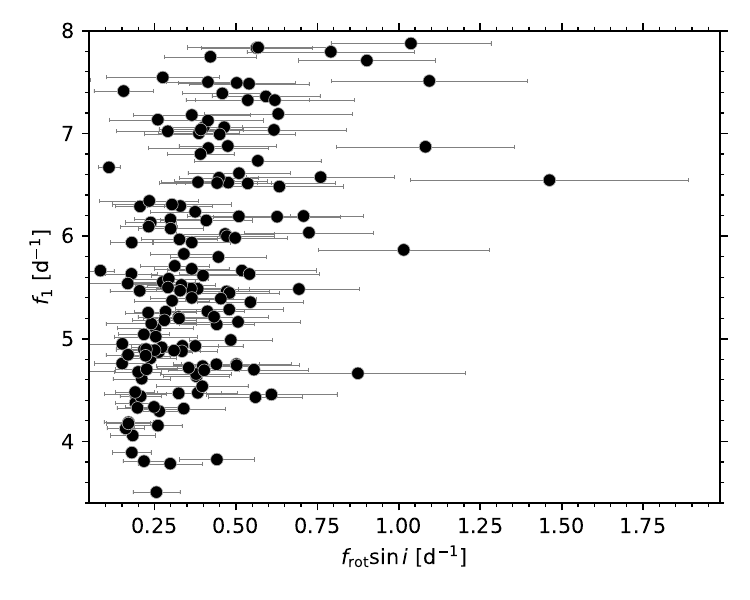}
    \caption{Main pulsation frequency $f_1$ against the projected surface rotation frequency $f_\mathrm{rot} \sin i$ of the 162 stars with available $v\sin i$ in the \emph{Gaia} ESP-HS data. Uncertainties in the pulsation frequency are typically within the symbol size.}
    \label{fig:frot_f1}
\end{figure}

Figure~\ref{fig:frot_f1} shows the main frequency observed in the TESS light curves against the estimated projected surface rotation frequency. We find a positive correlation between the two properties, as anticipated.  We only have a (rather uncertain) direct estimate of the projected value of the surface rotation frequency in the line-of-sight, while the pulsation frequency's shift is mainly determined by the rotation in the region of the star where the mode has most of its probing power.  The low-order modes in \bceph{} stars are dominant in the envelope and hence mainly probe the rotation in that region, denoted here as $\Omega_{\rm env}$. This, and the fact that the shift of any non-radial mode frequency in the line of sight is proportional to $m\Omega_{\rm env}$, implies that the main pulsation frequency $f_1$ is not tightly related to the projected surface rotation frequency, explaining the scatter in Fig.~\ref{fig:frot_f1}. 

\section{Mode identification from TESS and \emph{Gaia} data}

Identifying the degree $l$ of at least one pulsation mode is required to reduce degeneracy and perform meaningful asteroseismic modelling of \bceph{} stars \citep{Hendriks2019}. Additionally, an observational distribution of the mode degrees could advance our understanding of the mode selection mechanisms. Hence, we attempt to identify the pulsation modes of the \emph{Gaia} sample to (1) learn about the distribution of pulsation modes in \bceph{} stars from a large sample based on space photometry, (2) provide the input for detailed future modelling of the most  promising sample stars, (3) open pathways towards ensemble asteroseismology of \bceph{} pulsators observed across the entire sky from combined TESS and \emph{Gaia} space data.

\subsection{Multi-colour photometry}

How a mode's amplitude changes with wavelength depends on its degree, with low-degree modes showing a stronger decrease with wavelength. This has historically been used to identify pulsation modes of \bceph{} stars using ground-based multi-colour photometry or time-resolved spectroscopy (see Introduction, Sect.~\ref{sec:intro}). 
In multi-colour photometry the amplitude ratios are calculated from the amplitudes of the pulsation modes as measured in different photometric filters. The bluest filter is typically used as the reference filter because of its higher SNR. In our work, the highest SNR is provided by the TESS filter. Yet, we decided to use the bluest filter as reference band to enable comparison with earlier literature work on photometric mode identification \citep[cf.][]{Heynderickx1994}. As already discussed in \cite{Hey2024}, the \emph{Gaia} time series photometry offers multi-colour photometry which may help in mode identification.

\begin{figure}
    \includegraphics[width=\columnwidth]{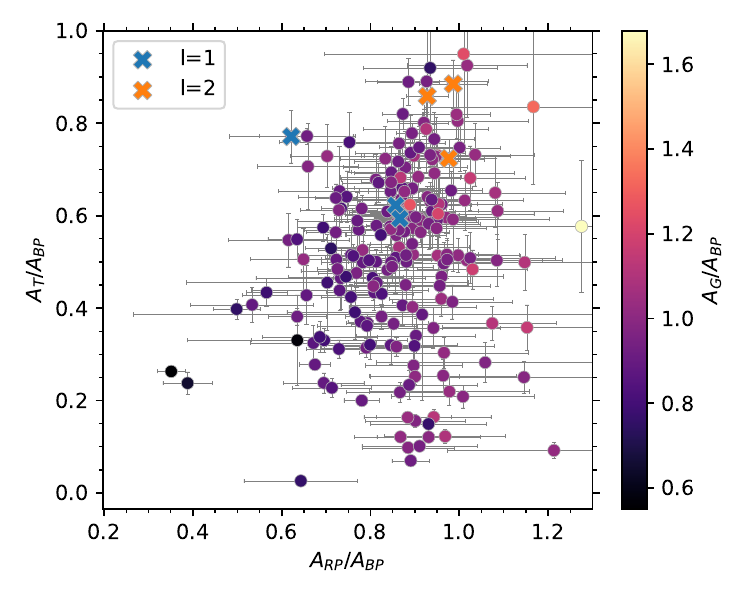}
    \caption{Pulsation amplitude ratio $A_T/A_\mathrm{BP}$ against $A_\mathrm{RP}/A_\mathrm{BP}$ for the main mode detected in the \emph{Gaia} photometry of 205 \bceph{} stars. The colour-coding is according to the amplitude ratio $A_G/A_\mathrm{BP}$. Crosses mark a few well-studied bright pulsators with secure mode identifications from the literature.}
    \label{fig:multicolor}
\end{figure}

Figure~\ref{fig:multicolor} shows amplitude ratios of the \emph{Gaia} $G_\mathrm{RP}$ and TESS $T$ band observations of stars with the same dominant pulsation frequency in \emph{Gaia} and TESS. The ratios are computed with respect to the $G_\mathrm{BP}$ amplitude. In order to understand the position of different stars in this diagram and assess the feasibility of our approach, we marked stars with unambiguous known mode identifications from the literature ($l=1$: TIC 128821888 (12 Lac), TIC 293680998 (IL Vel), TIC 436285033; $l=2$: TIC 169786455 (V856 Cen), TIC 288489491 (SY Equ), TIC 327656776 (KZ Mus)).Merely six stars in our sample have an identified mode from long-term or multi-site campaigns in the literature since these observations focussed on stars that are either too bright for \emph{Gaia} or have an amplitude below $4$\,mmag. Only TIC 419354107 (BW Vul) has an identified dominant radial pulsation in the literature, though its amplitude is so large that non-linear effects occur \citep{Aerts1995}, which may affect its amplitude ratios.  Hence no stars with known dominant linear radial mode are marked in Fig.~\ref{fig:multicolor}.

As expected the quadrupole ($l=2$) modes fall in the top right corner with amplitude ratios close to unity independent of the filter. Identified dipole ($l=1$) modes are found at lower amplitude ratios, while we expect to find the stars with dominant radial ($l=0$) mode in the lower left corner because their amplitudes decrease strongest with wavelength \citep[see][]{Heynderickx1994}.No clearly separated groups emerge from Fig.~\ref{fig:multicolor} in terms of the mode degree. Hence, we cannot achieve unambiguous identification for any given star from \emph{Gaia} and TESS multi-colour amplitude ratios alone. However, we also use the sample properties and additional information from the light curves. Particularly the trend of having $l=2$ in the upper right corner and $l=1$ in the centre of the plot does emerge and is encouraging.

\subsection{Rotationally split multiplets and stillstand stars in the TESS data}
\label{sec:splittings}

Given the high photometric precision of the TESS light curves of these high-amplitude \bceph{} pulsators, mode identification might also be possible by identifying rotationally split multiplets. Following the prototypical case of HD\,129929 \citep{Aerts2003}, we use the pre-whitened pulsation frequencies in the TESS data. A pulsation mode with $l>0$ can be split into $2l+1$ observable mode components of azimuthal order $m$. In the corotating frame, the offset from the $m=0$ central mode is proportional to $\langle\Omega\rangle$ when treating the Coriolis force perturbatively up to first order \citep{Ledoux1951}. 

\begin{figure*}
\sidecaption
    \includegraphics[width = 12cm]{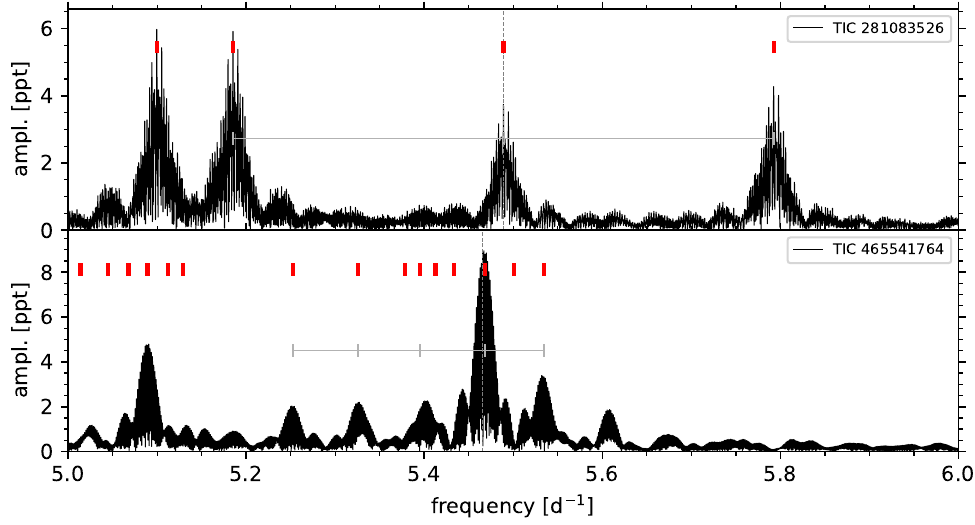}
    \caption{Examples of rotationally split multiplets found in the amplitude spectra of the TESS data. The top panel shows an $l=1$ mode split into three azimuthal components. The bottom panel shows one of the few identified complete $l=2$ multiplets. The splittings are marked with grey markers connected by a line. The red markers at the top of each panel indicate the extracted frequencies from pre-whitening. 
    The dashed lines mark the dominant frequency in the \emph{Gaia} DR3 light curve.}
    \label{fig:splittings}
\end{figure*}

We exploited the connection between the two measurements $f_1$ and $f_\mathrm{rot}\sin\,i$ and the unknown $\langle\Omega\rangle$ required to find rotational multiplets, by relying on the correlation seen in Fig.~\ref{fig:frot_f1}. Up to first order in $\langle\Omega\rangle$ the splittings are symmetrical, which is a good approximation for slow to modest rotators among the \bceph{} pulsators. This enabled us to distinguish between chance alignments of frequency peaks and true multiplets in the periodograms. 

Figure~\ref{fig:splittings} shows  two 
examples of rotationally split multiplets we found in the TESS data. For the example in the top panel, the dominant frequency in the {\it Gaia\/} DR3 light curve matches  a significant frequency in the TESS light curve, where it is not the dominant one due to strong beating effects. In total, we identified 19 stars with a rotational multiplet involving the dominant \emph{Gaia} frequency in the TESS data satisfying at least four of the following five criteria: 1) the multiplet is complete, 2) the splitting is on the same order of magnitude as $f_\mathrm{rot}\sin\,i$,  3) the splittings are symmetric up to $|\delta_A| < 0.1$ (see Sect.\,6.2), 4) the multiplet and its splittings do not conflict with any other potential multiplet in the Fourier spectrum, and 5) none of the frequencies in the multiplets originated from contaminating sources. The majority of these stars (17, representing 89\,\%) pulsate in a dominant $l=1$ mode and only two in a $l=2$ mode (representing 11\,\%). As we only seek three evenly spaced frequencies to identify an $l=1$ mode compared to five for an $l=2$ mode, the preponderance of $l=1$ modes may suffer from selection bias. In addition, we identified 14 stars with potential split multiplets that almost satisfied four conditions. We will come back to them after the mode identification in Sect.\,5.3.

\begin{figure}
    \includegraphics[width=\columnwidth]{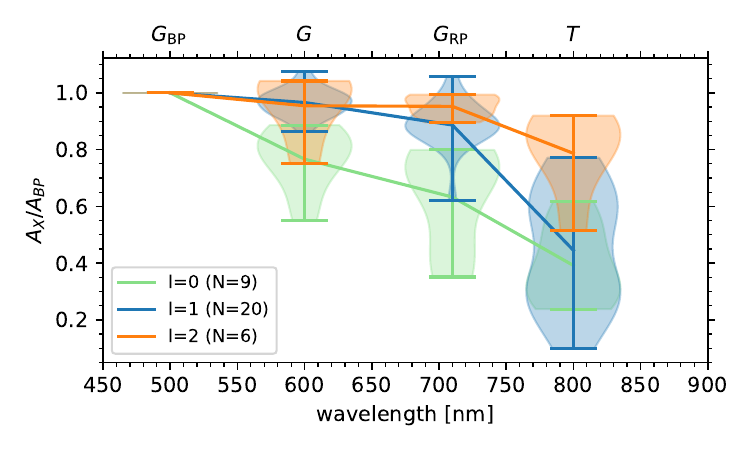}
    \caption{Violin plots of amplitude ratios for identified pulsation modes over the filter's effective wavelengths. The lines connect the mean values for each mode degree. The filter names are given at the top.}
    \label{fig:violin_manual}
\end{figure}

\begin{figure}
    \includegraphics[width=\columnwidth]{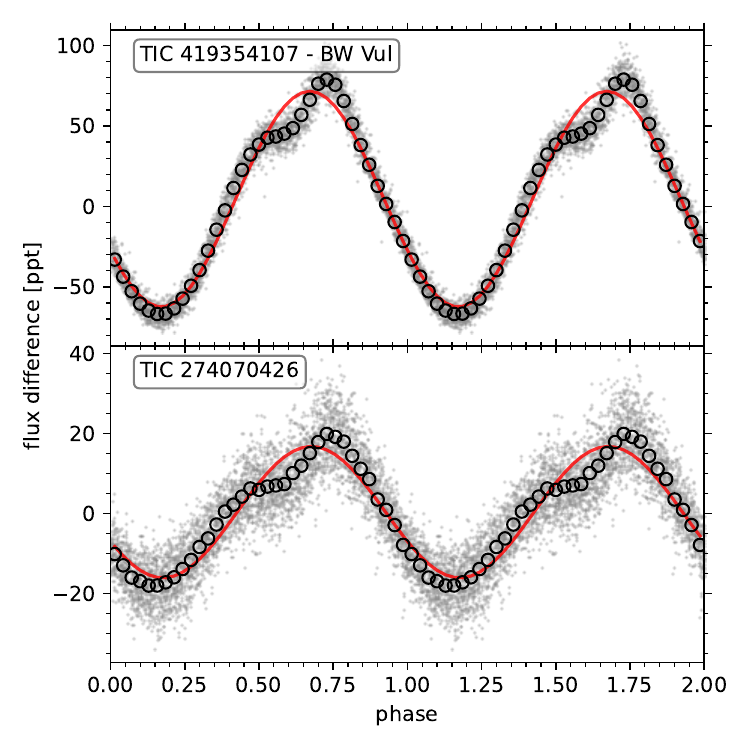}
    \caption{Light curves folded over the dominant frequency of two stars with clear stillstands in our sample (stalling of the light curve near phase 0.5). The folded light curve is binned into 35 bins, shown as black circles, to which we fitted a sinusoidal with a fixed frequency for reference (red). The top panel shows the archetypical stillstand star BW Vul \citep{Walker1954}.}
    \label{fig:stillstand_examples}
\end{figure}

Unlike split multiplets for pulsators with $l\ge1$, radial modes cannot easily be identified from the periodograms. However, the light curve of a high-amplitude radial pulsation in \bceph{} stars sometimes exhibits the stillstand phenomenon, just as for Cepheids \citep[e.g.,][]{Bersier1994}. During the rise of the light curve, at the time of maximum radial velocity, the brightness briefly stalls before increasing further and dropping rapidly after the maximum \citep{Walker1954}. The stillstand phenomenon stems from the development of shock fronts in the expanding photosphere \citep{Mathias1998,Fokin2004}. The most prominent example among the known $\beta\,$Cep stars revealing a stillstand in its photometric light and spectroscopic radial-velocity curves is the high-amplitude non-linear radial pulsator BW Vul \citep[TIC\,419354107, e.g.][]{Sterken1987,Aerts1995}. Based on phase-folded and binned TESS light curves, we identified five more stars with stillstands (TIC\,199208037, TIC\,274070426, TIC\,347520615, TIC\,395218466, TIC\,444357430). The folded TESS light curve of TIC\,274070426 is shown alongside the one of BW Vul in Fig.\,\ref{fig:stillstand_examples}, while those of the other four are displayed in Fig.~\ref{fig:stillstand_all}. To our knowledge, none of these stars have been reported before to show this phenomenon. TIC\,395218466 is of particular interest as it is a member of an eccentric eclipsing binary as detailed in Appendix~\ref{app:stillstands}.

Three additional stars have been identified to have a radial mode due to the position of its frequency with respect to those of split multiplets. In total, we thus identified nine radial mode pulsators. Figure~\ref{fig:violin_manual} shows the amplitude ratio distributions of all the mode degrees identified from rotational splitting or a stillstand in the TESS data, as well as from securely identified modes from the literature. We show the ratios as a function of the effective wavelength of the considered filters\footnote{All filters are broadband filters that overlap strongly. The overlap of the distributions is partly caused by this overlap in wavelength.}. For each photometric band and pulsation mode, we show the mean value of the amplitude ratio and a violin plot of the distribution. For all three considered degrees $l$ the distributions overlap strongly, which explains why we could not find a clean separation in Fig.~\ref{fig:multicolor} from the \emph{Gaia} data alone and had to work with extra information from the rotational splitting or stillstand in the TESS data. We find that the mean amplitude ratio values are distributed as expected, with a larger amplitude ratio for higher $l$. The radial modes show clearly smaller amplitude ratios than the non-radial modes. The distinction between $l=1$ and $l=2$ modes is small, following the results in Fig.~\ref{fig:multicolor}.

\subsection{Probabilistic mode identification from TESS and \emph{Gaia}}
\label{sec:mcphot}

With the secured mode identifications from rotational splittings and stillstands in the TESS data, we now return to the \emph{Gaia}-TESS multi-colour photometry for all pulsators, to identify the dominant mode of the stars without rotational splitting. Given the overlap of the distributions of the amplitude ratios for modes of $l=0, 1, 2$ in Fig.~\ref{fig:violin_manual} and the spread in Fig.~\ref{fig:multicolor}, we aimed to estimate the mode identifications for the full sample based on a probabilistic framework including all photometric measurements and their uncertainties. To find the probability of a given mode identification under the measured \emph{Gaia} and TESS amplitude ratios, we used the distributions shown in Fig.~\ref{fig:violin_manual} as our training set and modelled them with a 3D Gaussian kernel density estimator (KDE, $K_l$). As only very few dominant pulsation modes have been identified with a degree $l>2$ from ground-based observations \citep[e.g.][]{Heynderickx1994}, we assumed $l\in \{0,1,2\}$. By cross-validation, we ensured that a Gaussian KDE indeed describes our data best. Similarly, we used cross-validation to choose the optimal value for KDE bandwidth parameter.

Since our training set is sparse, we augmented our observations by bootstrapping. In order to account for correlations in the data, we resampled the underlying pulsation amplitudes using a Gaussian perturbation based on the measurement uncertainties. We then calculated the amplitude ratios for each augmented data point. For every 
securely identified mode, we added 1\,000 points to capture the uncertainty distribution properly.
Based on the KDE calculated over the augmented data set and for every star, we calculated the likelihood $L_l$ for a given mode $l$ in a Na\"ive Bayesian scheme \citep{Zhang2004} by integrating $K_l$ weighted by a Gaussian distribution $N$ with the uncertainties as its covariance $\mathbf{\sigma}$:
\begin{equation}
    L_l = \int\int\int K_l(\mathbf{A}) N(\mathbf{A}, \mathbf{\sigma}) \mathrm{d}\mathbf{A}.
\end{equation}

To estimate the probability for the mode identification, we initially chose a flat prior but noticed that many stars with clear $l=1$ splittings were assigned either the correct degree but with a low probability or the wrong degree. This is an effect of the overlap of the distributions which reduces the probability of $l=1$ modes. Given the overall clear separation of $l=1$ modes, we chose to use a non-flat prior. The best prior is the one that enables the correct recovery 
for all known mode identifications. From a grid search of possible priors, we found the weights 0.33, 0.47, 0.20, for radial, dipole, and quadrupole mode occurrences, respectively, to provide the best result.

With the weighted probabilities, we identified \nmc{} mode degrees ($l=0$: 23, $l=1$: 104, $l=2$: 16) with probability $p>60\,\%$. The large number of $l=1$ modes follows from the abundance of triplets in the training set and their prior probability. Five stars from the training set do not reach the chosen threshold of $p>0.6$ but only one of them is assigned the wrong degree. We have thus defined a robust probabilistic mode identification methodology suitable to treat all sample stars. For reference and to inform 
any future independent modelling efforts of the stars by community members, we provide all mode identifications (including those with $p<0.6$) and their probabilities in Table~\ref{tab:obs}.

\begin{table*}
    \centering
    \caption{Observational parameters and mode identifications. For simplicity, we omitted the uncertainties in this excerpt.}
    \label{tab:obs}
    \resizebox{\linewidth}{!}{%
    \begin{tabular}{rrlrrrrrrrrrrrrrrrrrrrrrr}
    \hline
    \hline
    TIC &\emph{Gaia}DR3 source ID & Name & RA & Dec. & G &$\log T_\mathrm{eff}$ & $\log L$ & Ref. & $f_\mathrm{rot} \sin i$ & $f_1$ & $f_\emph{Gaia}$ & $f_\mathrm{TESS}$ & LC type & $P_\mathrm{beat}$ & $A_\mathrm{var}$ & $a_G/a_\mathrm{BP}$ & $a_\mathrm{RP}/a_\mathrm{BP}$ & $a_T/a_\mathrm{BP}$ & $l$ & $p(l=0)$ & $p(l=1)$ & $p(l=2)$ & splitting & $\delta_A$\\
    &  & & (deg.) & (deg.) & (mag) & & & & (d$^{-1}$) & (d$^{-1}$) & (d$^{-1}$) & (d$^{-1}$) & & (d) & & & & & & & (d$^{-1}$) & \\
    \hline
4030832 & 5933776187446650880 & HD 147421 & 246.18895 & -53.46433 & 9.02 & 4.48 & 4.72 & 2 & 0.54 & 5.35610 & 5.35473 & 5.35610 & mono & 0.21 & 0.98 & 0.96 & 0.89 & 0.66 & 1 & 0.09 & 0.73 & 0.18 & \dots & \dots\\
7038369 & 5933980456100089856 & \dots & 246.58808 & -51.79895 & 9.58 & 4.35 & 3.93 &  & 0.19 & 4.46055 & 4.45906 & 4.46055 & multi & 10.37 & 0.89 & 0.91 & 0.77 & 0.57 & \dots & 0.45 & 0.52 & 0.03 & \dots & \dots\\
11478661 & 2060836535411106304 & \dots & 303.93684 & 37.71048 & 10.93 & 4.32 & 4.07 &  & 0.21 & 6.28892 & 6.28626 & 6.28892 & multi & 21.79 & 0.89 & 0.57 & 0.64 & 0.33 & 0 & 0.77 & 0.23 & 0.00 & \dots & \dots\\
11856739 & 5940654422886567936 & \dots & 247.12647 & -49.35147 & 10.78 & 4.45 & 4.55 &  & 0.38 & 4.47288 & 4.47143 & 4.47288 & hybrid & 0.37 & 0.91 & 0.89 & 0.78 & 0.50 & \dots & 0.52 & 0.48 & 0.00 & \dots & \dots\\
13332837 & 2061190956100233216 & HD 229085 & 305.39636 & 38.61325 & 9.74 & 4.37 & 4.17 & 3, 4 & 0.48 & 6.87714 & 6.87450 & 6.87714 & hybrid & 0.23 & 0.88 & 0.90 & 0.73 & 0.65 & \dots & 0.39 & 0.48 & 0.13 & \dots & \dots\\
13765605 & 465510069033121152 & \dots & 37.40739 & 61.20823 & 11.15 & 4.41 & 4.16 &  & 0.34 & 4.92968 & 4.92960 & 4.92968 & multi & 2.10 & 0.94 & 1.18 & 0.95 & 0.73 & 1 & 0.01 & 0.66 & 0.34 & \dots & \dots\\
13973539 & 2061131715608660736 & HD 229171 & 305.76197 & 38.45576 & 9.18 & 4.40 & 4.65 & 3 & 0.26 & 3.50524 & 3.50277 & 3.50524 & hybrid & 0.75 & 0.90 & 1.03 & 1.01 & 0.63 & 1 & 0.08 & 0.75 & 0.17 & \dots & \dots\\
14085632 & 2057943789022547968 & \dots & 305.72188 & 37.11278 & 10.67 & 4.53 & 4.94 &  & 0.38 & 4.63305 & 4.63295 & 4.63305 & multi & 7.21 & 0.88 & 0.94 & 0.93 & 0.89 & 2 & 0.02 & 0.18 & 0.80 & \dots & \dots\\
15166556 & 5990434159009247232 & HD 146442 & 244.59709 & -45.84034 & 8.98 & 4.36 & 3.73 & 2 & 0.76 & 6.57384 & 6.57244 & 6.57384 & multi & 4.83 & 0.87 & 1.01 & 0.95 & 0.73 & 1 & 0.07 & 0.47 & 0.46 & 0.20 & 0.03\\
\dots
    \end{tabular}}
\tablefoot{\textit{Name} gives the Simbad main identifier. \textit{Ref.} indicates stars listed in other works. $f_\emph{Gaia}$ and amplitudes in the \emph{Gaia} bands from \cite{DeRidder2023}. The full table, including uncertainties, is available at the CDS.}
\tablebib{(1) \cite{Stankov2005}; (2) \cite{Pigulski2008b}; (3) \cite{LabadieBartz2020}; (4) \cite{Shi2024}; (5) \cite{Eze2024}.}
\end{table*}

Some stars in Fig.~\ref{fig:multicolor} exhibit a low $A_T/A_\mathrm{BP}$ amplitude, while near unity in $A_\mathrm{RP}/A_\mathrm{BP}$. A detailed analysis revealed that the majority of these stars contain a fraction of contaminating flux in their light curves. Although we tried to minimise the impact of contaminating light, it cannot always be avoided, for example in cases of increased sky background in TESS observations or very nearby bright stars as found in open clusters. Thanks to the two \emph{Gaia} amplitude ratios included in the degree determination, the impact of contaminating light lowering the TESS amplitude is limited. Yet, it highlights the vulnerability of TESS amplitudes due to contamination from nearby sources and the sky background.

In summary, we are able to identify the dominant mode degree of \nmc{} \emph{Gaia} \bceph{} pulsators with a probability $p>60\,\%$ based on their TESS light curve, \emph{Gaia} multi-colour amplitude ratios, and the estimated surface rotation rates. For five additional stars, the multicolour photometry alone was inconclusive, yet we identified their mode degrees from splittings or a stillstand. This sample of \nid{} stars represents the largest sample of identified pulsation modes in \bceph{} stars to date. Even if each of these pulsators revealed a limited number of identified pulsation modes compared to the handful of $\beta\,$Cep stars modelled in great detail \citep[as summarised by ][]{Burssens2023}, our results offer a prime sample for further and more detailed asteroseismic population studies of \bceph{} stars. Only this type of sample studies can offer suitable input to calibrate stellar evolution theories for galactic massive field stars as a whole. In the following, we present our initial global seismic analysis and grid modelling of this unique and homogeneously composed sample.

\section{\bceph{} ensemble asteroseismology}

To gain deeper insights into the population of the \emph{Gaia}-observed \bceph{} stars, we used the probabilistic identification of the dominant modes. In the following, we characterise the ensemble and subject all sample stars to one homogenous type of basic modelling by relying on a grid of stellar evolution models. The aim of this work is to understand the properties of the population of field \bceph{} stars instead of doing detailed studies of only a few individual stars. Our approach is a more powerful way to understand global aspects of stellar evolution theory and pinpoint some of its possible shortcomings for the entire mass regime covered by the sample.

\subsection{Distributions of the identified modes}

The mean behaviour of the identified modes of the full sample over the wavelength is very similar to the hand-selected sample in Fig.~\ref{fig:violin_manual}. However, we find a slightly clearer separation between the two considered non-radial modes. Notably the distribution of $A_T/A_\mathrm{BP}$ for $l=1$ shown in Fig.~\ref{fig:violin_mc} has a heavy tail in the TESS amplitude but not in the other photometric bands.

\begin{figure}
    \includegraphics[width=\columnwidth]{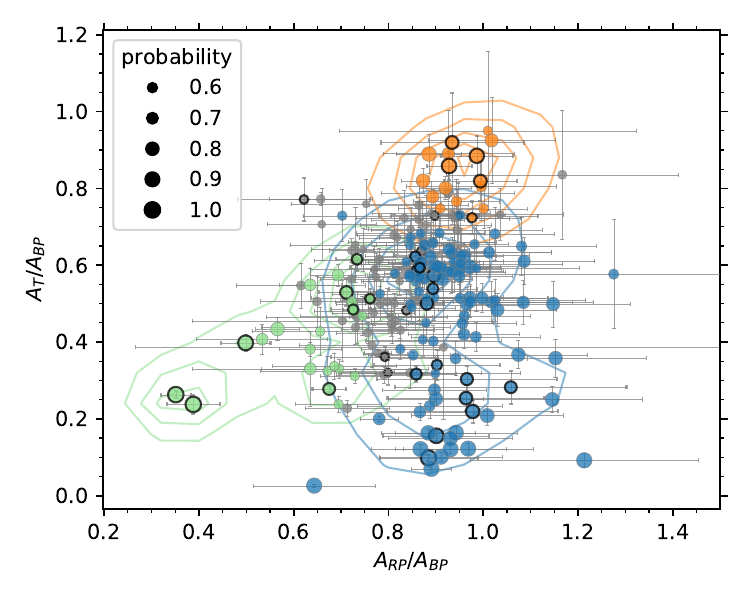}
    \caption{Amplitude ratio diagram based on $A_\mathrm{RP}/A_\mathrm{BP}$ and $A_T/A_\mathrm{BP}$ similar to Fig.~\ref{fig:multicolor}. The colour-code shows the mode degree as determined with our probabilistic framework: $l=0$ (green), $l=1$ (blue), $l=2$ (orange). Grey symbols denote stars with $p<0.6$. The symbol size is proportional to the probability of the mode identification. Stars marked with black outlines constitute the underlying sample from Fig.~\ref{fig:violin_manual}.}
    \label{fig:mc_ids}
\end{figure}

Figure~\ref{fig:mc_ids} shows an amplitude ratio diagram similar to Fig.~\ref{fig:multicolor} colour-coded by the determined mode degree. The distribution of mode degrees follows the expected distributions with $l=2$ modes in the top right (orange) at the highest amplitude ratios. Stars with $l=1$ as their primary mode degree are located in the central part (blue) and radial modes $l=0$ in the lower left (green).

When going into details, we find a complicated pattern with several overlapping regions. Due to the three dimensionality of the distributions these overlap regions create fuzzy borders in the two-dimensional projection in Fig.~\ref{fig:mc_ids}. Stars to which our algorithm assigned low probabilities in the mode identification belong to the overlapping regions between the different degrees as also seen from the shown contours. An interesting feature is  the group of high probability $l=1$ modes at the bottom of Fig.~\ref{fig:mc_ids}. This region creates the tail in the $A_\mathrm{T}/A_\mathrm{BP}$ amplitude ratios in Fig.~\ref{fig:violin_mc}. From the TESS data, we find that some of these stars have multiplets in which the \emph{Gaia} observed frequency (and often also the main TESS frequency) is not the central frequency. Indeed, the multi-colour photometry also carries information on the intrinsic mode excitation amplitudes, aside from the mode geometry and inclination angle of the star. Our findings illustrate that the mode amplitudes are not equal among the prograde and retrograde components of the same $|m|$ within a multiplet for these heat-driven modes. This is in contrast to the case for stochastically excited modes in slowly-rotating low-mass stars, where the amplitudes allow for the derivation of the star's inclination angle. Mode visibilities purely based on geometrical arguments are indeed not the only factor that determines the amplitudes (and hence detectability) within a multiplet for $\beta\,$Cep stars, as has been known since long from ground-based photometry.

When placing the stars with identified modes in the HRD, we find that the \bceph{} stars with a dominant radial mode are mainly of lower mass ($\lesssim 15\,M_\sun$). The radial pulsators seem to concentrate in the later half of the main sequence in the overlap region between the \bceph{} and SPB instability strips. In contrast, non-radial pulsators can be observed throughout the whole \bceph{} mass range (cf. Fig.~\ref{fig:HRD_mc}), although two stars with radial modes stand out due to their relatively high luminosity and mass.

\subsection{Rotationally split multiplets and asymmetries}

As introduced in Sect.~\ref{sec:splittings}, rotation splits non-radial modes into multiplets with splittings proportional to the averaged envelope rotation rate. We now return to the 14 stars with likely multiplets (not used in the definition set to construct the KDE in Sect.\,5.3). We now have their probabilistic mode identification. Adding them to the 19 stars with deterministic identification from the full multiplet structure we had already from Sect.\,5.2, they compose  a sample of 33 stars with multiplets which we now use to study the stellar rotation properties. The full sample of these 33 stars with multiplets is shown in Figs.~\ref{fig:train_multiplets} and \ref{fig:add_multiplets}.

\begin{figure}
    \includegraphics[width=\columnwidth]{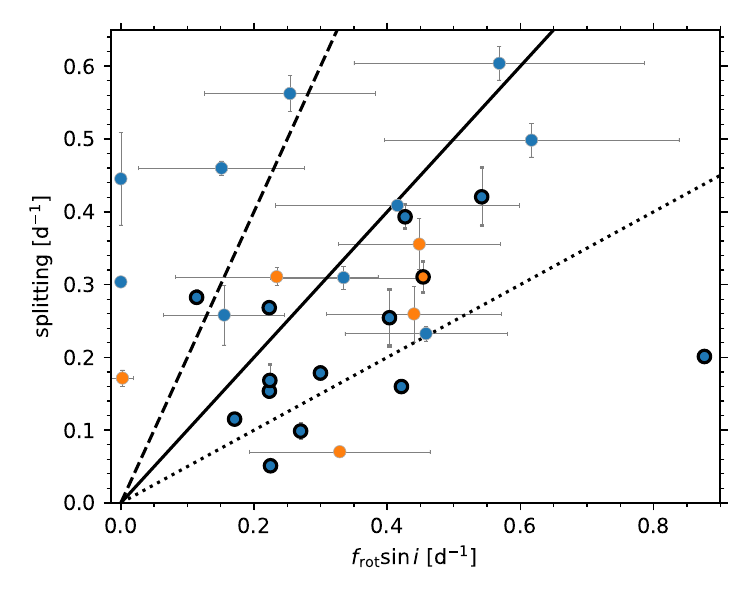}
    \caption{Observed mode splittings against the projected surface rotation rate ($f_\mathrm{rot} \sin i$, encircled in thin grey) or photometric surface rotation rate ($f_\mathrm{rot}$, encircled in thick solid black). The three lines show the splitting values when equal to the surface rotation rate (solid) as well as half (dotted) and twice (dashed) the surface rotation rate. The uncertainties on the splittings indicate the maximum difference of the individual splittings to the mean splitting of the multiplet and are a measure for the asymmetry. The colours show our deterministic identifications as $l=1$ (blue) and $l=2$ (orange) modes.}
    \label{fig:split_rot}
\end{figure}

In Fig.~\ref{fig:split_rot}, we show the observed splittings against the estimated (projected) surface rotation rate. As expected, a positive correlation between the rotation rate and the splitting is observed. Yet, we recall that \bceph{} stars can have strong radial differential rotation \citep{Aerts2003,Pamyatnykh2004,Burssens2023} and the surface rotation rate may differ from the internal rotation rate at the fractional radius where the mode is most sensitive. Stars at $v\sin i=0$\,km\,s$^{-1}$ have  undetectable line broadening in \emph{Gaia}, yet we include them here as they are included as such in \emph{Gaia} DR3.

We also mark the absolute asymmetries (maximum difference to the mean splitting) in Fig.~\ref{fig:split_rot}. For most of the stars this asymmetry is small, though it is pronounced for a minority. A purely rotationally split multiplet would consist of evenly and symmetrically spaced components if the Coriolis force due to the mean envelope rotation has only a small influence. This allows for a first-order perturbative approach to compute the oscillation mode frequencies. However, second-order effects are relevant for \bceph{} stars and include rotational deformation of the star \citep{Saio1981}. This effect shifts the split frequencies asymmetrically with respect to the central frequency \citep{Dziembowski1992, Goupil2000}. In addition, magnetic fields \citep[e.g.][]{Bugnet2021} or binarity \citep[e.g.][]{Guo2021} also influence the split multiplets, leading to asymmetries in the observed frequency spectra and enabling detailed asteroseismic inferences \citep{Guo2024}. 

\begin{figure}
    \includegraphics[width=\columnwidth]{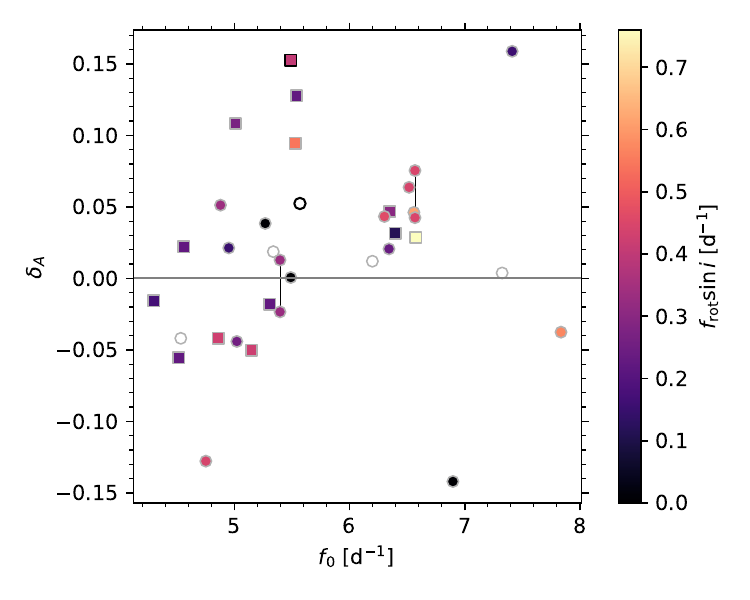}
    \caption{Asymmetries of 33 rotationally split multiplets $\delta_A$ against their central frequency $f_0$ colour-coded by the projected surface rotation rate. Square symbols mark stars with measured photometric surface rotation rates ($f_\mathrm{rot}$). Symbols connected by a line belong to the same ($l=2$) multiplet. Black outlines indicate stars where the \emph{Gaia} frequencies do not match the TESS frequencies within our threshold. The measurement uncertainties are within the symbol sizes.}
    \label{fig:asymmetries}
\end{figure}

\cite{Guo2024} \citep[see also][]{Goupil2000,Deheuvels2017} define the observed asymmetry as
\begin{equation}
    \delta_A= - \frac{f_{-m} + f_{+m} - 2f_0}{f_{+m} - f_{-m}},
\end{equation}
with $f_0$ the central frequency and $f_{\pm m}$ the frequencies of azimuthal order $m=\pm1$. In Fig.~\ref{fig:asymmetries}, we show the 33 most obvious asymmetries occurring in the multiplets detected in our sample stars against the central frequency $f_0$ of the multiplet. Most of these splittings are positive as expected for rotational deformation \citep{Guo2024}.
Negative asymmetries in \bceph{} stars can hint towards internal magnetic fields \citep{MathisBugnet2023}, especially for dipole p-modes. Our understanding of core magnetism in massive stars is still limited (\citealt{Lecoanet2022}; Vanlaer et al., in prep.), despite its potential impact on stellar evolution. A large sample of measured internal magnetic fields from asteroseismology would deliver constraints for evolutionary models, similarly to the recent advances in red giants \citep{GangLi2023,Deheuvels2023}.

\subsection{Modelling the population of \bceph{} stars}
\label{sec:model}

To understand the population of galactic \bceph{} stars better and gain insights into the distributions of their physical properties with respect to the selected main pulsation mode, we applied grid modelling. In doing so, we used extra seismic information compared to classical grid modelling where one just relies on evolution tracks and the measured effective temperature and luminosity to determine stellar properties. Here, we used the \bceph{} MESA/GYRE stellar evolution and pulsation grids presented by \cite{Burssens2023}. The evolution grid was calculated for masses between 9\,$M_\sun$ and 21.5\,$M_\sun$ at solar metallicity with envelope mixing by internal gravity waves set to $D=1000\,$cm$^2$\,s$^{-1}$ at the base of the envelope and increasing outwards as $\rho^{-1}$ and an exponentially decaying diffusive core overshoot with the parameter range $0.005 \le f_\mathrm{ov} \le 0.035$. For details of the physical implementation, we refer the reader to \cite{Burssens2023}.

The required mode identification for the modelling is not limited to the degree $l$ but we also need  the radial order $n$ and azimuthal order $m$. All models will be fit in the azimuthal order $m=0$ necessitating us to identify the central peak of split modes denoted here as $f_0$. For complete multiplets or radial pulsators, $f_0$ is measured. For the few quadrupole pulsators with incomplete multiplets, we assumed the highest peak of the central two peaks to be the one with $m=0$.

The radial order can neither be deduced from the observations nor from a direct fit of the model. Rather, we fitted models with the radial orders $n\in \{0, 1, 2\}$ for radial pulsators and $n\in \{-2, -1, 0, 1, 2\}$ for non-radial modes\footnote{The f-mode ($n=0$) only exists for $l=2$ as the frequency goes to zero for dipole modes \citep[e.g.][]{Takata2012}.} using a $\chi^2$ minimisation over $\log T_\mathrm{eff}$, $\log L$, and $f_0$. In principle, the radial order can be larger but higher order modes have not been securely established in \bceph{} stars, hence we did not probe them.

Stars without a model within the $1\sigma$ uncertainty ellipse were excluded from the subsequent analysis, reducing the number of modelled stars to \nmodel{}. For each star, we chose the best fitting model for each $n$ and compared the position of its other modes to the observed frequency spectra. Our focus laid on the relative positions of the main pulsation mode and other visible modes with respect to those predicted by the model. In most but not all cases our choice of $n$ coincides with the lowest $\chi^2$ value among the set of best fitting models. We show one illustrative example for the identification in Appendix~\ref{sec:findn}, including a detailed reasoning for our choice of $n$.

We find the majority of the radial pulsators to pulsate in their fundamental mode (eight out of twelve modelled radial pulsators). Three radial pulsators pulsate in their first overtone and one stars is likely a second overtone pulsator. A similar picture emerges for the 107 non-radial pulsators. The dominant pulsation mode in 50 stars is the $p_1$-mode ($n=1$) and 24 pulsate in $g_1$ ($n=-1$). Only 29 stars pulsate in their second-order mode with 28 identified $p_2$ and one $g_2$ pulsators. The remaining four stars were identified to be dominant f-mode pulsators.

For the final model selection and uncertainty estimation, we employed a MCMC approach with a $\chi^2$ merit function, similar to \cite{Fritzewski2024b}, to which we refer for details. Since we rely on a grid of stellar models with fixed input physics on some aspects (notably the level of envelope mixing, where the low-order modes have dominant probing power), we allow for an overall mismatch between observed and fitted frequencies \citep[see][for an extensive discussion]{Aerts2018}. In order to encapsulate this theoretical uncertainty caused by the fixed yet uncertain input physics adopted in the model grid, we follow the predictions for zonal p-modes by \citet{Aerts2018} in the $\beta\,$Cep mass regime and allow for a typical frequency uncertainty of 0.1\,d$^{-1}$. This captures any systematic errors in the predicted frequencies while allowing to recognise the correspondence between the observed and predicted frequency pattern (cf.\, the procedure explained in in Appendix~\ref{sec:findn}).

\begin{figure}
    \includegraphics[width=\columnwidth]{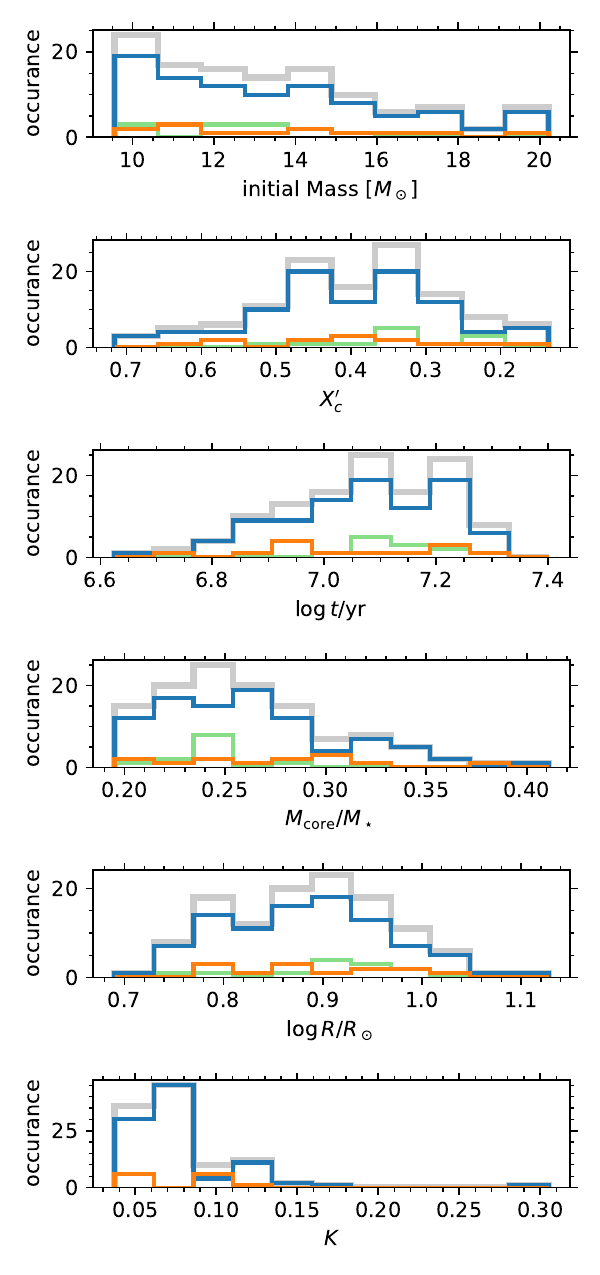}
    \caption{Distribution of \bceph{} parameters for the modelled stars. Each panel contains the histograms for the full sample (grey) and each degree $l=0$ (green), $l=1$ (blue), and $l=2$ (orange). From top to bottom, these panels show the initial mass distribution, the normalised core hydrogen fraction $X_c'$, the stellar age in years, the relative convective core mass, the radius and the K-value.}
    \label{fig:model_hist}
\end{figure}

The histograms in Fig.~\ref{fig:model_hist} show distributions of the obtained masses, normalised core hydrogen fraction ($X_c'$), age, relative convective core mass, radius, and horizontal versus radial displacement of the dominant mode. As expected from the distribution of \bceph{} stars in the HRD in Fig.~\ref{fig:HRD_all}, the number of \bceph{} stars declines with mass, with most \bceph{} stars found in the range $10\leq M_\mathrm{ini}/M_\sun\leq15$. Eight stars could not be modelled because they lie above the upper mass limit of $21.5\,M_\sun$, as shown by the evolutionary tracks in Fig.~\ref{fig:HRD_all}. In agreement with Fig.~\ref{fig:HRD_mc}, radial pulsators are mostly found among the lower mass \bceph{} stars. 

The distribution of $X_c'$ shows more structure when focusing on the pulsation degree. Overall, the majority of our sample is in their second half of the main-sequence evolution. In particular, the small group of radial pulsators are found in the second half of their main sequence evolution, while non-radial pulsators are found over the whole range.

We do not show a distribution of the core overshoot because most of the stars are modelled with $f_\mathrm{ov}\approx0.02$, which corresponds to the central value in the grid. Yet, the uncertainties span the whole range of $f_\mathrm{ov}$ covered by the grid, which can be explained by the limited sensitivity of the observed modes to the near-core region and the degeneracies between the mass, metallicity, and core overshoot when relying on just one mode \citep[e.g.,][]{Ausseloos2004}. The estimated core overshoot values result in a value for the mass-normalised convective mass ($M_\mathrm{core}/M_\star$). This inferred parameter is the more interesting quantity from a stellar evolution point of view and it is a function of both the luminosity and age. The majority of stars has a value in the range $0.2\leq M_\mathrm{core}/M_\star\leq0.3$, in agreement with the distribution of initial mass shown in the top panel and the HRD. Notably the radial pulsators are concentrated around $M_\mathrm{core}/M_\star\approx 0.25$ due to their common evolutionary status. The few stars with a relatively high convective core mass belong to the higher-mass young O-stars well separated in the HRD (cf.\,Fig.~\ref{fig:HRD_all}). The radius distribution is peaked at intermediate radii, especially for dominant radial mode pulsators.

\begin{table}
    \centering
    \caption{Modelled properties of the \bceph{} sample.}
    \label{tab:tab_model}
    \resizebox{\columnwidth}{!}{%
    \begin{tabular}{rrlrrrr}
    \hline
    \hline
    TIC &\emph{Gaia} DR3 source ID & Name & $X_c'$ & $M_{ini}$ & $\log t$ & $M_\mathrm{core}/M$ \\
    & & & &($M_\sun$) & &\\
    \hline
4030832 & 5933776187446650880 & HD 147421 & 0.49 & 17.91 & 6.84 & 0.333\\
11478661 & 2060836535411106304 & \dots & 0.21 & 10.33 & 7.31 & 0.195\\
13765605 & 465510069033121152 & \dots & 0.46 & 11.01 & 7.16 & 0.241\\
13973539 & 2061131715608660736 & HD 229171 & 0.14 & 15.80 & 7.04 & 0.240\\
14085632 & 2057943789022547968 & \dots & 0.58 & 20.22 & 6.71 & 0.383\\
15166556 & 5990434159009247232 & HD 146442 & 0.44 & 9.56 & 7.29 & 0.212\\
18827544 & 5941164183970835456 & \dots & 0.44 & 9.67 & 7.24 & 0.215\\
42365645 & 2061531358019903488 & HD 227977 & 0.60 & 10.48 & 7.12 & 0.261\\
42534550 & 5978123369668545536 & \dots & 0.28 & 10.16 & 7.30 & 0.201\\
\dots
    \end{tabular}}
\tablefoot{The full table, including uncertainties, is available at the CDS.}
\end{table}

As a final distribution, we show in the bottom panel the distribution for the $K-$value, defined as $K \equiv G M / (2\pi f_0)^2 R^3$, for the non-radial modes. This dimensionless quantity is connected with the boundary conditions for the pulsation equations \citep[][Chapter\,3]{Aerts2010}. In the case of non-radial modes, it represents the ratio between the horizontal and radial displacement of the fluid elements at the stellar surface caused by the mode (as this ratio is zero for radial modes, these are not included in the bottom panel of Fig.~\ref{fig:model_hist}). This parameter has high values above unity (typically between 10 and 100) for high-order g-modes of observed SPB stars \citep{DeCatAerts2002}, as their modes dominantly cause horizontal displacements. Here, as expected for low-order modes, we find the opposite with dominantly radial displacements but with non-negligible contributions from horizontal motions. The values for $K$ range from about 0.03 to about 0.2 for the dominant mode of most of the {\it Gaia\/} \bceph\ stars in our sample.

\begin{figure}
    \includegraphics[width=\columnwidth]{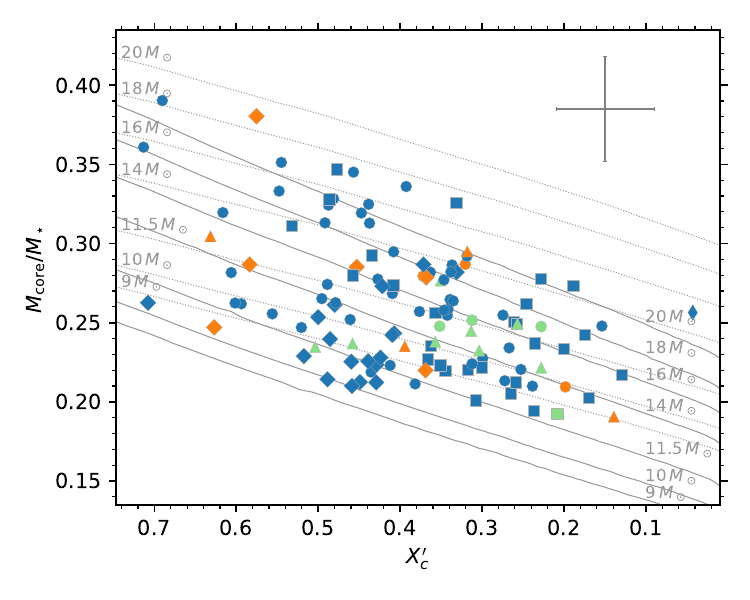}
    \caption{Normalised convective core mass $M_\mathrm{core}/M_\mathrm{ini}$ over the normalised core hydrogen mass fraction $X_c'$. The stars are colour-coded by their primary degree: $l=0$ (green), $l=1$ (blue), and $l=2$ (orange). The shape represents the radial order: g$_2$-modes are represented by thin diamonds, g$_1$ by thin triangles, radial fundamental modes by triangles, p$_1$ by circles and p$_2$ by squares. The grey tracks in the background show the evolution for the indicated masses with $f_\mathrm{ov}=0.005$ (solid, labelled at low $X_c'$) and $f_\mathrm{ov}=0.035$ (dashed, labelled at high $X_c'$). The typical uncertainty is indicated in the upper right corner.}
    \label{fig:mcc}
\end{figure}

To unravel the correlation between the convective core mass and the core hydrogen fraction, we show these two parameters in Fig.~\ref{fig:mcc}. The overall distribution shows clearly the distinction between stars with $M_\mathrm{core}/M_\star\leq0.3$ and these with more massive convective cores. The stars with higher relative core masses are typically more massive and younger as expected from stellar evolution theory. As seen from the evolutionary tracks of different masses, only the lowest mass \bceph{} stars can be found among full breadth of the main sequence evolution, a consequence of the position of the instability strip lying closer to the zero-age main sequence for these stars.

In summary, we successfully modelled \nmodel{} \bceph{} stars with a grid modelling approach 
adding the seismic information on the dominant mode to the two classical observables, effective temperature and luminosity. We provide distributions of their masses, ages, convective core masses, radii. As such, these distributions provide a first insight into the properties of the population of \bceph{} stars. Of particular interest for future investigations on the mode excitation of this population is the connection between the stellar age and the degree of the dominant mode, if any. We give the best fitting model properties in Table~\ref{tab:tab_model}. A discussion of the accuracy of the asteroseismic parameters in comparison with the input values from \emph{Gaia} can be found in Appendix~\ref{app:model_details}.

\subsection{Period-luminosity relations}

Radial and low-order non-radial p-modes probe the internal  density of a star. For a given type of pulsator this translates into a connection between a mode's frequency and the luminosity as established for Cepheids by \citet{Leavitt1912}.  For the $\delta\,$Sct stars, which are situated in the low-luminosity part of the classical instability strip, this also gives rise to a period-luminosity relation \citep{DeRidder2023} but it is not so clean as the one found for Cepheids because a fraction of these p-mode pulsators have a dominant non-radial mode \citep{Ziaali2019}.

While the $\delta\,$Sct stars occupy a small region in mass and radius, this is not the case for the $\beta\,$Cep stars. Given their large spread in mass and radius along the main sequence, one does not expect a clear period-luminosity relation for this class as a whole. Nevertheless, several authors attempted to find such a relation in the past \citep[e.g.][]{Blaauw1953, Jakate1980, Waelkens1981}. However, the uncertain mode identifications led to a large scatter in these relations \citep[see review by][]{Lesh1978}. 

\begin{figure}
    \includegraphics[width=\columnwidth]{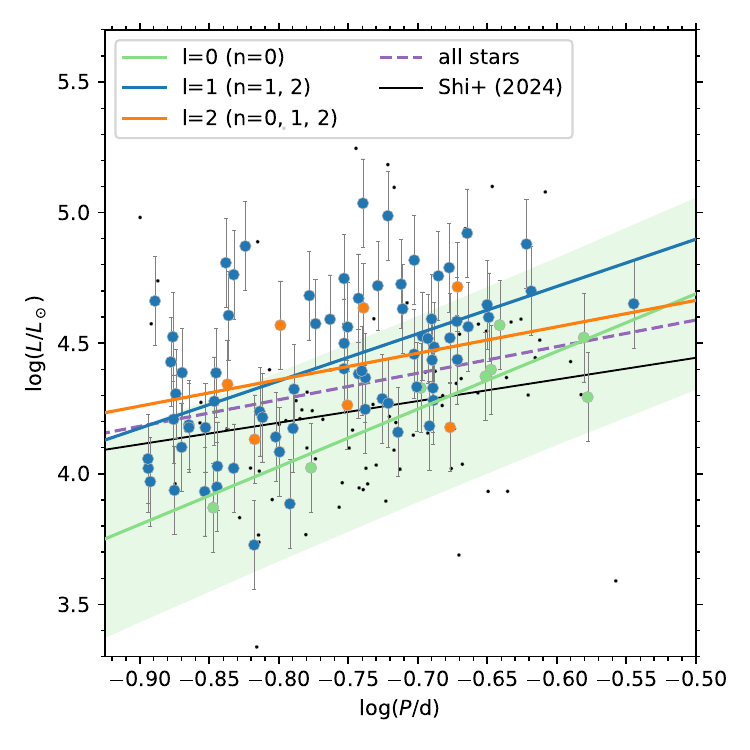}
    \caption{Period-luminosity relations for stars with modes of different degrees ($l=0$ (green), $l=1$ (blue), and $l=2$ (orange)). The period-luminosity relations include only the radial fundamental mode ($n=0$) for radial pulsators and p-modes for the non-radial pulsators. The black line shows the relation of the \cite{Shi2024} sample. For comparison, a similar period-luminosity relation for our sample is plotted in purple (dashed). The green shaded area shows the uncertainty band of the fundamental radial-mode pulsators. The small black data points in the background show dominant g-mode and higher-order radial pulsators excluded from the fits.}
    \label{fig:pl}
\end{figure}

Here we show the relations and their scatter occurring for our sample of $\beta\,$Cep star when using only those with identified pulsation modes. Figure~\ref{fig:pl} shows the period-luminosity relations in the $\log P$-$\log L/L_\sun$ plane, colour-coded by the mode degree. Using a linear regression, we arrived at a period luminosity relation for each sub-set of pulsators according to their dominant mode:
\begin{align}
\log L/L_\sun & = (2.21\pm0.52)\log P + (5.8\pm0.36)\quad (l=0)\\
\log L/L_\sun & = (1.81\pm0.36)\log P + (5.8\pm0.27)\quad (l=1)\\
\log L/L_\sun & = (0.94\pm1.34)\log P + (5.1\pm1.02)\quad (l=2)
\end{align}
with $P$ the mode period in days. Figure~\ref{fig:pl} shows only the $1\sigma$-uncertainty band for the stars with a dominant fundamental radial mode. Those with a dominant dipole p-mode have a similar uncertainty band while it is larger for the quadrupole mode pulsators.

We find the three relations to be stacked by their mode degree, such that the lowest degree modes (radial) follow a period-luminosity relation with the lowest luminosity for a given period. The relation for the quadrupole modes intersects with the relation for the dipolar modes due to fewer stars in the sample and larger spread in the radial order. \cite{Shi2024} recently provided a similar period-luminosity relation for their sample of \bceph{} stars. In comparison to a linear regression of all 155 stars from \cite{Shi2024}, regardless of their mode degree, our period-luminosity relations are steeper. When including all stars of our sample, we find a similar slope to theirs, highlighting the importance of mode identification. 

\subsection{Average envelope rotation}

Radial differential rotation was observed in the handful of \bceph{} stars with a measurement of the internal rotation \citep[e.g.][Vanlaer in prep.]{Aerts2003, Pamyatnykh2004,Dziembowski2008,Burssens2023}. Based on the observed splittings (Sect.~\ref{sec:splittings}) and our modelling results for the {\it Gaia\/} \bceph\ stars, we calculated the internal envelope rotation rate $\Omega_\mathrm{env}$ from the Ledoux constant \citep{Ledoux1951} for the best grid model for the star. This rotation rate then also allows us to calculate the level of differential rotation between the envelope and the surface for stars with a measured $f_\mathrm{rot}\sin i$, as well as the spin parameter of the mode.

\begin{figure}
    \includegraphics[width=\columnwidth]{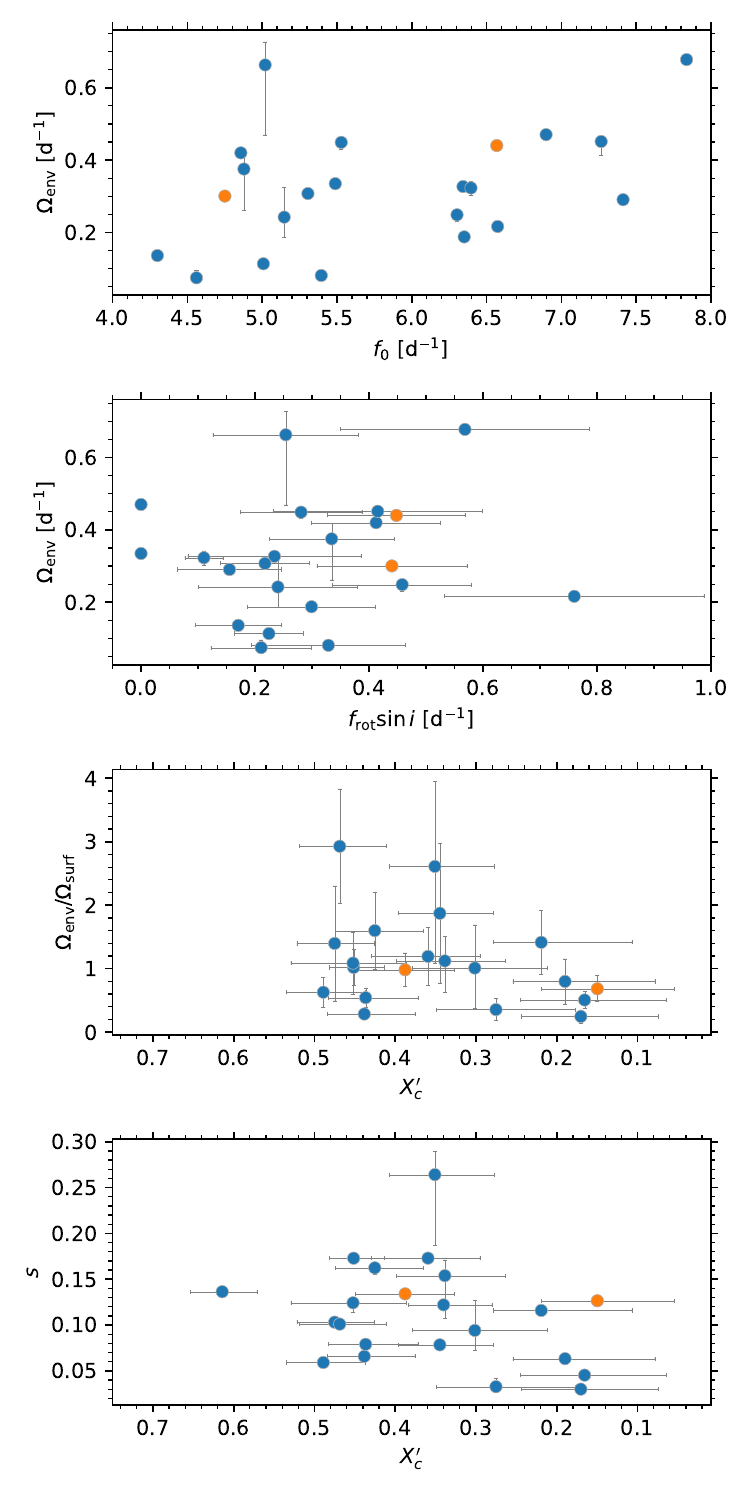}
    \caption{Envelope rotation $\Omega_{\rm env}$ and derived parameters against observed frequencies and estimated core hydrogen fractions for 25 stars. The top panel shows the envelope rotation rate against the zonal mode frequency of the multiplet ($f_0$) and the second panel against the projected surface rotation rate $f_\mathrm{rot}\sin i$ for the stars having this quantity available. The third and fourth panels display derived rotational properties against the evolutionary stage. In the former, we show an upper limit for the level of differential rotation ($\Omega_\mathrm{env}/\Omega_\mathrm{surf}$) and in the bottom panel the spin parameter ($s$) of the mode in the envelope. The colour-code corresponds to the mode degree: $l=1$ in blue and $l=2$ in orange.}
    \label{fig:env_rotation}
\end{figure}

In Fig.~\ref{fig:env_rotation}, we show the estimated $\Omega_{\rm env}$ and derived parameters. Unlike the observed projected surface rotation and as expected, the envelope rotation deduced from the modelling does not have any correlation with the pulsation frequency of the zonal mode, as $f_0$ is unaffected by $\Omega_{\rm env}$ when treating the rotation perturbatively and only up to first order  as we do here (top panel). The highest envelope rotation rate is not found for the fastest surface rotators but for intermediately-fast rotating stars (second panel), reflecting the effect of the unknown factor $\sin\,i$ and/or differential rotation between the envelope and the surface. This is visually represented in the third panel of Fig.~\ref{fig:env_rotation}, which shows the level of differential rotation assuming an almost equator-on view ($\sin\,i\simeq 1$). This lower limit of the true surface rotation frequency is denoted here as $\Omega_{\rm surf}$. It is used to estimate an upper limit for the differential rotation between envelope and surface, which is plotted against the evolutionary stage $X_c'$. Aside from the projection effect, we find the most differential rotators to have an internal rate that is below four times the surface rotation. This is completely in line with the first such measurement ever done for a \bceph\ star \citep{Aerts2003}, which relied on rotational splitting of both a dipole and quadrupole mode. Yet, the majority of stars reveal $\Omega_{\rm env} / \Omega_{\rm surf} \lesssim 2$. Hence, despite the few cases of strong radial differential rotation as summarised by \cite{Burssens2023}, the majority of \bceph{} stars have a level of envelope-to-surface rotation between 0.3 and 2.

We also present the spin-parameter, $s$, of the dominant mode as a function of $X_c'$. It is defined here as $s \equiv 2\Omega_\mathrm{env}/f_0$ and characterises the regime in which the pulsation mode occurs \citep{Aerts2024}. All the {\it Gaia\/} \bceph\ stars have their dominant mode in the super-inertial regime with $s<1$. Noticeably, the star with the largest spin parameter (TIC\,282881728) also has the highest level of differential rotation and is in the later half of its main-sequence phase.

\begin{figure}
    \includegraphics[width=\columnwidth]{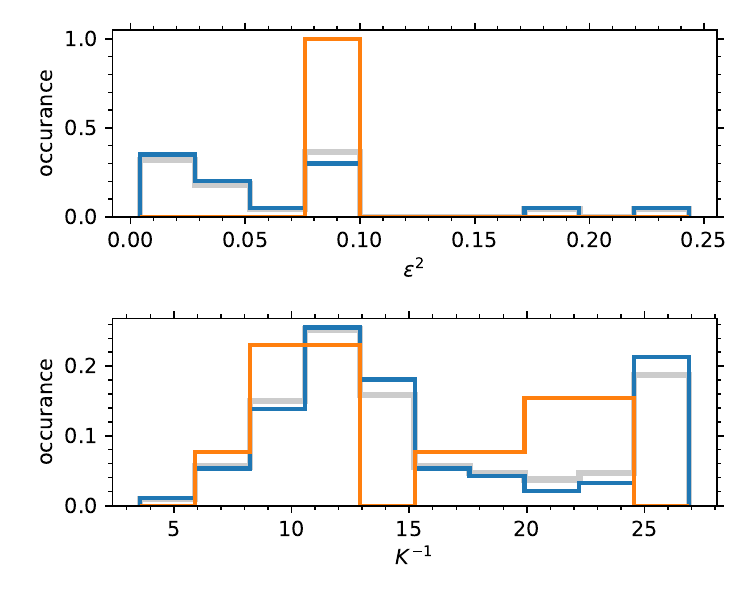}
    \caption{Histograms of $\epsilon^2$ (top) giving the influence of the centrifugal force with respect to gravity and the inverse of the $K$-value (bottom).}
    \label{fig:epsilon}
\end{figure}

Finally, we show in Fig.~\ref{fig:epsilon} histograms of the dimensionless quantities $\epsilon^2 \equiv (2\pi\Omega_{\rm env})^2 R^3 / GM$ and $K^{-1}=(2\pi f_0)^2 R^3 / G M$, following \citet{Goupil2000}. The latter quantity is the inverse of $K$ already introduced above. It represents an assessment of the importance of the centrifugal force with respect to the 2nd-order Coriolis force in the treatment of the pulsations, while the former evaluates the effect of the centrifugal force with respect to gravity. \citet{Goupil2000} reported the centrifugal force (and hence the stellar oblateness) to be more important for the computation of the p-modes of rapidly rotating $\delta\,$Sct stars compared to the 2nd-order rotational effect caused by the Coriolis force. While the  mode frequencies of \bceph{} stars are in general lower than those of $\delta\,$Sct stars, we still find the second-order Coriolis force to be less important than the centrifugal force. However, the difference in importance between the two forces is smaller than for $\delta\,$Sct stars.

\section{Conclusions}

We analysed the TESS light curves of \nlc{} \emph{Gaia}-detected \bceph{} candidates \citep{DeRidder2023}, including 145 new detections. By placing these stars in a \emph{Gaia} HRD, we confirmed that all of the stars are late O or early B stars near the \bceph{} instability region. Yet, a detailed light curve analysis shows that at least three stars are short period eclipsing binaries. By categorising the TESS light curves, we find the majority of \bceph{} stars in our sample to be multi-periodic.

The combination of the \emph{Gaia} and TESS time series photometry, allowed us to determine the mode degrees for \nid{} stars in our sample from their amplitude ratios. Most observed \bceph{} stars pulsate in non-radial modes with $l=1$. Our identified modes constitute the largest set of identified modes of \bceph{} stars available to date.

From the TESS light curves, we found \nsplit{} rotationally split multiplets accompanying the main pulsation mode. The majority of these multiplets have positive asymmetries in their splittings. Yet, several stars show negative asymmetries and are prime targets for detailed follow-up analyses. We provide the first distributions of the envelope rotation rates and of upper limits for the level of differential rotation for a meaningful sample of \bceph\ stars. We find that most stars have envelope-to-surface ratios less than two and many of them rotate almost rigidly. We do point out that our sample only consists of \bceph\ stars with a dominant mode in the super-inertial regime and thus lacks representation of the fastest rotating class members due to our sample selection from {\it Gaia}. 

Further, we performed the first ensemble modelling of \nmodel{} \bceph{} stars based on the grid calculated by \cite{Burssens2023}. We provide distributions in age, convective core mass, and mass in order to inform population studies of \bceph{} stars. With the large set of identified modes, we are in a position to provide these distributions for each mode degree. As expected from instability calculations, we find the stars to be at least one third through their main sequence evolution. The radial mode pulsators in particular occur mostly in the intermediate main-sequence stage.
The period-luminosity relations for the different mode degrees are steeper than one obtained for the whole ensemble.

With the upcoming \emph{Gaia} data releases and the on-going TESS mission our approach will be applicable to the majority of known and still to be discovered \bceph{} stars. With a longer time base and improved instrumental corrections, \emph{Gaia} will deliver many more new \bceph{} stars with future data releases. But even with the current data, expanding our work to all known \bceph{} stars by searching for lower amplitude pulsations in the \emph{Gaia} data is possible. Identifying the majority of main modes in \bceph{} stars will further improve our understanding of \bceph{} stars from a population point-of-view while delivering the highly needed input to ensemble forward modelling. Complementary to our photometric analysis of fainter \bceph{} stars, the upcoming CubeSpec mission \citep{Raskin2018, Bowman2022CubeSpec} will target bright \bceph{} stars with high-resolution time series spectroscopy from space to determine their pulsation modes, 
while the Global Asteroseismology Project is targetting mode identification from long-term multi-colour continuous ground-based observations \citep{Shitrit2024}. These ongoing and future projects will
revive studies of massive pulsators and unlock their full asteroseismic potential.

\begin{acknowledgements}
We thank the referee for the detailed and critical review that helped to improve the paper.
We thank Tim Bedding, Dominic Bowman, Siemen Burssens, Luc IJspeert, Kalarickal Sreenivas, and Vincent Vanlaer for useful discussions.
The research leading to these results has received funding from the KU\,Leuven Research Council (grant C16/18/005: PARADISE).
CA and MV acknowledge financial support from the European Research Council (ERC) under the Horizon Europe programme (Synergy Grant agreement N$^\circ$101071505: 4D-STAR). While partially funded by the European Union, views and opinions expressed are however those of the author(s) only and do not necessarily reflect those of the European Union or the European Research Council. Neither the European Union nor the granting authority can be held responsible for them.
JDR and CA acknowledge the Belgian Federal Science Policy Office (BELSPO) for the provision of financial support in the framework of the PRODEX Programme of the European Space Agency (ESA)
    This research has made use of NASA's Astrophysics Data System
    Bibliographic Services and of the SIMBAD database and the VizieR
    catalogue access tool, operated at CDS, Strasbourg, France.
    This paper includes data collected by the TESS mission. Funding for the TESS mission is provided by the NASA's Science Mission Directorate.
    This work has made use of data from the European Space Agency
    (ESA) mission \emph{Gaia} (\url{https://www.cosmos.esa.int/gaia}),
    processed by the \emph{Gaia} Data Processing and Analysis
    Consortium (DPAC,
    \url{https://www.cosmos.esa.int/web/gaia/dpac/consortium}). Funding
    for the DPAC has been provided by national institutions, in
    particular the institutions participating in the \emph{Gaia}
    Multilateral Agreement.
    \newline
    \textbf{Software:}
    This work made use of \texttt{Topcat} \citep{2005ASPC..347...29T}.
    This research made use of \texttt{Lightkurve}, a Python package for Kepler and TESS data analysis \citep{lightkurve}.
    This research made use of the following \texttt{Python} packages:
    \texttt{corner} \citep{corner};
    \texttt{emcee} \citep{ForemanMackey2013};
    \texttt{IPython} \citep{ipython};
    \texttt{MatPlotLib} \citep{Hunter:2007};
    \texttt{NumPy} \citep{numpy2020};
    \texttt{Pandas} \citep{pandas};
    \texttt{SciPy} \citep{scipy};
    \texttt{scikit-learn} \citep{scikitlearn}.

\end{acknowledgements}

\bibliographystyle{aa} 
\bibliography{BCephs.bib} 


\begin{appendix}
\section{Detrending and special photometry}
\label{app:photometry}

\begin{figure}
    \centering
    \includegraphics[width=\columnwidth]{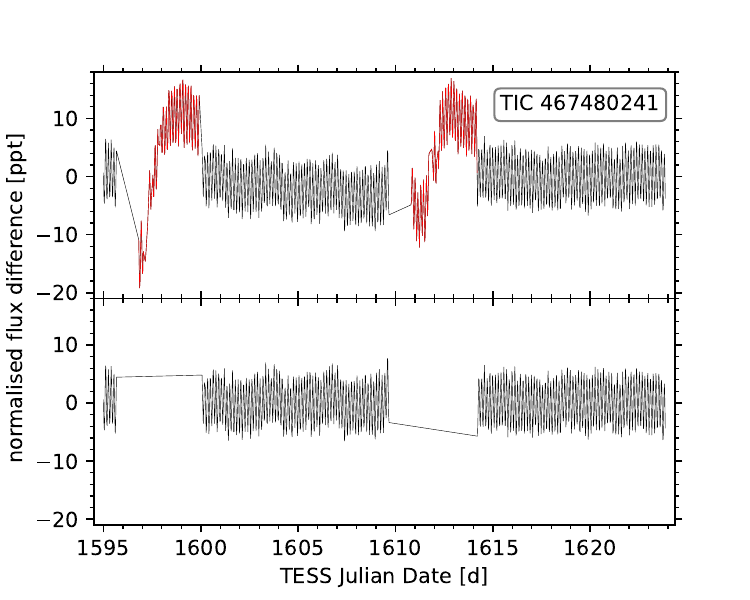}
    \caption{Example light curve before (top) and after (bottom) the final detrending. Sections highlighted in red in the top panel have been removed.}
    \label{fig:example_detrending}
\end{figure}

In order to use the pulsation modes of the $\beta\,$Cep stars in our sample we ensured 1) minimal contribution by contaminating stars (Sect.~\ref{sec:reduction}) and 2) a clean light curve containing as few instrumental signals as possible. We removed instrumental and long-term trends after excluding intervals which proved problematic in multiple stars. We preferred excluding these intervals over complicated detrending steps to avoid overcorrection. Further, some weak long-term trends sometimes remained and the rising or falling curves at the sector edges were not always easily corrected. Hence, we detrended blocks of data separated by a gap of at least three hours with second-order polynomials. In the example shown in Fig.~\ref{fig:example_detrending}, more than ten stars in our sample display similar instrumental effects at the selected times. We note that in the top panel, a downward slope is present in the first half of the sector which has been successfully removed by our detrending. Afterwards, we $\sigma$-clipped the light curves with a threshold of 5\,$\sigma$. Finally, in 18 light curves, hand-picked time intervals or even entire sectors were removed if they displayed unusual behaviour likely caused by instrumental effects or contamination and $\sigma$-clipped them again.

\begin{figure*}
    \sidecaption
    \includegraphics[width=12cm]{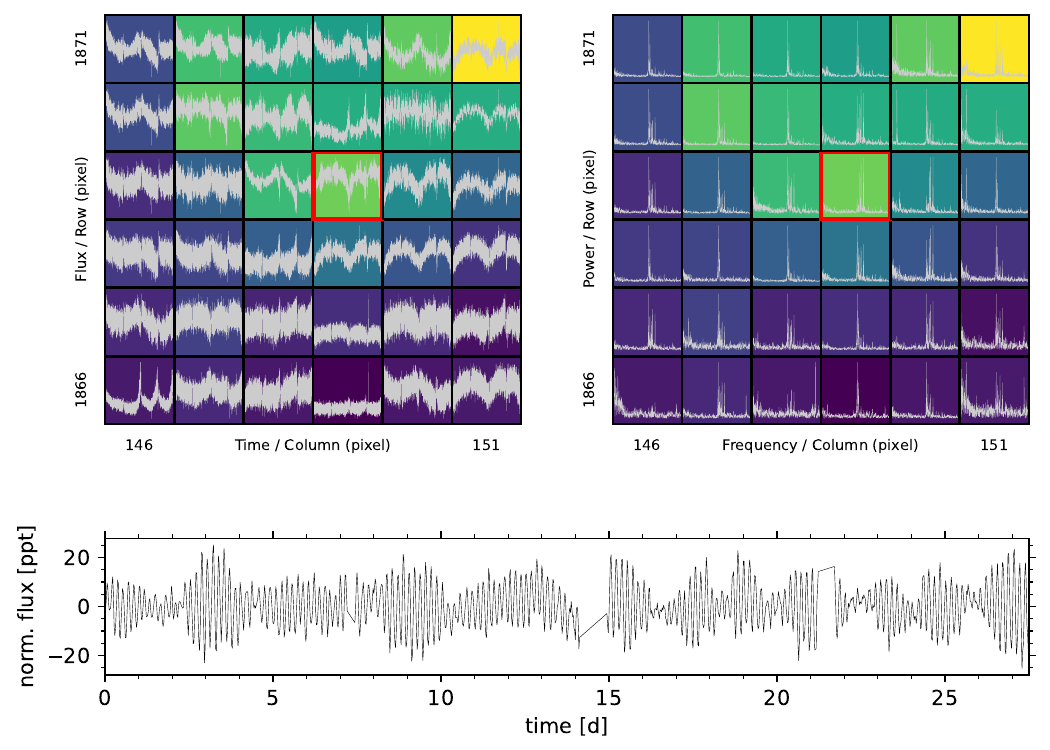}
    \caption{Example of a single pixel TESS light curve. The target TIC 264026227, member of the open cluster NGC\,5606, is contaminated by several nearby stars. Top left: The TESS FFI cut-out of the field around TIC 264026227, with the pixel used for or final light curve marked in red. The colour scale shows the flux intensity with yellow being brighter pixels. Over-plotted in each pixel is its light curve. Top right: Same as left but with the periodogram over-plotted. Bottom: Final light curve shows clear \bceph{} variability after additional cleaning and detrending beyond what is shown in the top left panel.}
    \label{fig:singlepix}
\end{figure*}

For 15 stars, we used single-pixel photometry, in order to avoid contamination from nearby sources. One example is given in Fig.~\ref{fig:singlepix}. Although the main \bceph{} pulsation is clearly visible in the periodograms of nearby pixels, several additional peaks emerge for these stars too, indicating a periodic contamination from other sources. The particular case of TIC 264026227 is a member of the open cluster NGC\,5606, which host other massive stars in close proximity of TIC 264026227. We note that the \emph{Gaia} detected frequency does correspond to the main TESS frequency, although the light curves in nearby pixels have other periodic components that are more prominent. Hence, using a single pixel for the photometry of this star ensures the cleanest possible extraction.

\section{Additional Hertzsprung-Russell diagrams and mode identification}

\begin{figure}
    \includegraphics[width=\columnwidth]{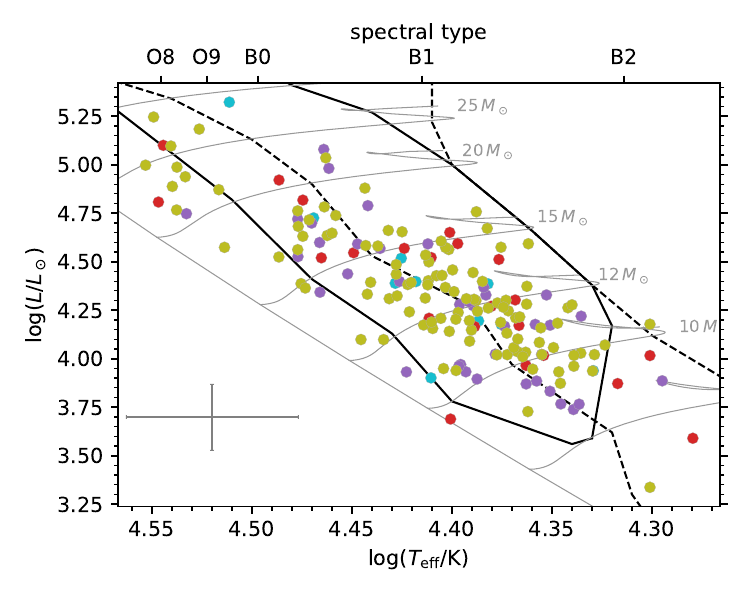}
    \caption{Hertzsprung-Russell diagram similar to Fig.~\ref{fig:HRD_all} but including only the \nlc{} stars from our \emph{Gaia}-detected sample having TESS light curves. Each star is colour-coded according to its light curve shape and content in the Fourier domain with purple indicating mono-periodic stars, olive multi-periodic, and red hybrid pulsators. The stars classified as \emph{other} are shown in cyan and are discussed in the main text.}
    \label{fig:HRD_class}
\end{figure}

\begin{figure}
    \includegraphics[width=\columnwidth]{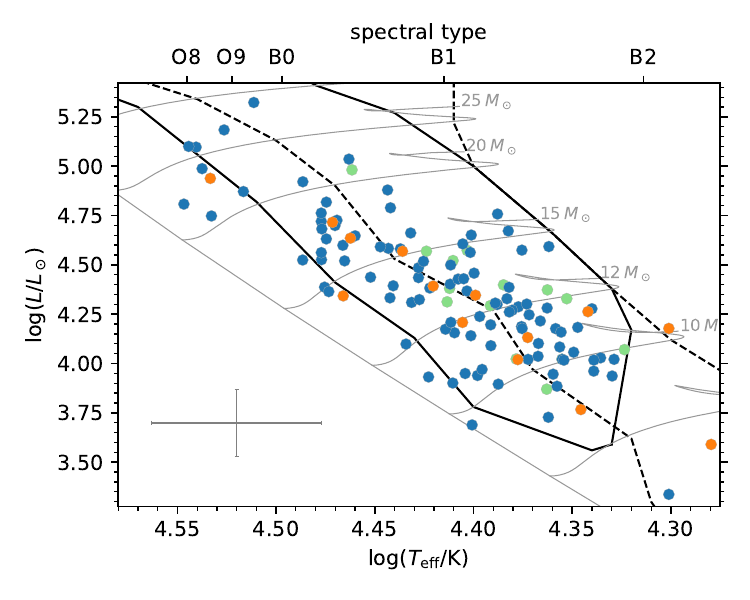}
    \caption{Hertzsprung-Russell diagram similar to Fig.~\ref{fig:HRD_all} featuring only the stars with identified modes from multi-colour photometry and colour-coded by their mode degree: $l=0$ (green), $l=1$ (blue), $l=2$ (orange).}
    \label{fig:HRD_mc}
\end{figure}

For convenience of the interested reader, we provide some extra HRDs with colour-coding by the light curve classification (Fig.~\ref{fig:HRD_class}, Sect.~\ref{sec:classification}) and mode degree as identified from multi-colour photometry (Fig.~\ref{fig:HRD_mc}, Sect.~\ref{sec:mcphot}). Both diagrams show no obvious structure in the light of the current study.

\begin{figure}
    \includegraphics[width=\columnwidth]{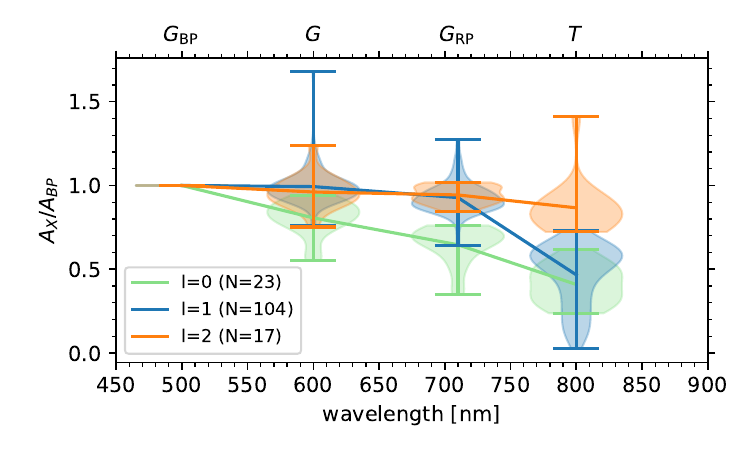}
    \caption{Same as Fig.~\ref{fig:violin_manual} but based on the probabilistic assignment of mode identifications from multi-colour photometry for all modes with $p>0.6$.}
    \label{fig:violin_mc}
\end{figure}

Figure~\ref{fig:violin_mc} shows the distribution of amplitude ratios for the set of mode degrees as determined by our probabilistic approach. Qualitatively, the diagram resembles Fig.~\ref{fig:violin_manual}. 
While the \emph{Gaia} amplitude ratios of the non-radial pulsators overlap significantly,
the TESS amplitudes for the non-radial modes are well separated. As expected, the amplitude ratios of the radial modes drop stronger over the whole wavelength range than the amplitude ratios of the non-radial modes.

\section{Identification of the radial order}
\label{sec:findn}

The grid modelling presented in Sect.~\ref{sec:model} needs a full mode identification in terms of $n$, $l$, and $m$. Since our model selection is based on only the dominant mode degree, the radial order is still ambiguous and cannot be fitted in our scheme. To find the most likely radial order, we search the best fitting model for $n \in \{-2, -1, 0, 1, 2\}$ given the mode degree as determined by the multi-colour photometry. 

\begin{figure*}
    \includegraphics[width=\textwidth]{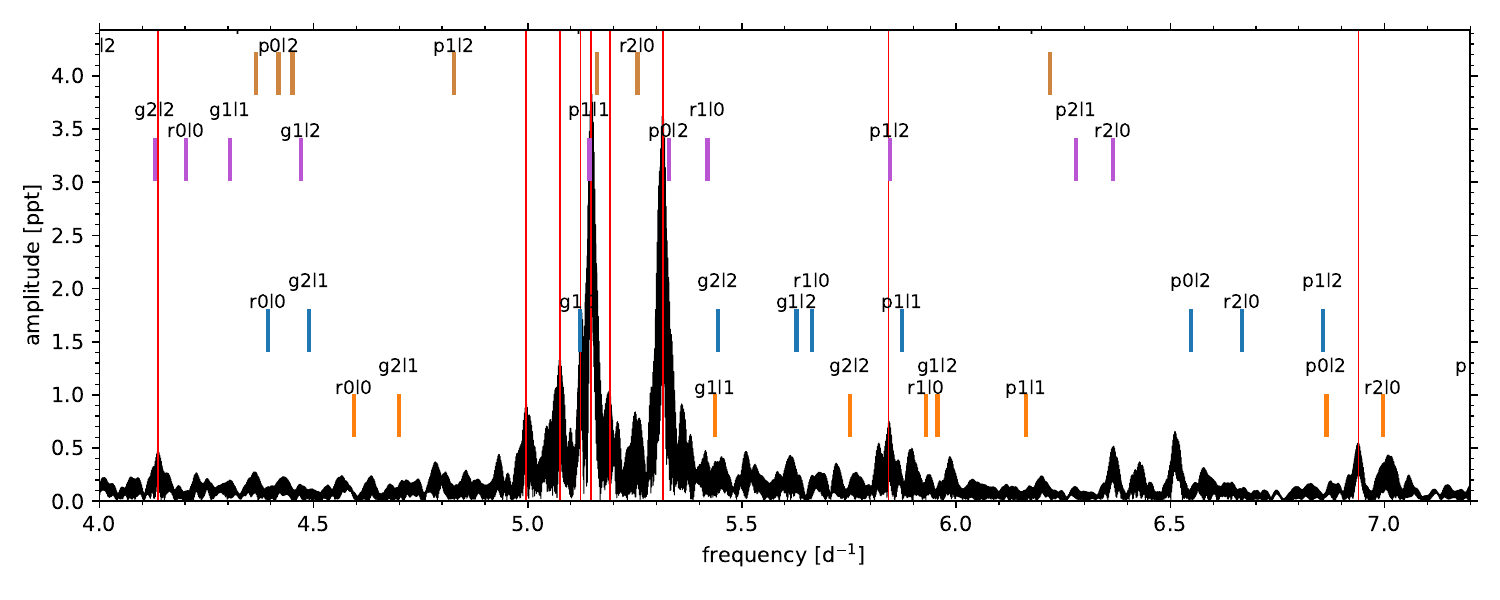}
    \caption{Diagnostic plot to identify the radial order $n$. In the background, we show in black the frequency spectrum of TIC\,105756939. The thin red vertical lines indicate the positions of extracted frequencies. The short coloured lines show the positions of the model predictions for the modes of different radial orders. From bottom to top, we show we show $n=-2, -1, 1, 2$ in orange, blue, purple, brown, respectively.}
    \label{fig:findn}
\end{figure*}

For each star, we produced a diagnostic plot similar to Fig.~\ref{fig:findn}. It shows for all relevant radial order out of the selected five radial orders the positions of all twelve considered low order radial and zonal p- and g-modes over-plotted on the frequency spectrum. In these diagrams, we examined the positions of these modes with respect to all identified (and unidentified low-SNR) mode frequencies.

In the chosen example, the radial order $n=1$ is able to approximately explain detected modes at frequencies both higher and lower than the identified dominant mode. However, it fails to reproduce the highest detected frequency in Fig.~\ref{fig:findn}. Hence, we can conclude that TIC\,105756939 is pulsating with $(n, l)=(1,1)$.

\section{Properties of the modelling results}

\subsection{Recovered values}
\label{app:model_details}

\begin{figure}
    \includegraphics[width=\columnwidth]{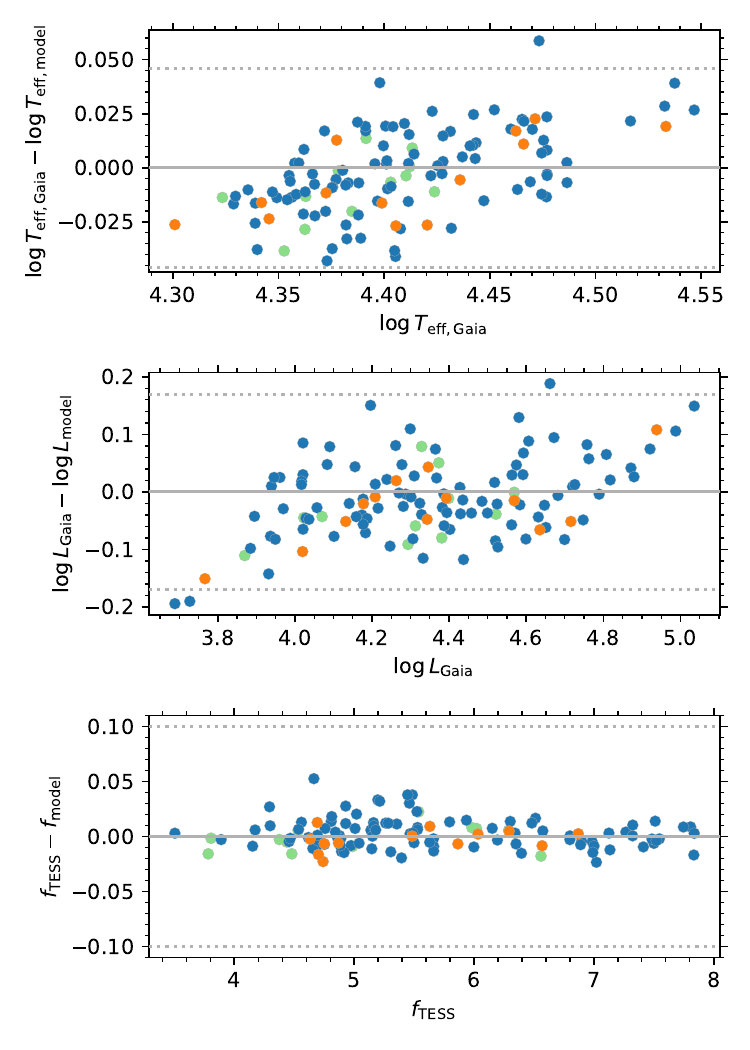}
    \caption{Comparison between the measured values and the properties recovered from the grid modelling. From top to bottom we show the effective temperature, the luminosity, and the pulsation frequency of the dominant mode. The dotted lines mark the uncertainty interval used in our fitting procedure. The colour code indicates the main pulsation mode: $l=0$ green, $l=1$ blue, $l=2$ orange.}
    \label{fig:recovered}
\end{figure}

By comparing the input and recovered values from the grid modelling, we can assess the accuracy and understand potential biases of the modelling. The three panels of Fig.~\ref{fig:recovered} show the measured and the differences to the recovered $\log T_\mathrm{eff}$, $\log L$, and $f_0$. There is a slight bias towards higher effective temperatures as the majority of best fitting modes are located below zero. A similar effect can be found for the luminosity, where the lower luminosity stars tend to be fitted with higher luminosity models. We also note that stars close to the edges of the model grid are affected by the lack of models and their luminosity is shifted towards the model grid, leading to the structure seen at low and high luminosities. However, all models are within the estimated uncertainty of the measurements. The frequencies are well recovered and deviate less than the uncertainty interval (0.1\,d$^{-1}$) from the measurements.

\subsection{Ages}

Due to their short lifetimes, massive stars are young in a Galactic context and can often still be found in their birth environment such as OB associations or open clusters. The \emph{Gaia} mission has also enabled the large scale identification and description of open clusters, as well the attribution of individual stars to these groups. We use the catalogues from \cite{CantatGaudin2020}, \cite{Kounkel2020}, and \cite{Hunt2023} to identify \bceph{} stars in open clusters and get an independent handle on their ages. We find 34 (15\,\%) of our sample stars to be hosted by an open cluster or co-moving group.

\begin{figure}
    \includegraphics[width=\columnwidth]{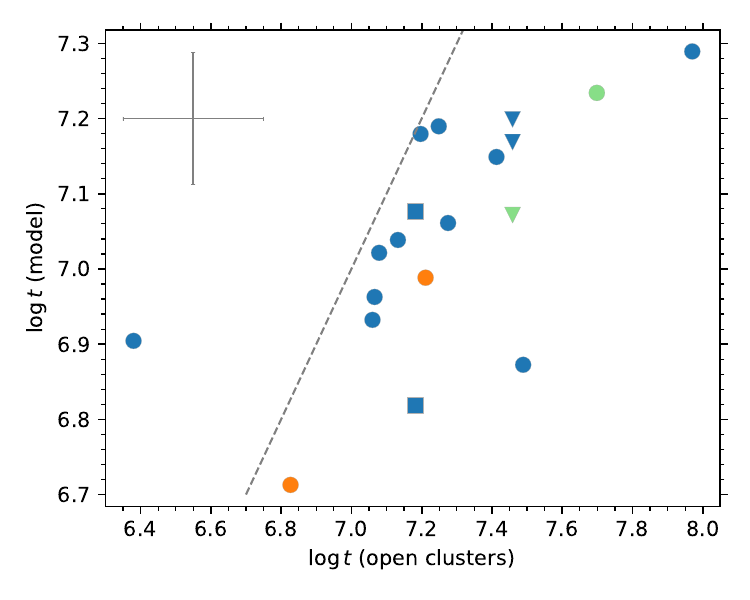}
    \caption{Comparison between ages for stars in open clusters and our asteroseismic ages. Members of NGC\,637 are shown with squares and those of NGC\,5606 with triangles. The typical uncertainty is given in the upper left corner.}
    \label{fig:age_compare}
\end{figure}

All the three open cluster catalogues infer the cluster ages based on their colour-magnitude diagrams and isochrones. Fig.~\ref{fig:age_compare} compares our asteroseismic age to these isochronal ages. Overall our asteroseismic ages are younger than the isochronal estimates, pointing to systematic uncertainty caused by the different input physics of the isochrone and asteroseismic models in the adopted grids. However, taking into account the typical ages uncertainty of 0.2\,dex for the isochronal measurements \citep{Hunt2023} and 0.1\,dex for the asteroseismic age, the ages are mostly in agreement.

For the two open cluster NGC\,637 and NGC\,5606 multiple \bceph{} members are in our sample. We show their members with distinct symbols in Fig.~\ref{fig:age_compare}. As these stars are part of coeval populations, their ages should agree within the uncertainties, unless they would be binary or multiple products. We find their asteroseismic ages exhibiting a large spread even beyond the uncertainties in the case of NGC\,5606. Detailed modelling of these stars should take into account their coevality to better constrain the asteroseismic inferences and investigate if some of them could be multiples or merger products.

\section{Light curves and frequency spectra}

\subsection{Light curves for all studied stars}
\label{app:all_lc}
\begin{figure*}
    \includegraphics[width=\textwidth]{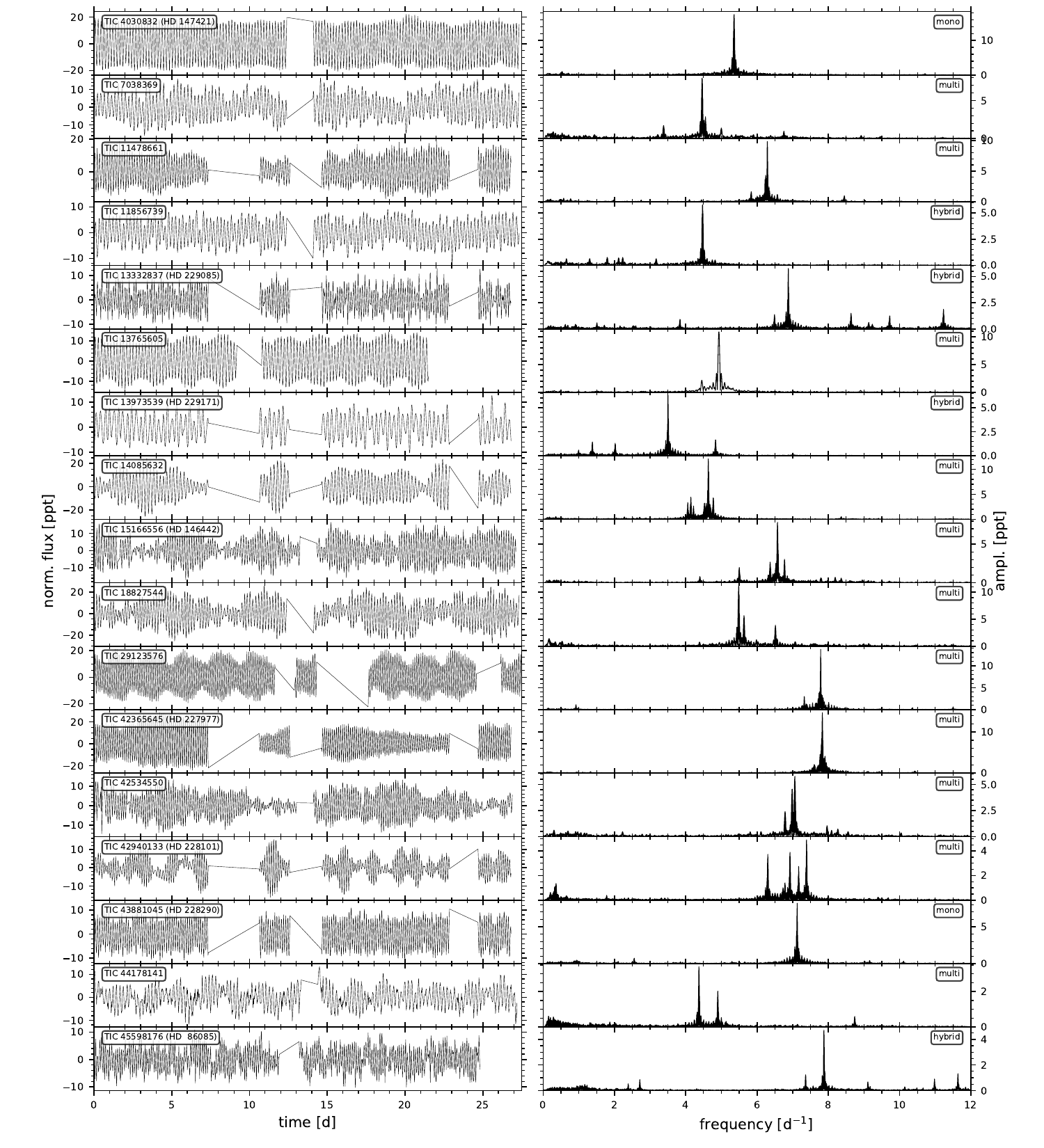}
    \caption{Light curves (left) and periodograms (right) for all studied stars. The light curves show only the first available TESS sector for every star for illustrative purpose. The periodogram is based on all available TESS data. For each star, we indicate its light curve classification in the top right.}
    \label{fig:all_lcs}
\end{figure*}

\begin{figure*}\ContinuedFloat
    \centering
    \includegraphics[width=\textwidth]{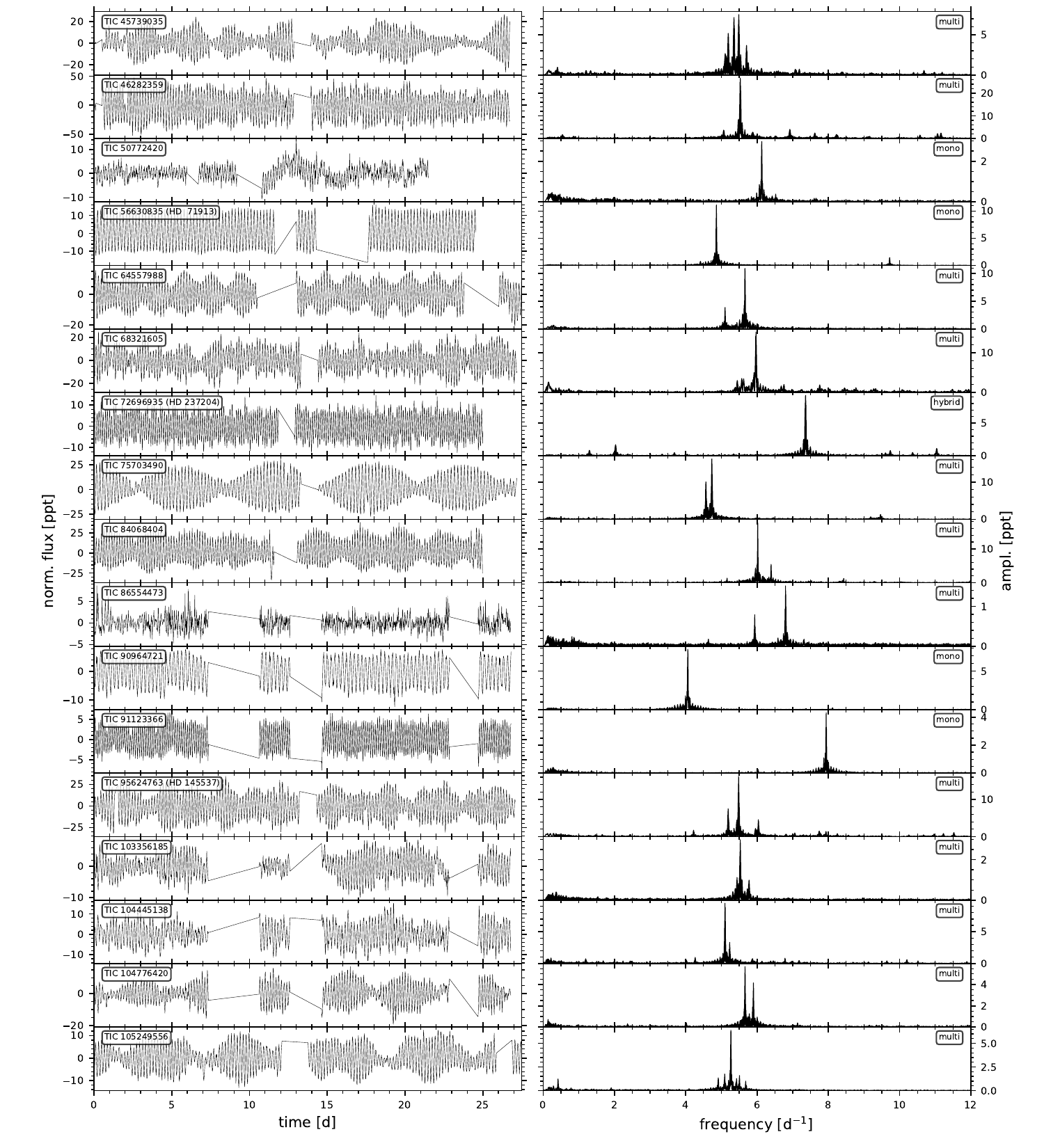}
    \caption{continued.}
\end{figure*}

\begin{figure*}\ContinuedFloat
    \centering
    \includegraphics[width=\textwidth]{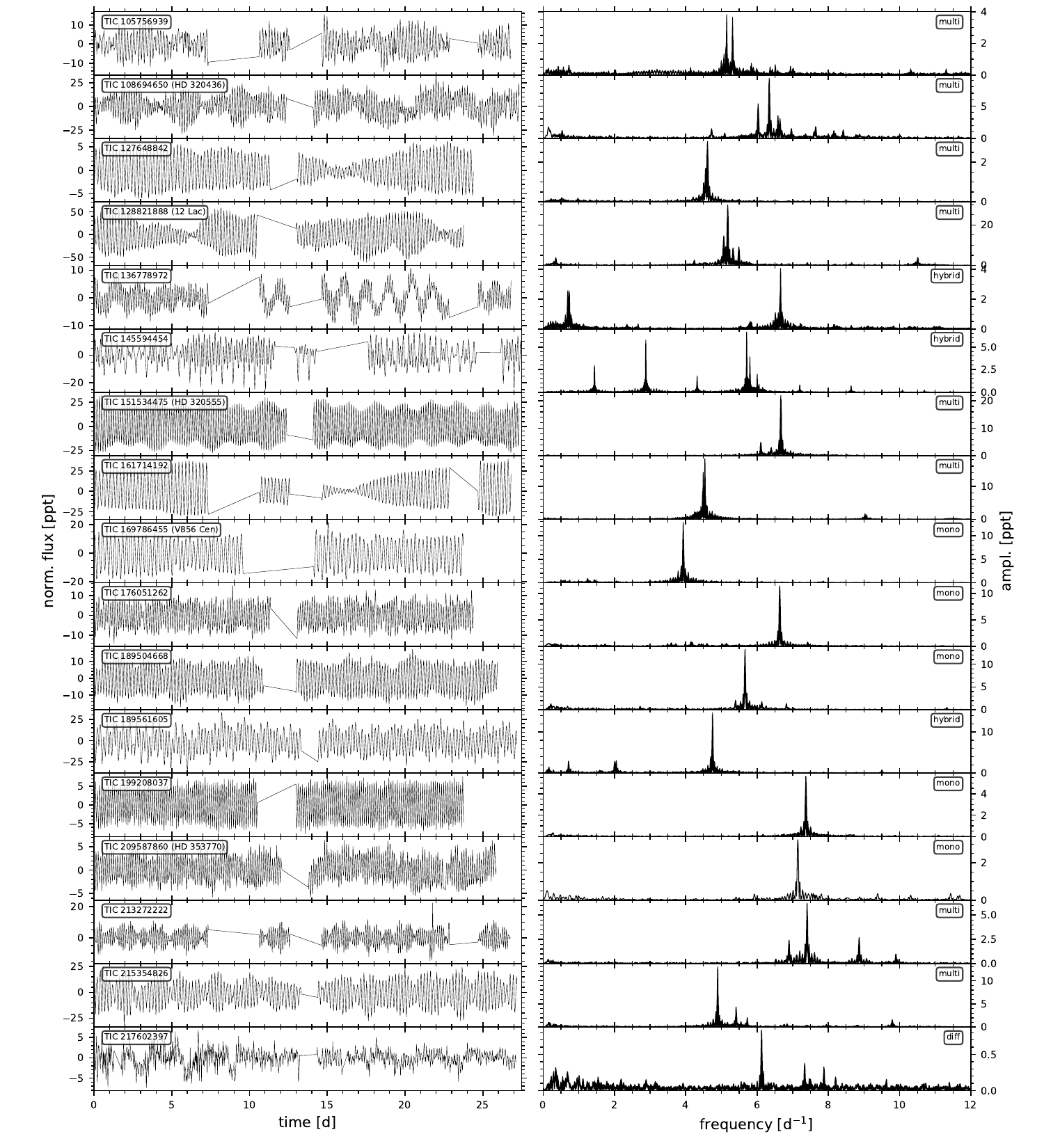}
    \caption{continued.}
\end{figure*}

\begin{figure*}\ContinuedFloat
    \centering
    \includegraphics[width=\textwidth]{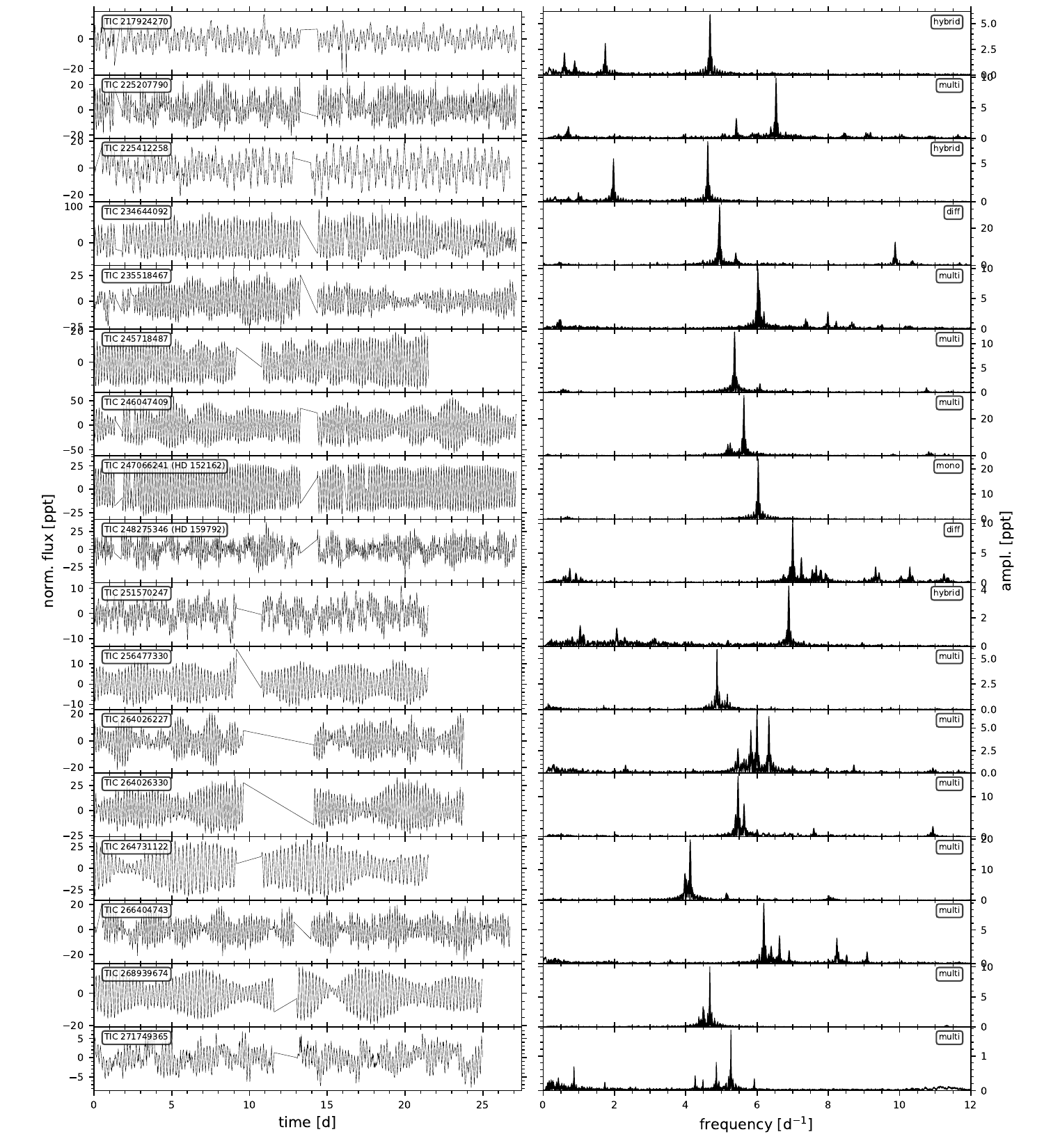}
    \caption{continued.}
\end{figure*}

\begin{figure*}\ContinuedFloat
    \centering
    \includegraphics[width=\textwidth]{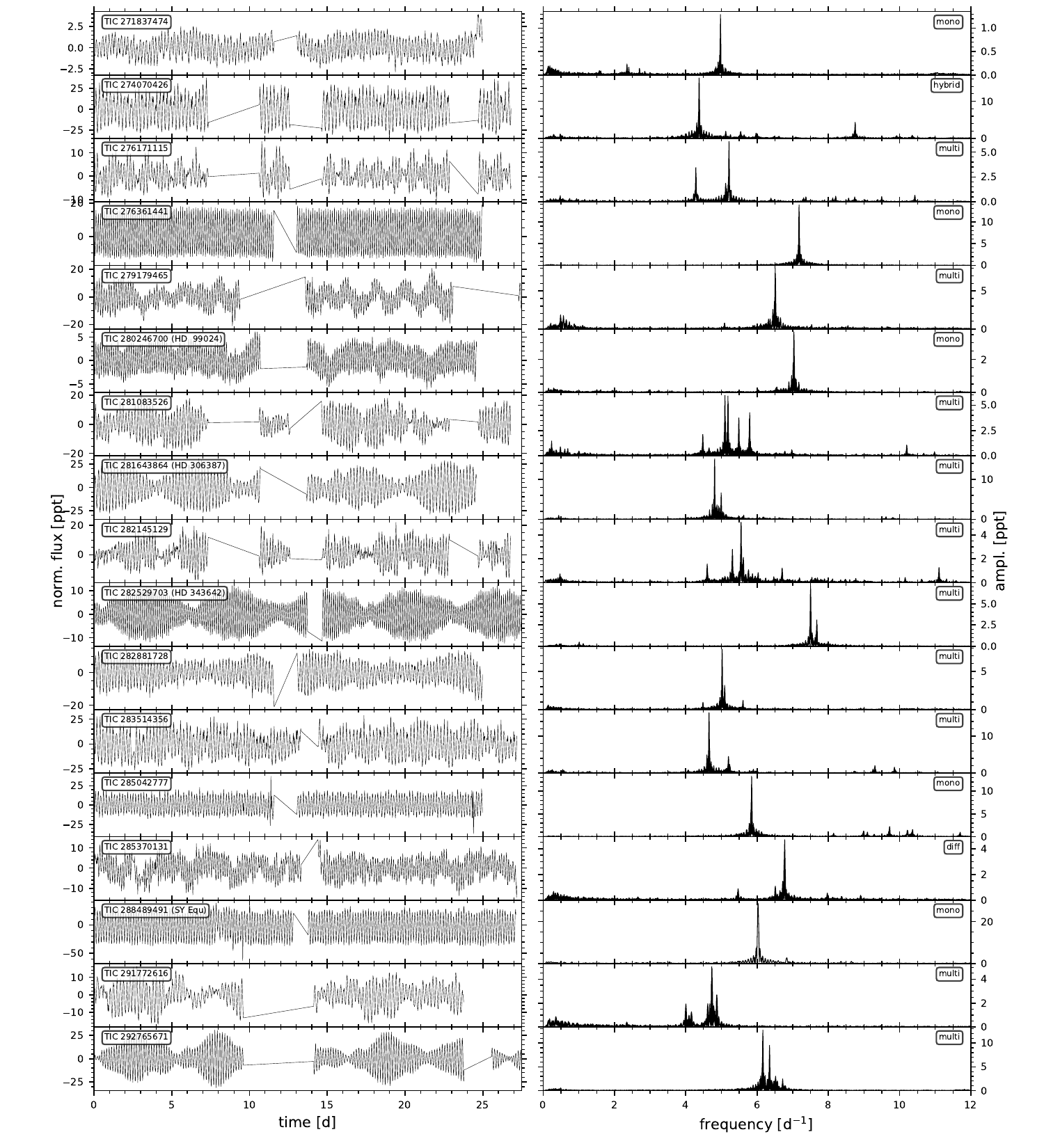}
    \caption{continued.}
\end{figure*}

\begin{figure*}\ContinuedFloat
    \centering
    \includegraphics[width=\textwidth]{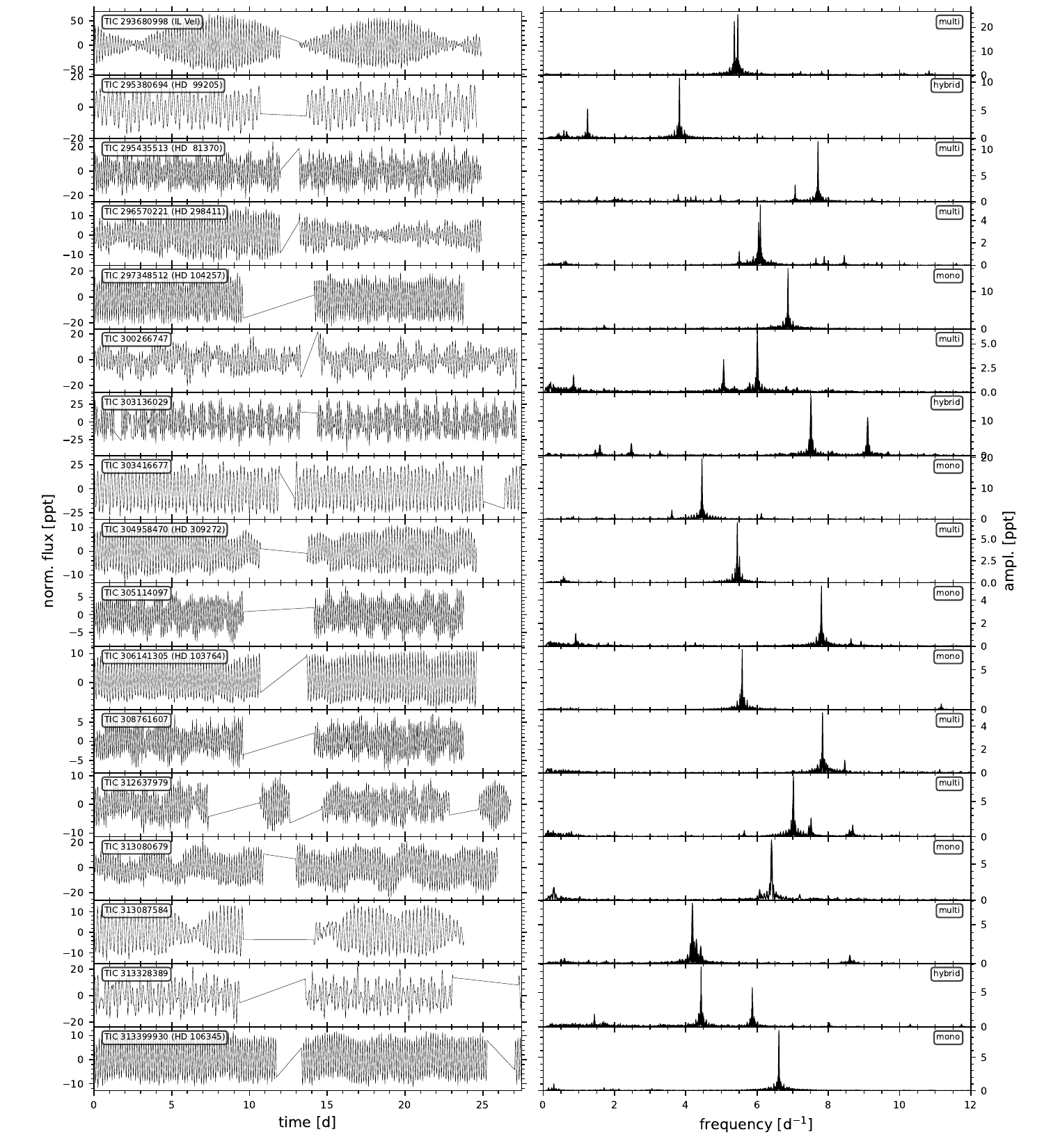}
    \caption{continued.}
\end{figure*}

\begin{figure*}\ContinuedFloat
    \centering
    \includegraphics[width=\textwidth]{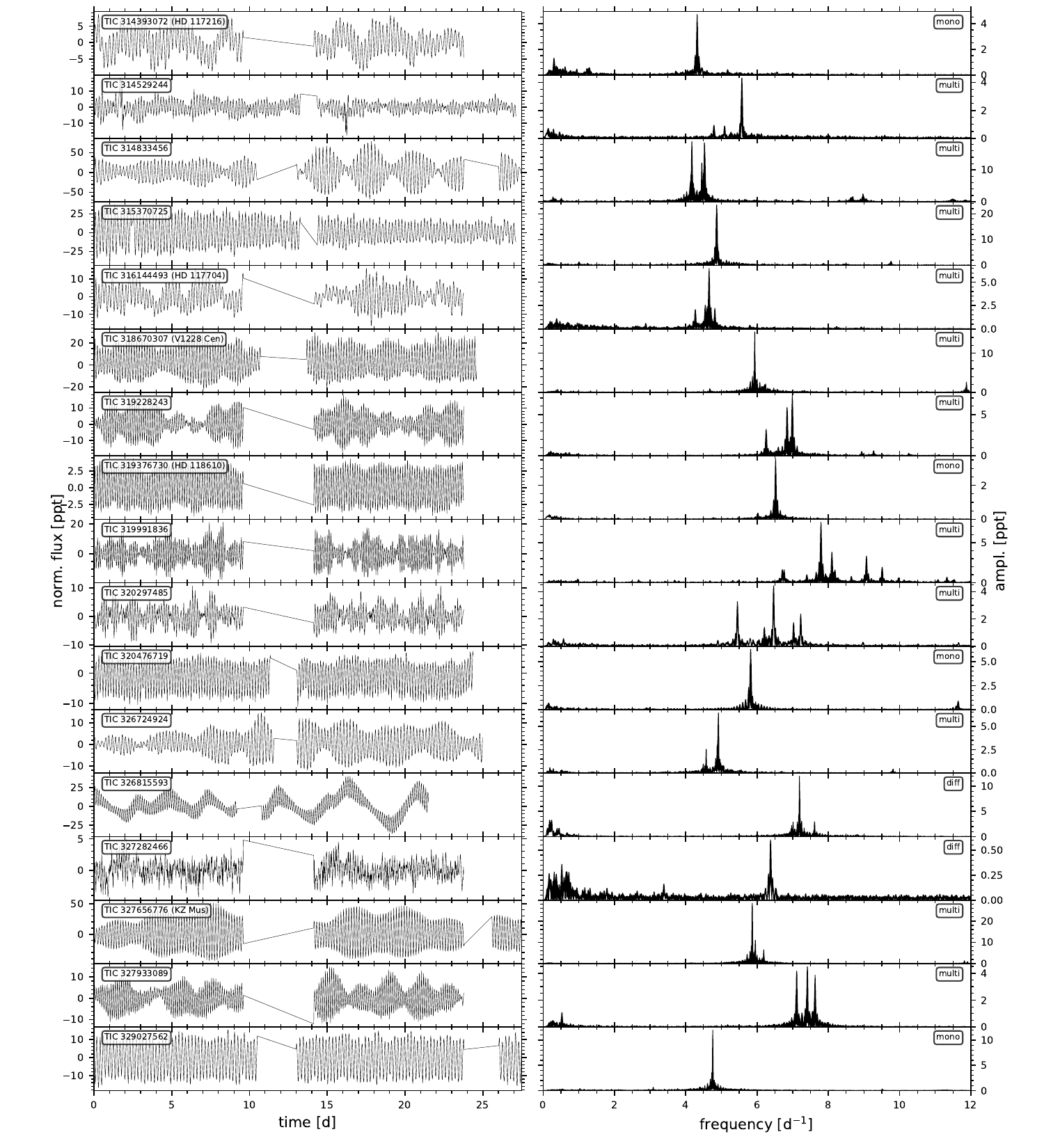}
    \caption{continued.}
\end{figure*}

\begin{figure*}\ContinuedFloat
    \centering
    \includegraphics[width=\textwidth]{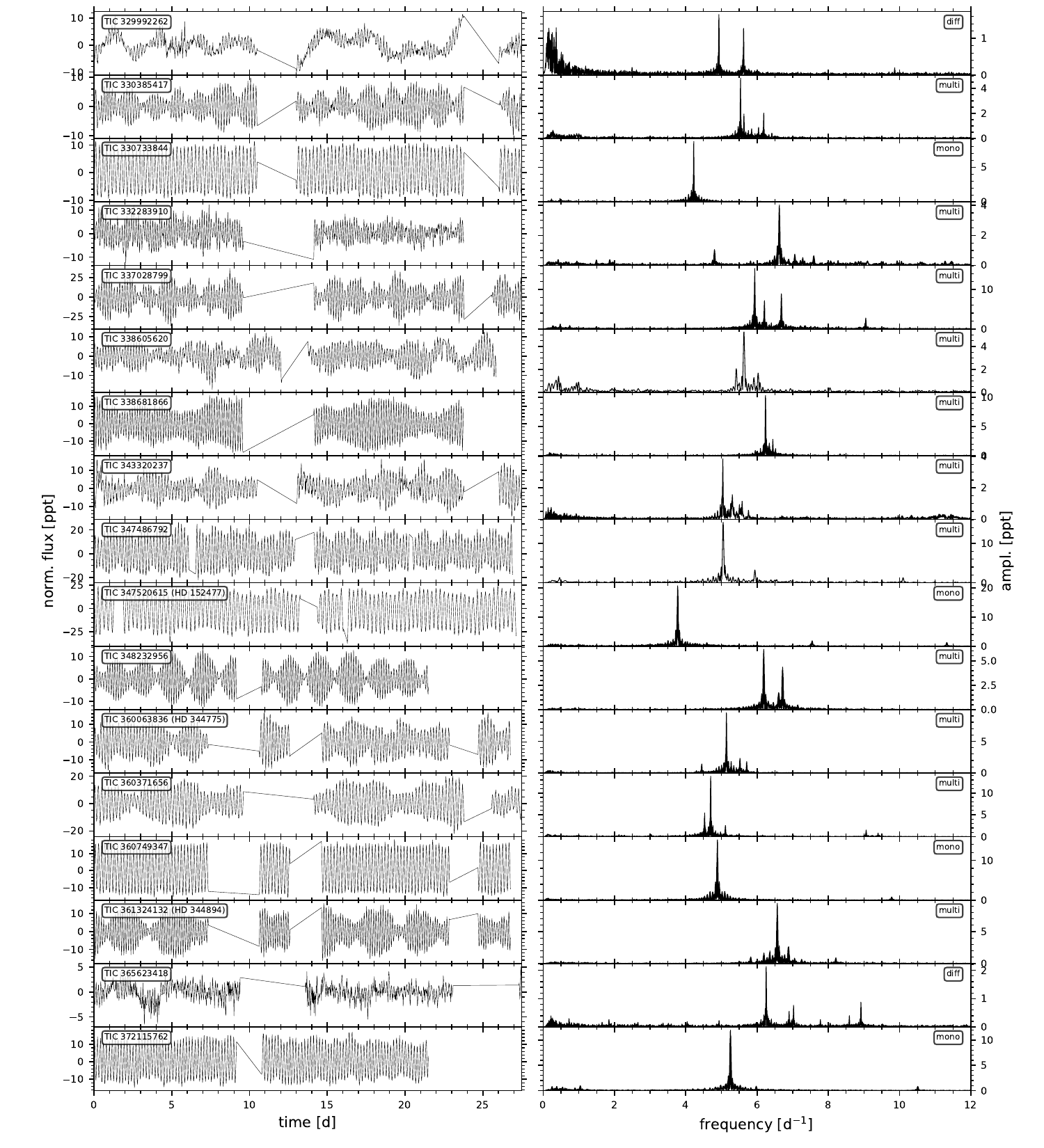}
    \caption{continued.}
\end{figure*}

\begin{figure*}\ContinuedFloat
    \centering
    \includegraphics[width=\textwidth]{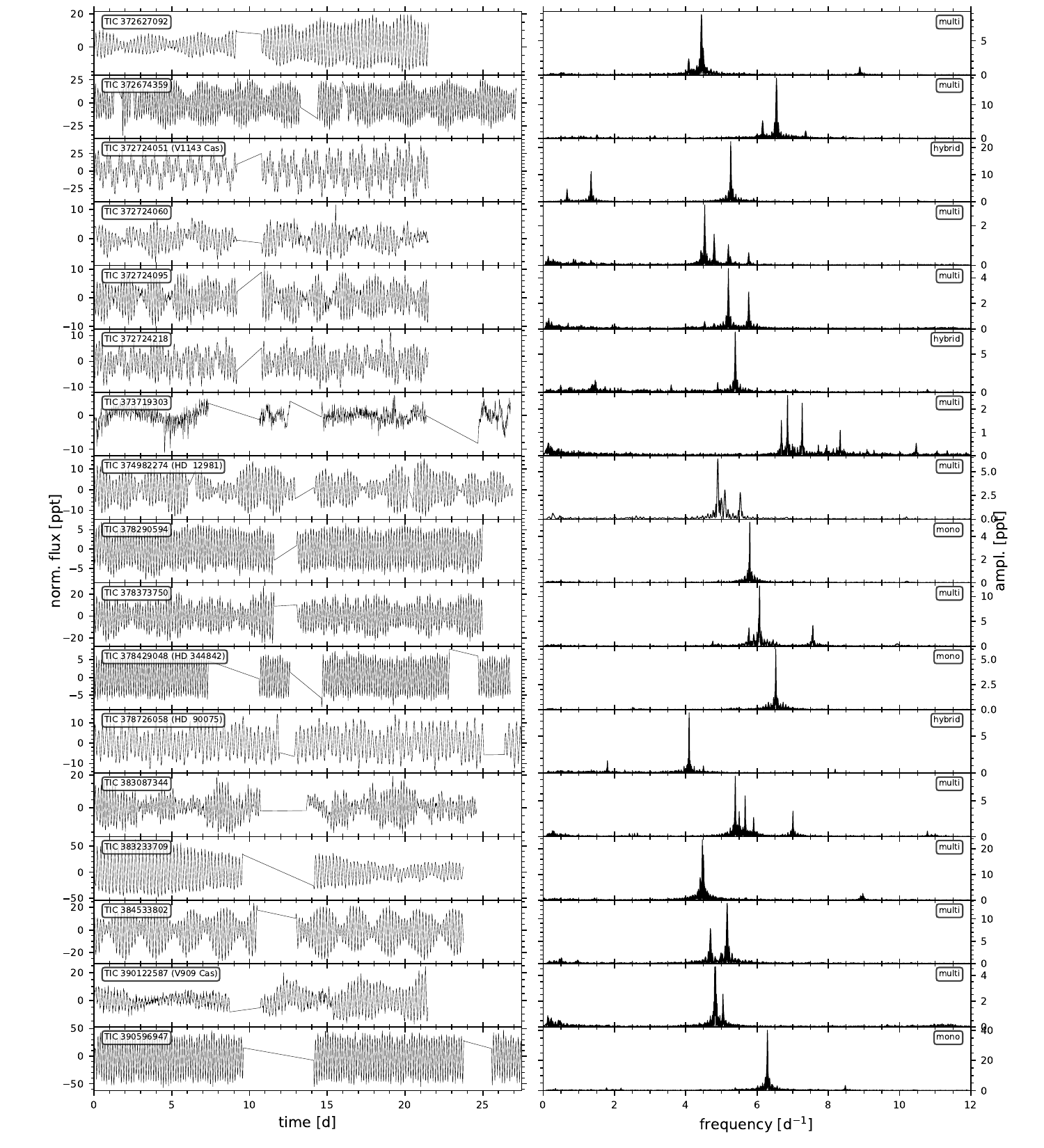}
    \caption{continued.}
\end{figure*}

\begin{figure*}\ContinuedFloat
    \centering
    \includegraphics[width=\textwidth]{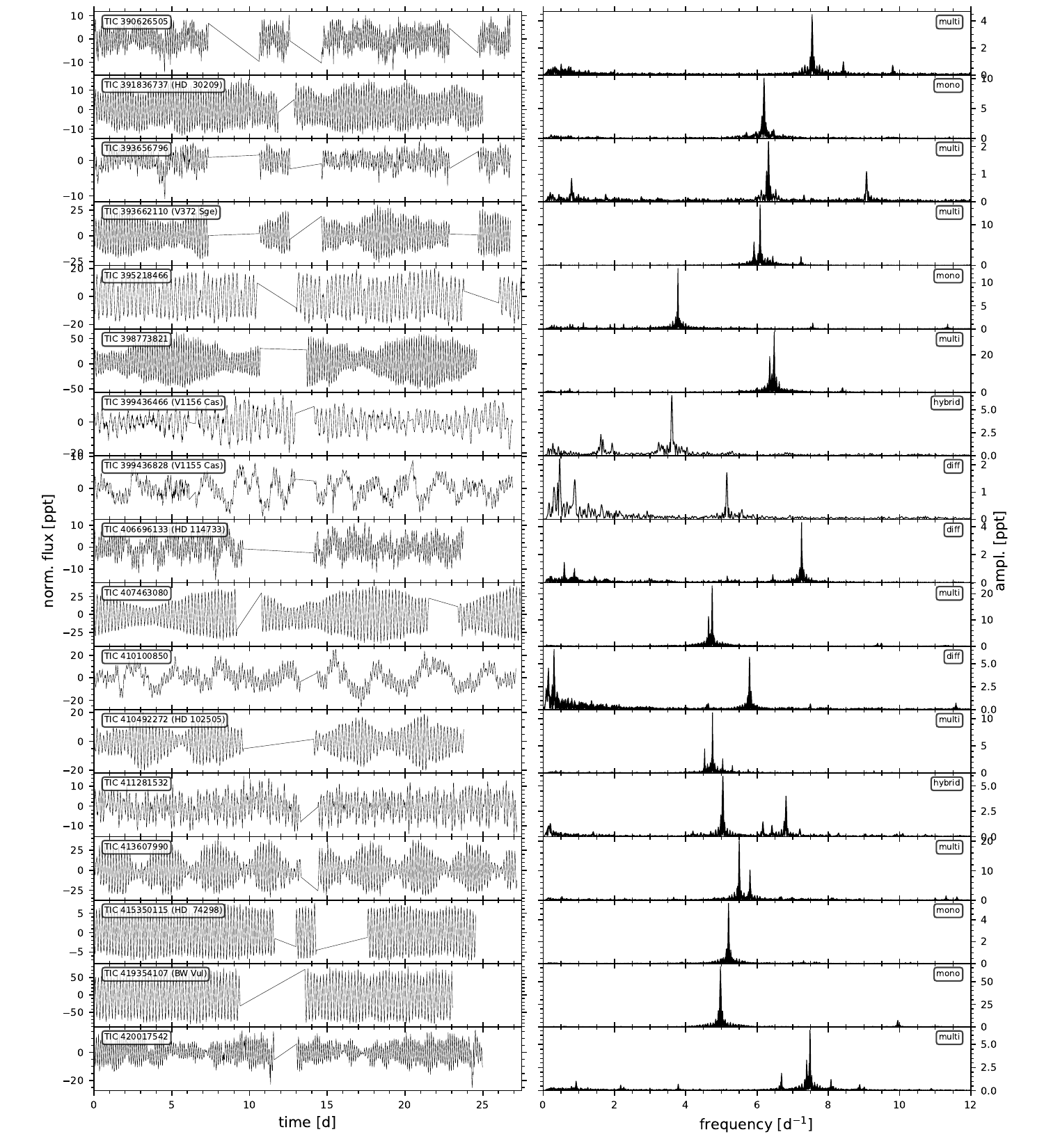}
    \caption{continued.}
\end{figure*}

\begin{figure*}\ContinuedFloat
    \centering
    \includegraphics[width=\textwidth]{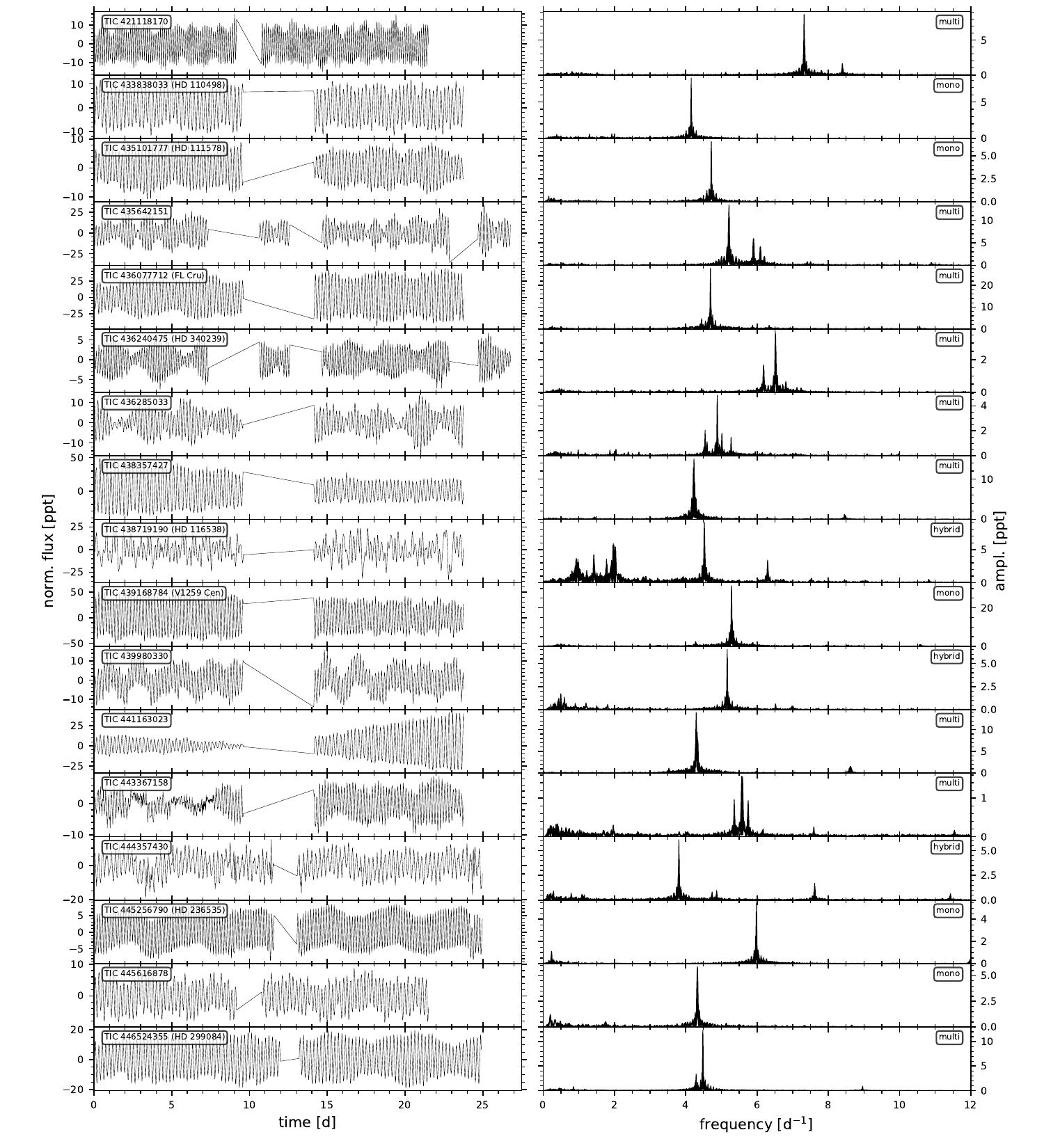}
    \caption{continued.}
\end{figure*}

\begin{figure*}\ContinuedFloat
    \centering
    \includegraphics[width=\textwidth]{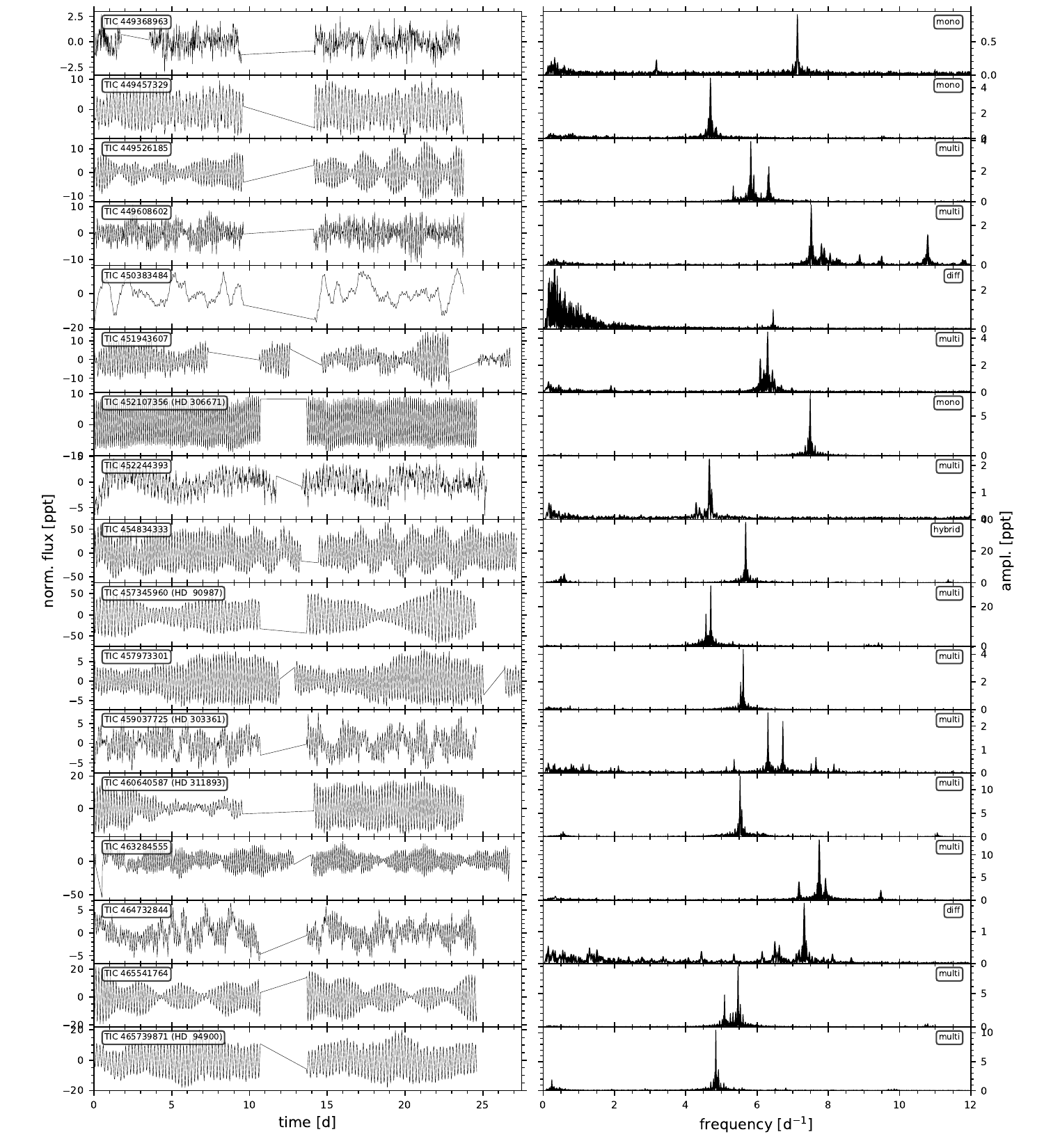}
    \caption{continued.}
\end{figure*}

\begin{figure*}\ContinuedFloat
    \centering
    \includegraphics[width=\textwidth]{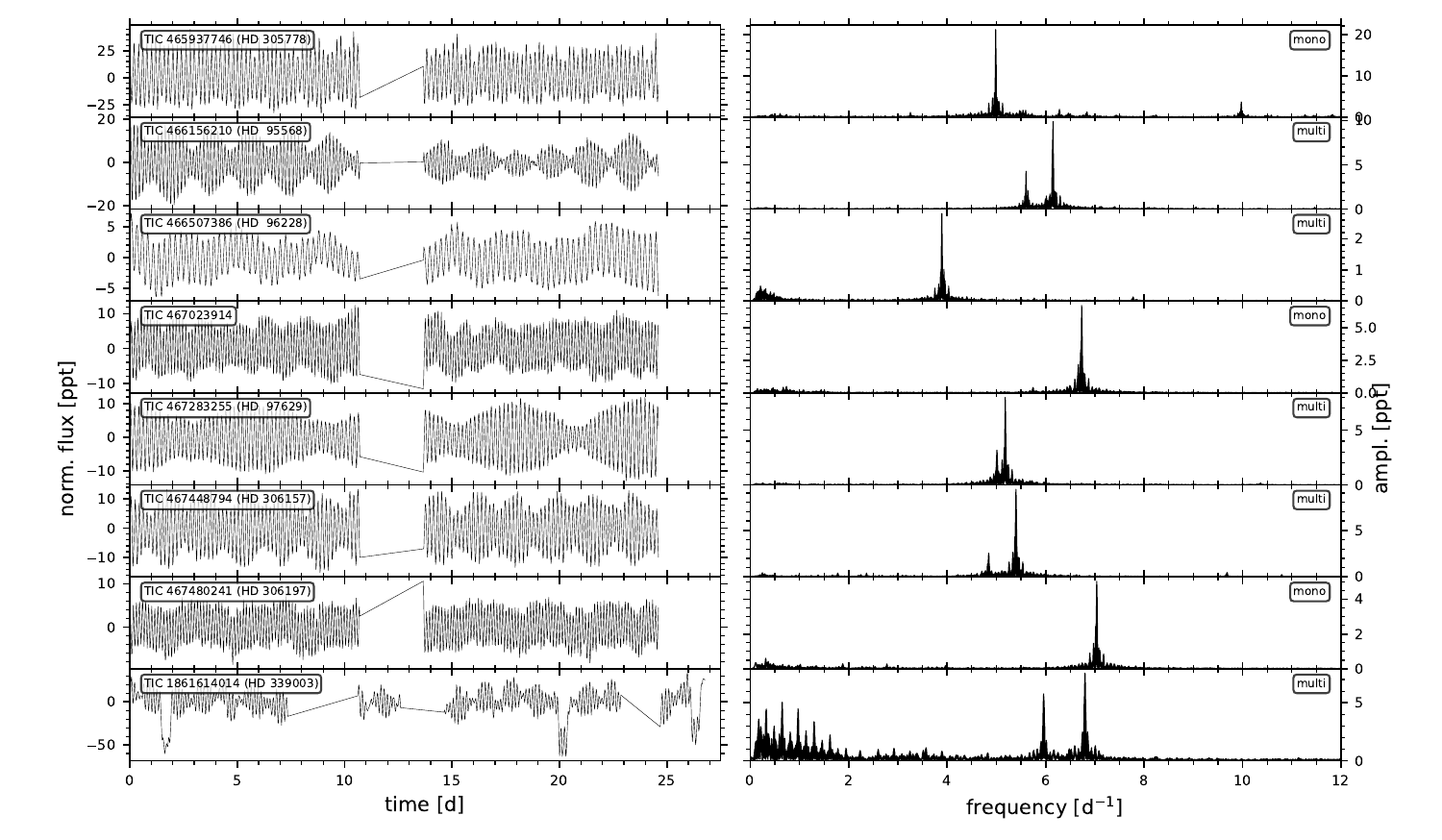}
    \caption{continued.}
\end{figure*}

\subsection{Stillstand stars} \label{app:stillstands}

\begin{figure*}
    \includegraphics[width=\textwidth]{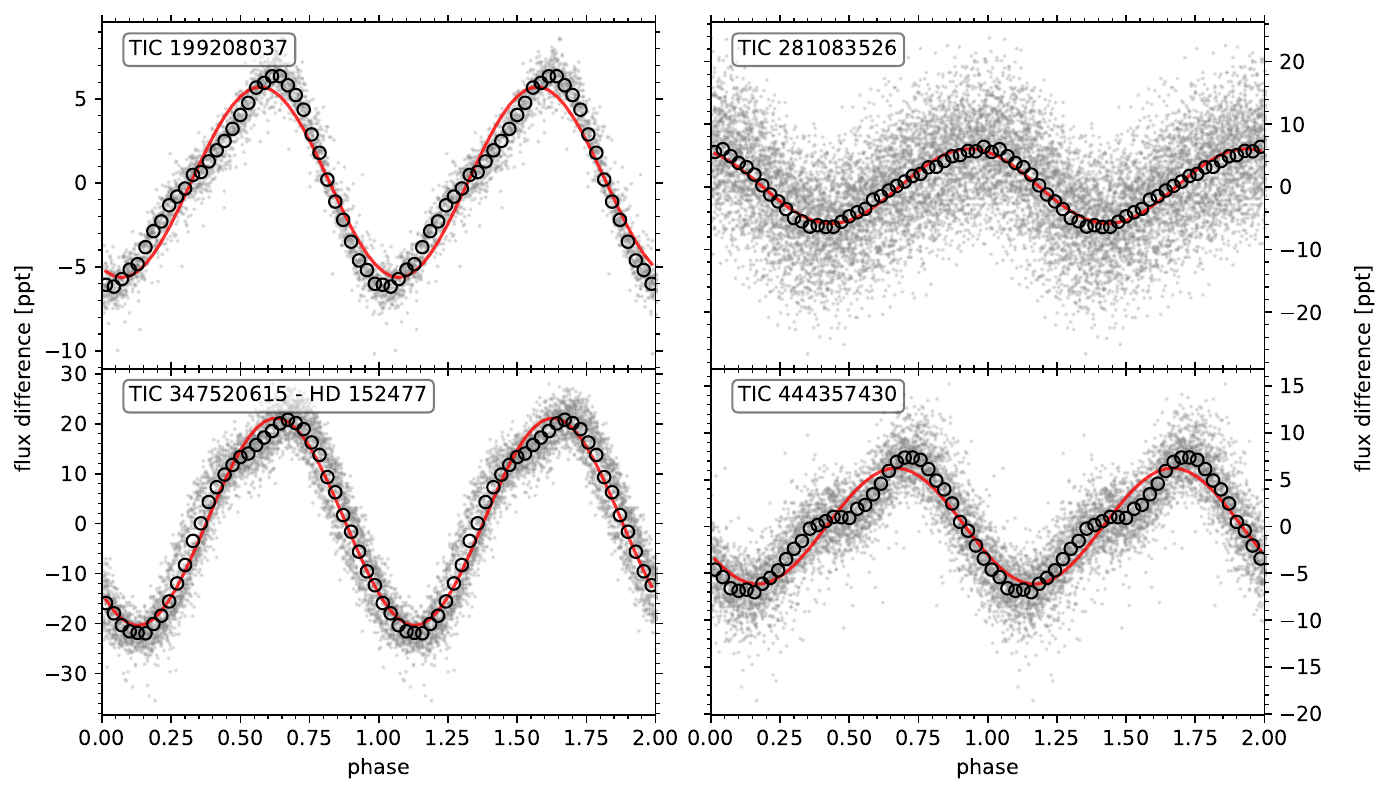}
    \caption{Same as Fig. \ref{fig:stillstand_examples}, showing an additional four stillstand stars not yet shown in the main text.}
    \label{fig:stillstand_all}
\end{figure*}

In this section, we show the remaining stars that display a stillstand in their dominant pulsation mode not yet shown in the main text. Four of these are displayed in Fig.~\ref{fig:stillstand_all}.

\begin{figure*}
    \includegraphics[width=\linewidth]{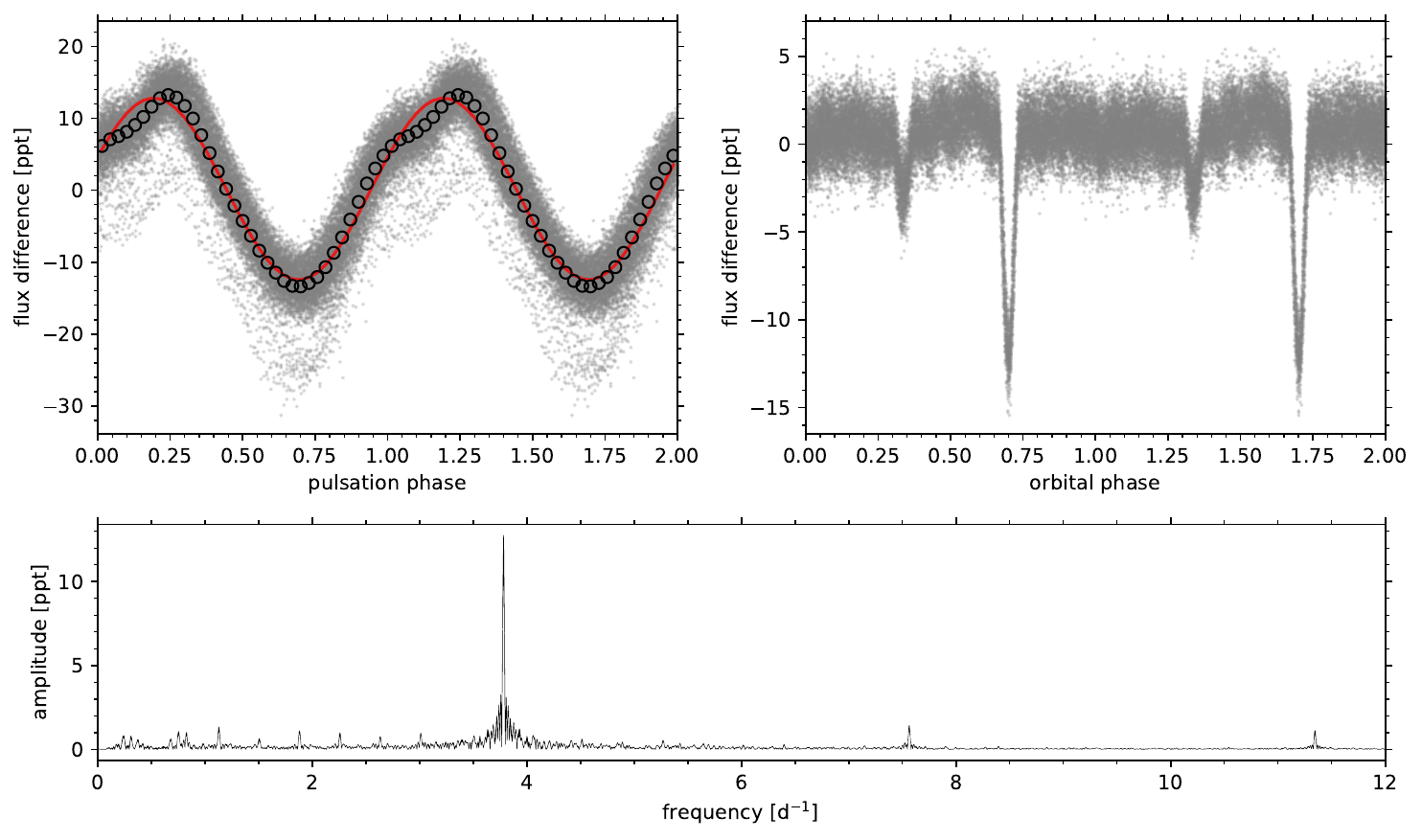}
    \caption{TIC 395218466 is an eccentric eclipsing binary with a dominant radial mode displaying a stillstand. In the upper left panel, the unbinned light curve is phase folded over its dominant pulsation period of 0.26433\,d like in Fig. \ref{fig:stillstand_examples}. In the upper right panel, the light curve is instead folded over the orbital period of 2.657983\,d after all extracted signals but the orbital harmonics were removed. The light curve shown in this figure is not binned to a 30 minute cadence as the eclipses are more clearly defined in the unbinned light curve. The bottom panel shows its original Fourier spectrum.}
    \label{fig:TIC395218466_myBeloved}
\end{figure*}

We also highlight the stillstand star TIC 395218466, a member of an eccentric, short-period eclipsing binary. Its light curve folded against the pulsation and orbital periods and corresponding Fourier transform are shown in Fig.~\ref{fig:TIC395218466_myBeloved}. By taking the median of the frequencies of the orbital harmonics divided by the harmonic number, we find it has an orbital period of $2.657983\, \pm \, \,0.000041\,$d. \texttt{STAR SHADOW}'s analysis of the timings of the eclipses \citep[see][]{IJspeert2024} suggests an orbital eccentricity of $0.5413\, \pm \, 0.0066$. Interestingly, the dominant pulsation frequency is only 0.0146\,d$^{-1}$ away from the tenth harmonic of the orbital frequency which might indicate that tides play a role in modifying or even exciting the pulsation. To our knowledge, no star displaying a stillstand has ever been observed in an eclipsing binary, let alone on an eccentric orbit with likely tidally-modified pulsations, making TIC 395218466 an especially promising target for more detailed studies of its binary and asteroseismic properties.

\subsection{Splittings}

Here, we present all identified complete split multiplets of the pulsations modes of the studied \bceph{} stars. As seen from in Sect.~\ref{app:all_lc}, some stars show more such multiplets that will be discussed elsewhere.

\begin{figure*}
    \includegraphics[width=\textwidth]{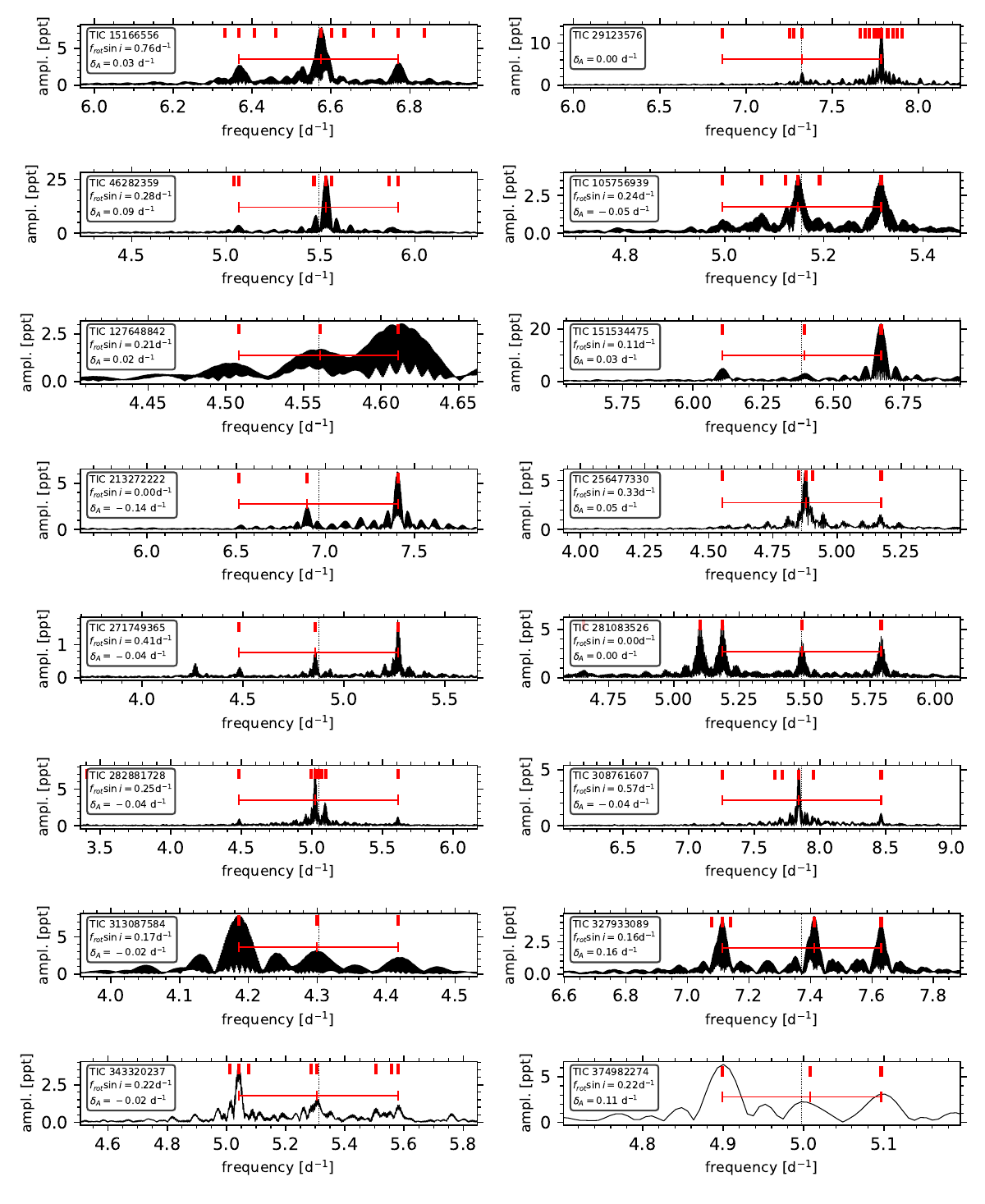}
    \caption{Splittings identified in the sample of \bceph{} stars and used in the multicolour photometry. The connected, red markers indicate the splitting and the red markers at the top show all extracted frequencies. Information on the stars can be found in the box in each panel.}
    \label{fig:train_multiplets}
\end{figure*}

\begin{figure*}\ContinuedFloat
    \centering
    \includegraphics[width=\textwidth]{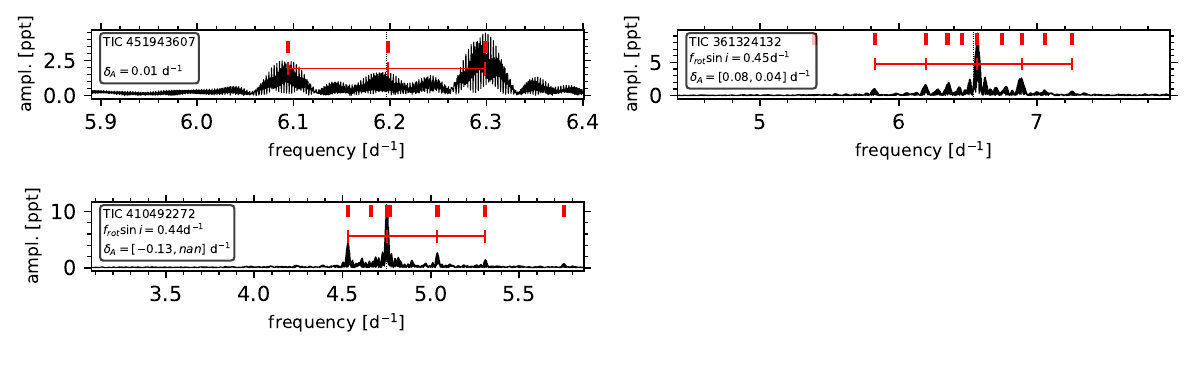}
    \caption{continued.}
\end{figure*}

\begin{figure*}
    \includegraphics[width=\textwidth]{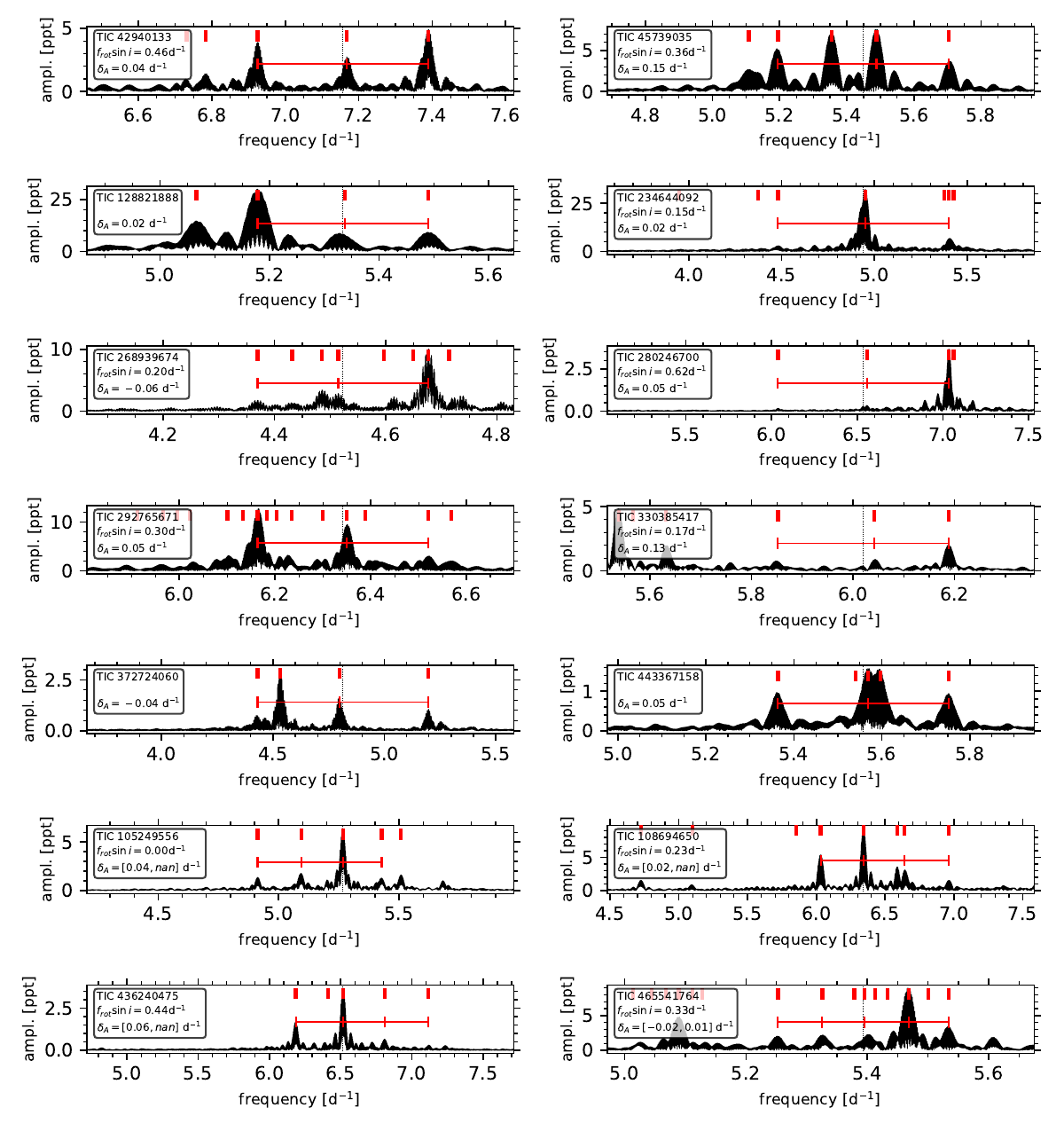}
    \caption{Additional splittings found in the sample of \bceph{} stars. The symbols are identical to Fig.~\ref{fig:train_multiplets}.}
    \label{fig:add_multiplets}
\end{figure*}

\end{appendix}

\end{document}